\documentclass[twocolumn]{aastex631}

\usepackage{savesym}
\savesymbol{tablenum}
\usepackage{siunitx}
\restoresymbol{SIX}{tablenum}

\usepackage{graphicx,epsf}
\usepackage{subfigure}
\usepackage{graphicx}	
\usepackage{amsmath}	
\usepackage{amssymb}	
\usepackage{units}
\usepackage{newtxtext,newtxmath}
\usepackage{siunitx}
\usepackage{footnote}
\usepackage{slantsc}

\newcommand{\Msun}{{\rm M}_\odot}
\newcommand{\Rsun}{{\rm R}_\odot}
\newcommand{\Lsun}{{\rm L}_\odot}
\newcommand{\kms}{\textrm{km}\,\textrm{s}^{-1}}
\newcommand{\niVI}{${}^{56}\textrm{Ni}$}
\newcommand{\coVI}{${}^{56}\textrm{Co}$}

\newcommand{\Einstein}{NASA Einstein Fellow}

\def\arcsec{\hbox{$^{\prime\prime}$}}
\def\ergs{erg\,s$^{-1}$}

\def\one{\textts{{\,\textsc{i}}}}
\def\two{\textts{ {\,\textsc{ii}}}}
\def\three{\textts{ {\,\textsc{iii}}}}
\def\four{\textts{ {\,\textsc{iv}}}}
\def\five{\textts{ {\,\textsc{v}}} }
\def\six{\textts{ {\,\textsc{vi}}} }

\def\sn{{SN~2020tlf}}

\def\msunyr{M$_{\odot}$\,yr$^{-1}$}
\newcommand{\code}[1]{\texttt{#1}}
\def\heracles{{\code{HERACLES}}}
\def\cmfgen{{\code{CMFGEN}}}

\DeclareRobustCommand{\ion}[2]{\relax\ifmmode\ifx\testbx\f@series{\mathbf{#1\,\mathsc{#2}}}\else{\mathrm{#1\,\mathsc{#2}}}\fi\else\textup{#1\,{\mdseries\textsc{#2}}}\fi}

\shorttitle{Supernova 2020\lowercase{tlf}}
\shortauthors{Jacobson-Gal\'an et al.}

\begin{document}

\title{Final Moments I: Precursor Emission, Envelope Inflation, and Enhanced Mass loss Preceding the Luminous Type II Supernova 2020tlf}

\correspondingauthor{Wynn Jacobson-Gal\'{a}n (he, him, his)}
\email{wynnjg@berkeley.edu}

\author[0000-0002-3934-2644]{W.~V.~Jacobson-Gal\'{a}n}
\affil{Department of Astronomy and Astrophysics, University of California, Berkeley, CA 94720, USA}
\affil{Center for Interdisciplinary Exploration and Research in Astrophysics (CIERA), and Department of Physics and Astronomy, Northwestern University, Evanston, IL 60208, USA}

\author[0000-0003-0599-8407]{L.~Dessart}
\affil{Institut d’Astrophysique de Paris, CNRS-Sorbonne Université, 98 bis boulevard Arago, F-75014 Paris, France}

\author[0000-0002-6230-0151]{D.~O.~Jones}
\affiliation{Department of Astronomy and Astrophysics, University of California, Santa Cruz, CA 95064, USA}
\affiliation{\Einstein}

\author[0000-0003-4768-7586]{R.~Margutti}
\affil{Department of Astronomy and Astrophysics, University of California, Berkeley, CA 94720, USA}

\author[0000-0001-5126-6237]{D.~L.~Coppejans}
\affiliation{Center for Interdisciplinary Exploration and Research in Astrophysics (CIERA), and Department of Physics and Astronomy, Northwestern University, Evanston, IL 60208, USA}

\author[0000-0001-9494-179X]{G.~Dimitriadis}
\affiliation{School of Physics, Trinity College Dublin, The University of Dublin, Dublin, Ireland}

\author[0000-0002-2445-5275]{R.~J.~Foley}
\affiliation{Department of Astronomy and Astrophysics, University of California, Santa Cruz, CA 95064, USA}

\author[0000-0002-5740-7747]{C.~D.~Kilpatrick}
\affil{Center for Interdisciplinary Exploration and Research in Astrophysics (CIERA), and Department of Physics and Astronomy, Northwestern University, Evanston, IL 60208, USA}

\author[0000-0002-4513-3849]{D.~J.~Matthews}
\affiliation{Department of Astronomy and Astrophysics, University of California, Berkeley, CA 94720, USA}

\author[0000-0002-3825-0553]{S.~Rest}
\affiliation{Department of Physics and Astronomy, The Johns Hopkins University, Baltimore, MD 21218, USA}

\author[0000-0003-0794-5982]{G.~Terreran}
\affil{Center for Interdisciplinary Exploration and Research in Astrophysics (CIERA), and Department of Physics and Astronomy, Northwestern University, Evanston, IL 60208, USA}
\affil{Las Cumbres Observatory, 6740 Cortona Dr, Suite 102, Goleta, CA 93117-5575, USA}

\author[0000-0002-6298-1663]{P.~D.~Aleo}
\affil{Department of Astronomy, University of Illinois at Urbana-Champaign, 1002 W. Green St., IL 61801, USA}
\affil{Center for Astrophysical Surveys, National Center for Supercomputing Applications, Urbana, IL, 61801, USA}

\author[0000-0002-4449-9152]{K.~Auchettl}
\affil{School of Physics, The University of Melbourne, VIC 3010, Australia}
\affil{ARC Centre of Excellence for All Sky Astrophysics in 3 Dimensions (ASTRO 3D)}
\affil{Department of Astronomy and Astrophysics, University of California, Santa Cruz, CA 95064,
USA}
\affil{DARK, Niels Bohr Institute, University of Copenhagen, Jagtvej 128, 2200 Copenhagen, Denmark}

\author[0000-0003-0526-2248]{P.~K.~Blanchard}
\affiliation{Center for Interdisciplinary Exploration and Research in Astrophysics (CIERA), and Department of Physics and Astronomy, Northwestern University, Evanston, IL 60208, USA}

\author[0000-0003-4263-2228]{D.~A.~Coulter}
\affil{Department of Astronomy and Astrophysics, University of California, Santa Cruz, CA 95064,
USA}

\author{K.~W.~Davis}
\affil{Department of Astronomy and Astrophysics, University of California, Santa Cruz, CA 95064,
USA}

\author[0000-0001-5486-2747]{T.~J.~L.~de Boer}
\affil{Institute for Astronomy, University of Hawaii, 2680 Woodlawn Drive, Honolulu, HI 96822, USA}

\author[0000-0003-4587-2366]{L.~DeMarchi}
\affiliation{Center for Interdisciplinary Exploration and Research in Astrophysics (CIERA), and Department of Physics and Astronomy, Northwestern University, Evanston, IL 60208, USA}

\author[0000-0001-7081-0082]{M.~R.~Drout}
\affiliation{David A. Dunlap Department of Astronomy and Astrophysics, University of Toronto, 50 St. George Street, Toronto, Ontario, M5S 3H4, Canada}

\author[0000-0003-1714-7415]{N.~Earl}
\affil{Department of Astronomy, University of Illinois at Urbana-Champaign, 1002 W. Green St., IL 61801, USA}

\author[0000-0003-4906-8447]{A.~Gagliano}
\affil{Department of Astronomy, University of Illinois at Urbana-Champaign, 1002 W. Green St., IL 61801, USA}
\affil{Center for Astrophysical Surveys, National Center for Supercomputing Applications, Urbana, IL, 61801, USA}

\author[0000-0002-8526-3963]{C.~Gall}
\affil{DARK, Niels Bohr Institute, University of Copenhagen, Jagtvej 128, 2200 Copenhagen, Denmark}

\author[0000-0002-4571-2306]{J.~Hjorth}
\affil{DARK, Niels Bohr Institute, University of Copenhagen, Jagtvej 128, 2200 Copenhagen, Denmark}

\author[0000-0003-1059-9603]{M.~E.~Huber}
\affil{Institute for Astronomy, University of Hawaii, 2680 Woodlawn Drive, Honolulu, HI 96822, USA}

\author[0000-0003-2405-2967]{A.~L.~Ibik}
\affiliation{David A. Dunlap Department of Astronomy and Astrophysics, University of Toronto, 50 St. George Street, Toronto, Ontario, M5S 3H4, Canada}

\author[0000-0002-0763-3885]{D. Milisavljevic}
\affil{Department of Physics and Astronomy, Purdue University, 525 Northwestern Avenue, West Lafayette, IN 47907, USA}

\author[0000-0001-8415-6720]{Y.-C.~Pan}
\affil{Graduate Institute of Astronomy, National Central University, 300 Zhongda Road, Zhongli, Taoyuan 32001, Taiwan}

\author[0000-0002-4410-5387]{A.~Rest}
\affil{Space Telescope Science Institute, Baltimore, MD 21218}
\affiliation{Department of Physics and Astronomy, The Johns Hopkins University, Baltimore, MD 21218, USA}

\author[0000-0003-1724-2885]{R.~Ridden-Harper}
\affiliation{School of Physical and Chemical Sciences | Te Kura Mat\={u}, University of Canterbury, Private Bag 4800, Christchurch 8140, New Zealand}

\author[0000-0002-7559-315X]{C.~Rojas-Bravo}
\affil{Department of Astronomy and Astrophysics, University of California, Santa Cruz, CA 95064,
USA}

\author[0000-0003-2445-3891]{M.~R.~Siebert}
\affiliation{Department of Astronomy and Astrophysics, University of California, Santa Cruz, CA 95064, USA}

\author[0000-0001-9535-3199]{K.~W.~Smith}
\affil{Astrophysics Research Centre, School of Mathematics and Physics, Queen’s University Belfast, Belfast BT7 1NN, UK}

\author[0000-0002-5748-4558]{K.~Taggart}
\affil{Department of Astronomy and Astrophysics, University of California, Santa Cruz, CA 95064,
USA}

\author[0000-0002-1481-4676]{S.~Tinyanont}
\affil{Department of Astronomy and Astrophysics, University of California, Santa Cruz, CA 95064,
USA}

\author[0000-0001-5233-6989]{Q.~Wang}
\affiliation{Department of Physics and Astronomy, The Johns Hopkins University, Baltimore, MD 21218, USA}

\author[0000-0002-0632-8897]{Y.~Zenati}
\affiliation{Department of Physics and Astronomy, The Johns Hopkins University, Baltimore, MD 21218, USA}

\begin{abstract}

We present panchromatic observations and modeling of supernova (SN) 2020tlf, the first normal Type II-P/L SN with confirmed precursor emission, as detected by the Young Supernova Experiment (YSE) transient survey. Pre-SN activity was detected in $riz-$bands at -130 days and persisted at relatively constant flux until first light. Soon after discovery, "flash" spectroscopy of SN 2020tlf revealed narrow, symmetric emission lines that resulted from the photo-ionization of circumstellar material (CSM) shedded in progenitor mass loss episodes before explosion. Surprisingly, this novel display of pre-SN emission and associated mass loss occurred in a RSG progenitor with ZAMS mass of only 10-12~$\Msun$, as inferred from nebular spectra. Modeling of the light curve and multi-epoch spectra with the non-LTE radiative transfer code CMFGEN and radiation-hydrodynamical code HERACLES suggests a dense CSM limited to $r \approx 10^{15}$~cm, and mass loss rate of $10^{-2}~\Msun$~yr$^{-1}$. The luminous light-curve plateau and persistent blue excess indicates an extended progenitor, compatible with a RSG model with $R_{\star} = 1100~\Rsun$. Limits on the shock-powered X-ray and radio luminosity are consistent with model conclusions and suggest a CSM density of $\rho < 2 \times 10^{-16}$~g cm$^{-3}$ for distances from the progenitor star of $r \approx 5 \times 10^{15}$~cm, as well as a mass loss rate of $\dot M<1.3 \times 10^{-5}\,\rm{M_{\sun}\,yr^{-1}}$ at larger distances. A promising power source for the observed precursor emission is the ejection of stellar material following energy disposition into the stellar envelope as a result of gravity waves emitted during either neon/oxygen burning or a nuclear flash from silicon combustion.

\end{abstract}

\keywords{supernovae:general --- 
supernovae: individual (SN~2020tlf) --- mass loss --- eruptions --- massive stars}

\section{Introduction} \label{sec:intro}

The behavior of massive stars in their final years of evolution is almost entirely unconstrained. However, we can probe these terminal phases of stellar evolution prior to the core-collapse of massive stars >8~$\Msun$ by understanding the composition and origin of the high-density, circumstellar material (CSM) surrounding these stars at the time of explosion \citep{smith14}. This CSM can be comprised of primordial stellar material or elements synthesized during different stages of nuclear burning, and is enriched as the progenitor star loses mass via wind and violent outbursts (\citealt{smith14} and references therein).

Early-time optical observations of young ($t<10$~days since shock breakout; SBO) Type II supernovae (SNe II) is one such probe of the final stages of stellar evolution. In the era of all-sky transient surveys, rapid (``flash'') spectroscopic observations have become a powerful tool in understanding the very nearby circumstellar environment of pre-SN progenitor systems in the final days to months before explosion (e.g., \citealt{galyam14, groh14, Khazov16, bruch21}). Obtaining spectra of young SNe~II in the hours to days following shock breakout  allows us to identify prominent emission lines in very early-time SN spectra that result from the recombination of unshocked, photo-ionized CSM. However, because the recombination timescale of ionized H-rich CSM is inversely related to the number density of free  electrons $t_{\rm rec}\propto n_e^{-1}$ \citep{osterbrock06}, ``flash'' ionization from radiation associated with SBO is not responsible for the persistence of these narrow ($v_w \lesssim 500~\kms$), CSM-derived spectral features at $\gtrsim$~1~day after explosion (e.g., $t_{\rm rec}\le$ a few hours for H-rich gas with $T \approx$~$10^{5}-10^{6}$ K and $n_e\ge10^{8}\,\rm{cm^{-3}}$). The conversion of shock kinetic energy into high-energy radiation as it advances into the CSM provides a persistent source of ionizing photons that keep the CSM ionized for significantly longer timescales (e.g., $>>t_{\rm rec}$). The prominent, rapidly fading emission lines in the photo-ionization spectra of young SNe~II are direct evidence of dense and confined CSM surrounding the progenitor star, comprised of elements ejected during episodes of enhanced mass loss days-to-months before explosion. The strength/brightness of these features is derived from the CSM density and chemical abundances at the time  of explosion. This is a direct tracer of the progenitor’s chemical composition (CNO abundances specifically) and recent mass loss at small distances $r <10^{15}$~cm, as well as an indirect probe of progenitor identity.

Combining early-time spectroscopy with non-Local Thermal Equilibrium (non-LTE) radiative transfer modeling codes such as \cmfgen\ \citep{Hillier96} have been a successful tool in constraining the progenitor systems responsible for a growing number of supernovae that undergo a relatively flat (Type II-P) or linear (Type II-L) fading during the photospheric phase in their optical light curve evolution. The latter may be the result of massive star progenitors that have lost more of their H-rich envelope in episodes of enhanced mass loss \citep{Hillier19}. For such objects, radiative transfer modeling indicates that a dense ($\dot{M} = 10^{-4} - 10^{-2}$~$\Msun$ yr$^{-1}$; $v_w \sim 100-200 ~ \kms$) and compact ($r \lesssim 10^{15}$~cm) CSM is present in order to produce the observed spectral profiles of high-ionization species such as \ion{He}{ii}, \ion{N}{iii}, \ion{C}{iii/iv}, or \ion{O}{iv/v} in the early-time SNe~II spectra \citep{shivvers15, terreran16, Dessart16, dessart17, yaron17, Boian20, Tartaglia21, terreran21}. However, mass loss rates derived from SN spectral modeling are much larger than the generally inferred steady-state mass loss rates (e.g., $\lesssim 10^{-6}$~$\Msun$ yr$^{-1}$; \citealt{Beasor20}) observed in galactic, quiescent Red Supergiants (RSGs), which are considered the likely stellar type responsible for SNe~II \citep{smartt09}. In extreme cases, some RSGs, such as VY Canis Majoris, are estimated to be losing mass at enhanced rates of $\sim 10^{-3}$~$\Msun$ yr$^{-1}$ \citep{smith09}, which could match some lower mass loss estimates derived from \cmfgen\ modeling. However, VY CMa is more massive ($\sim25-30~\Msun$) than typical SN~II RSG progenitors and contains a much more extended CSM ($\sim 2\times 10^{16}$~cm). Overall, this deviation between theory and observation suggests that some RSGs must undergo enhanced mass loss in the final years before core-collapse. Furthermore, the identification and modeling of photo-ionization features in other objects such as Type IIb SN~2013cu \citep{galyam14}, Calcium-strong SN~2019ehk \citep{jacobson-galan20}, Type Ibn SN~2010al \citep{Pastorello15}, and electron-capture SN candidate 2018zd \citep{Hiramatsu21} represents a burgeoning technique for constraining the progenitor properties in a variety of SN sub-types beyond normal SNe~II. 

Indirect evidence of enhanced mass loss in SNe~II progenitors is also shown through the non-LTE modeling of multi-band and bolometric SN optical light curves. Based on recent studies, the presence of dense, confined CSM around a RSG progenitor at the time of explosion manifests in a few key light curve properties. First, SBO into dense CSM can produce a longer-lasting, and thus potentially easier to observe, as well as more luminous SBO signature, peaking in UV bands of the spectral energy distribution \citep{Chevalier11, Moriya11, Haynie21}. Modeling of early-time SNe~II light curves also revealed the need for local CSM ($r \lesssim 10^{15}$~cm) in order to reproduce the rapid rise time and brighter emission at peak observed in some objects \citep{dessart17, Moriya17, Morozova17, Morozova18} as well as the long plateau duration, delayed  photometric decline rate, and \ion{H}{i} line profile morphology \citep{Hillier19}. 

An additional observational probe of stellar behavior in the late-stage evolution of core-collapse SN progenitors is the detection of precursor emission prior to the terminal explosion. Optical flux has been observed as the precursors to a number of Type IIn supernovae (e.g., SN~2009ip, PTF~10bjb, SN~2010mc, PTF~10weh, SN~2011ht, PTF~12cxj, LSQ13zm, iPTF13z, SN~2016bdu, SN~2018cnf; \citealt{ofek13, ofek14, tartaglia16, Nyholm17, Pastorello18, Pastorello19}), which show persistent spectral signatures of CSM interaction for all of their evolution, as well as H-poor, interacting Type Ibn supernovae (SNe Ibn) \citep{Pastorello07, foley07}. The months-long, pre-SN flux observed in such SNe is typically found in the range of $M \approx -13 ~ {\rm  to} ~ -17$~mag and can occur anywhere from years to days prior to explosion. These eruptive events can also repeat in the years before explosion (e.g., SN~2009ip; e.g., \citealt{Mauerhan13, Pastorello13, ofek13a, margutti14}) or be one-time events, some of which are sustained for hundreds of days before core-collapse. In a recent sample study of precursor emission in ZTF-discovered SNe, \cite{Strotjohann21} found that $\sim$25\% of SNe~IIn have detectable pre-SN flux for $\sim$months prior to explosion associated with the ejection of $\sim$~1~$\Msun$ of material into the local progenitor environment. Unfortunately, no SNe~II with photo-ionization spectra were detected in their search for precursor emission from massive star progenitors.

In recent years, there have been a number of theoretical explanations put forth to explain eruptive or heightened mass loss in core-collapse SN progenitors that could then be responsible for detectable precursor emission and/or photo-ionization features in early-time spectra. Enhanced mass loss observed in these progenitor stars cannot be explained by line-driven winds and thus more exotic scenarios are needed to drive off a considerable amount of material from the stellar surface. In lower mass RSGs ($\sim$~8-12~$\Msun$), it is possible that nuclear flashes that ignite dynamical burning of oxygen, neon or silicon could lead to the ejection of outer layers of the stellar envelope in the final years to months before explosion \citep{Woosley80, meakin07, arnett09, dessart10, woosley15}. Alternatively, late-stage burning phases can induce gravity waves that propagate outwards and inject energy into the stellar envelope, leading to eruptions of $\sim 1 \Msun$ worth of material in the final months before explosion \citep{Quataert12, Shiode14, fuller17, Wu21}. Additionally, super-Eddington continuum-driven winds can be induced at the stellar surface during late-stage nuclear burning, which can then cause enhanced mass loss and detectable pre-SN emission \citep{shaviv01a, shaviv01b, ofek16}. However, this mechanism is unlikely to be present in RSGs and is more suited to super-massive ($M_{\rm ZAMS} \gtrsim 30~\Msun$) Luminous Blue Variable (LBV) stars.

\begin{figure}[t!]
\centering
\includegraphics[width=0.49\textwidth]{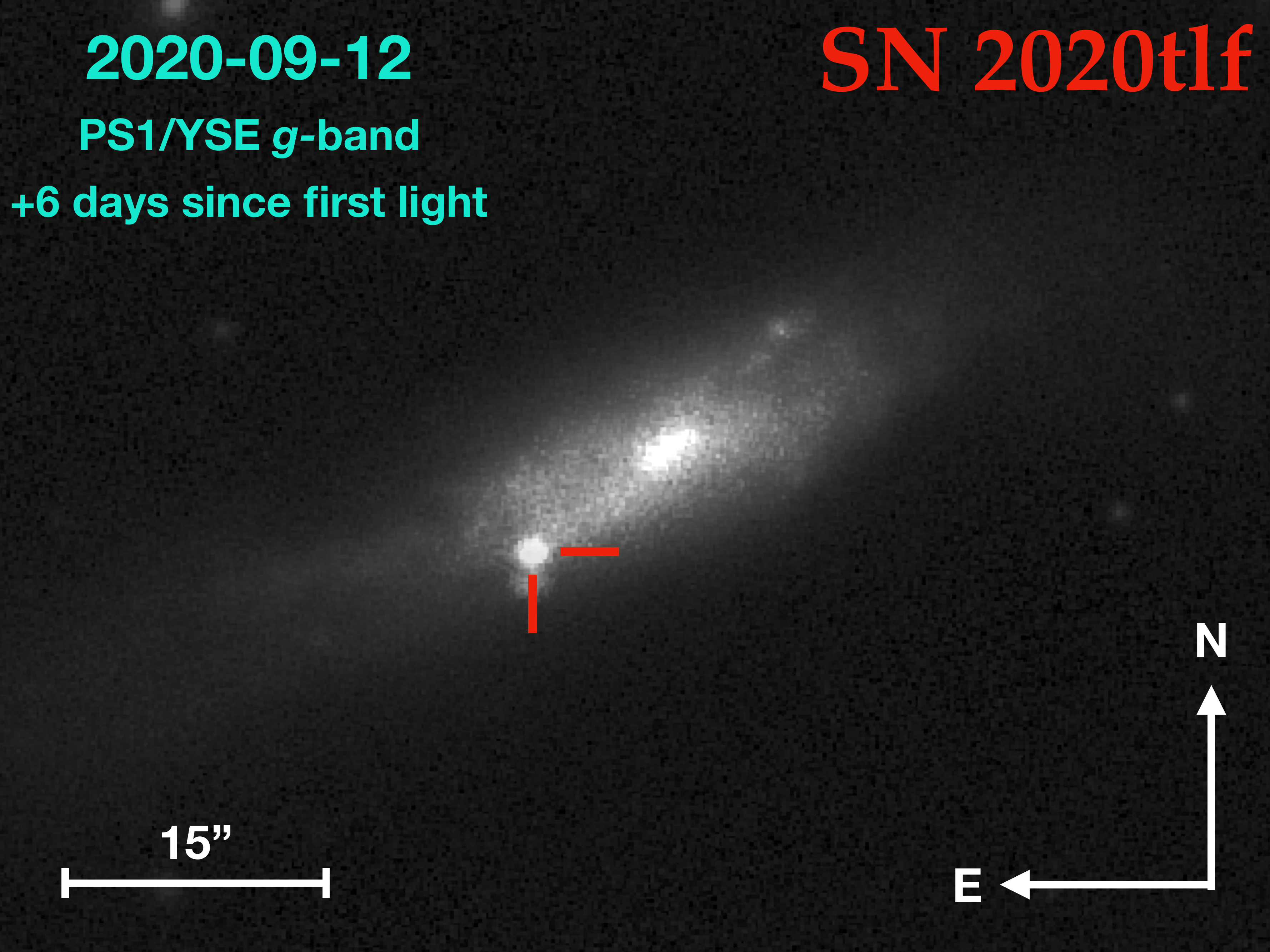}
\caption{PS1/YSE $g-$band explosion image of Type II SN~2020tlf in host galaxy NGC~5731. \label{fig:sn_image}}
\end{figure}

In this paper we present, analyze, and model multi-wavelength observations (X-ray to radio) of the Type II \sn{} (shown in Figure \ref{fig:sn_image}), discovered by the Asteroid Terrestrial-impact Last Alert System (ATLAS) on 16 Sept. 2020 (MJD 59108.72) in the $c-$band filter \citep{tonry20}. \sn{} has an ATLAS discovery apparent magnitude of 15.89~mag and is located at $\alpha = 14^{\textrm{h}}40^{\textrm{m}}10.03^{\textrm{s}}$, $\delta = +42^{\circ}46'39.45^{\prime \prime}$. As shown in \S\ref{sec:preSNobs}, the Pan-STARRS1 (PS1) telescope detected significant pre-explosion flux for $\sim130$~days prior to the discovery date reported above by ATLAS. We define the time of first light as the phase at which the observed magnitudes increased beyond the threshold of the pre-explosion PS1 detections. This results in a time of first light of MJD $59098.7 \pm 1.5$~days (06 Sept. 2020).

\sn{} was classified as a young SN~IIn with ``flash-ionization'' spectral features by \cite{Dimitriadis20} and \cite{Balcon20} on 17 Sept. 2020. Following its classification, \sn{} became sun-constrained for ground-based observatories. Once visible again at +95~days since first light, spectroscopic observations of \sn{} revealed that the narrow, photo-ionized emission features had disappeared (unlike typical SNe~IIn) and the SN had evolved into a normal Type II-like object. 

\sn{} is located 9.3\arcsec\:east and 6.9\arcsec\:south of the nucleus of the SABcd galaxy NGC~5731. In this paper, we use a redshift $z = 0.008463 \pm 0.0003$ \citep{Oosterloo93}, which corresponds to a distance of $36.8 \pm 1.29$~Mpc for standard $\Lambda$CDM cosmology ($H_{0}$ = 70 km s$^{-1}$ Mpc$^{-1}$, $\Omega_M = 0.27$, $\Omega_{\Lambda} = 0.73$); unfortunately no redshift-independent distance is available. Possible uncertainties on the distance could be the choice of $H_{0}$ and/or peculiar velocities of the host galaxy, the uncertainty on the former can, for example, contribute to $\lesssim 5$\% uncertainty of the SN luminosity. The main parameters of \sn{} and its host-galaxy are displayed in Table \ref{tbl:params}. This paper represents the first installment in a series of studies that will focus on constraining the ``final moments'' of massive star evolution through the derivation of progenitor properties from precursor activity and ``flash'' spectroscopy. 

\section{Pre-Explosion Observations} \label{sec:preSNobs}

\subsection{Young Supernova Experiment Observations}
\label{subsec:YSE}

SN~2020tlf was first reported to the Transient Name Server by ATLAS \citep{Tonry18} on 16 Sept. 2020, but the earliest detections of the SN are from the Young Supernova Experiment \citep[YSE;][]{Jones2021} with the PS1 telescope \citep{Kaiser2002} on 5 Sept. 2020. YSE began monitoring the field in which SN~2020tlf was discovered on 18 Jan. 2020.

\begin{figure*}
\centering
\includegraphics[width=\textwidth]{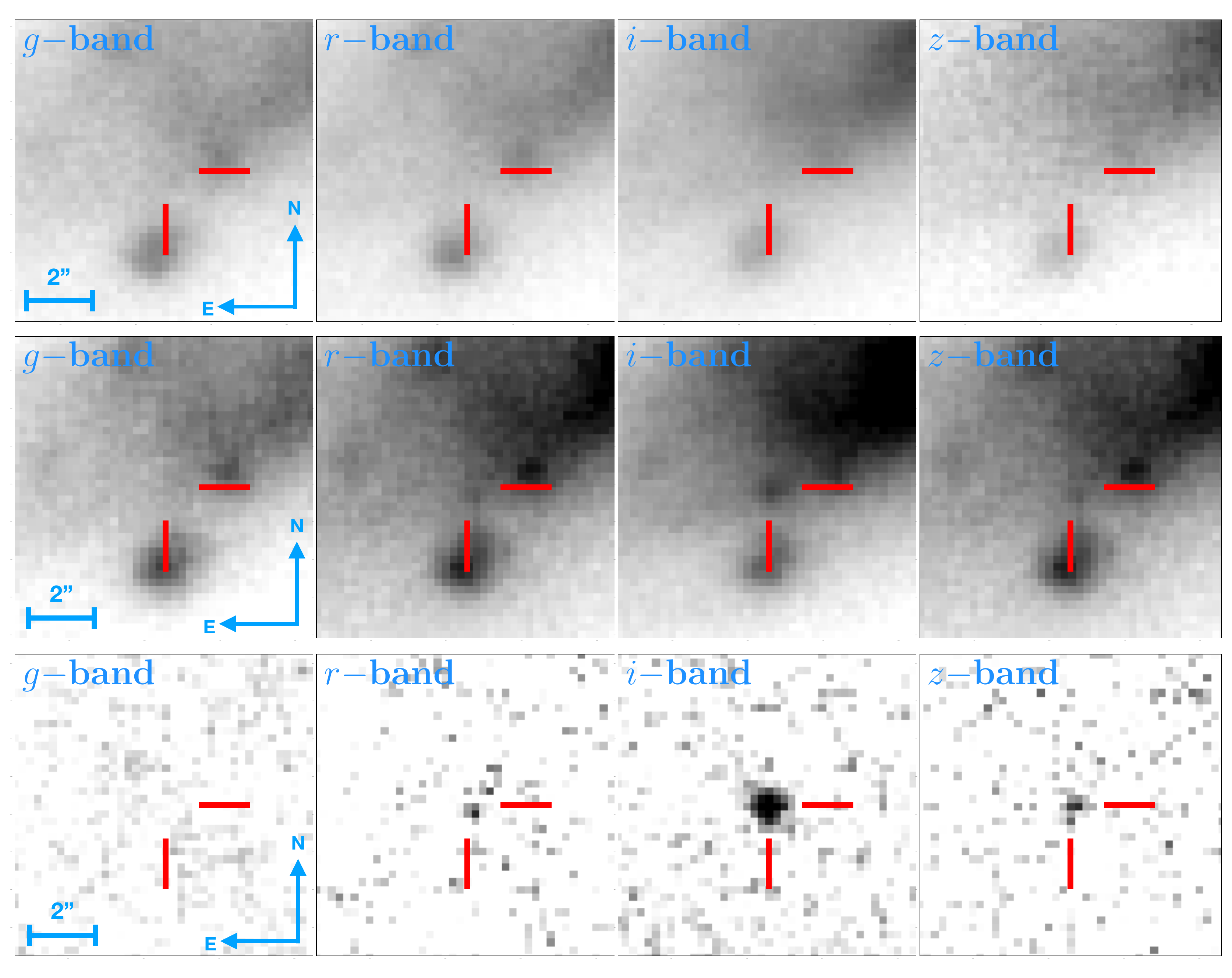}
\caption{Pre-explosion PS1/YSE stacked $griz-$band template (top), detection (middle), and difference (bottom) images of progenitor precursor emission preceding SN~2020tlf. Stacked images were created from 13 $z-$band, 45 $i-$band, 23 $r-$band and 22 $g-$band pre-explosion observations spanning a phase range of $\delta t = -169.7 \ \rm{to} \ -3.7$~days since first light (MJD 58929-59095). PS1 $g-$band is not detected. \label{fig:preSN_stack} }
\end{figure*}

YSE data is initially processed by the Image Processing Pipeline (IPP), described in \citet{magnier13}, including difference imaging and photometry. Those data are passed to the Transient Science Server \citep{Smith20}, where catalog cross-matching and machine learning tools are used to identify potential transients in each image. The YSE team performs manual vetting of potential transients to remove artifacts, asteroids, and other contaminating sources, and finally sends new transient discoveries and initial photometric epochs to the Transient Name Server for followup by the community. We then load the transient data into YSE's transient management system, ``YSE-PZ", which allows us to view Pan-STARRS data with that of other ongoing surveys and schedule follow-up observations.  Further detail on this procedure is given in \citet{Jones2021} and references therein.

This process allows for identification and follow-up of fast-rising transients. For SN~2020tlf, we re-measured the pre-explosion photometry using {\tt Photpipe} \citep{Rest+05} to ensure highly accurate photometric measurements that took into account pixel-to-pixel correlations in the difference images and host galaxy noise at the SN location. {\tt Photpipe} is a well-tested pipeline for measuring SN photometry and has been used to perform accurate measurements from Pan-STARRS in a number of previous studies (e.g., \citealt{Rest14,Foley18,Jones18,Scolnic18,Jones19}). In brief, {\tt Photpipe} takes as input IPP images that have been re-sampled and astrometrically aligned to match skycells in the PS1 sky tessellation and measures their zeropoints by using {\tt DoPhot} \citep{Schechter+93} to measure the photometry of stars in the image and comparing to stars in the PS1 DR2 catalog \citep{Flewelling+16}. Then, {\tt Photpipe} convolves a template image from the PS1 3$\pi$ survey \citep{Chambers2017} with data taken between the years 2010 and 2014, using a kernel that consists of three superimposed Gaussian functions, to match the point spread function (PSF) of the survey image and subtracts the template from the image.  Finally, {\tt Photpipe} uses {\tt DoPhot} again to measure fixed-position photometry of the SN at the weighted average of its location across all images.  Further details regarding this procedure are given in \cite{Rest14} and \cite{Jones19}.

To account for the bright host galaxy of SN~2020tlf, which could cause larger-than-expected pre-explosion photometric noise in the difference image \citep{Kessler15,Doctor17,Jones17}, we estimate the noise in the photometry by adding the Poisson noise at the SN location in quadrature to the standard deviation of fluxes measured in random difference-image apertures at coordinates with no pre-SN (or SN) light but approximately the same underlying host galaxy surface brightness as exists at the SN location. These apertures are placed in an annulus at the same elliptical radius from the center as SN~2020tlf to ensure similar surface brightness to the SN location.  We find that the SN host galaxy does not contribute significantly to the uncertainty in the photometry ($\lesssim 15\%$ of the total error budget). We can also rule out contributions from a possible Active Galactic Nucleus (AGN) in NGC~5731 to fluxes at the SN location given the significant offset of \sn{} from host center.

\begin{table}[t!]
\begin{center}
\caption{Main parameters of SN\,2020tlf and its host galaxy \label{tbl:params}}
\vskip0.1in
\begin{tabular}{lccc}
\hline
\hline
Host Galaxy &  &  NGC~5731 \\ 
Galaxy Type &  &  SAcd\footnote{\cite{deV91}} \\
Host Galaxy Offset &  &    $11.6\arcsec (2.10 ~ \rm kpc)$ \\
Redshift &  &  $0.008463 \pm 0.0003$\footnote{\cite{Oosterloo93}}\\  
Distance &  &  $36.8 \pm 1.29$~Mpc\\ 
Distance Modulus, $\mu$ &  &  $32.83 \pm 0.10$~mag\\ 
$\textrm{RA}_{\textrm{SN}}$ &  &   $14^{\textrm{h}}40^{\textrm{m}}10.03^{\textrm{s}}$\\
$\textrm{Dec}_{\textrm{SN}}$ &  &  $+42^{\circ}46'39.45^{\prime \prime}$\\
Time of First Light (MJD) &  &  59098.7 $\pm$ 1.5\\
Time of $B-$band Maximum (MJD) &  & 59117.6$\pm$0.2\\
$E(B-V)_{\textrm{MW}}$ &  &  0.014 $\pm$ 0.001~mag\footnote{\cite{schlegel98,schlafly11}}\\
$E(B-V)_{\textrm{host}}$ &  &  0.018 $\pm$ 0.010~mag\\
$m_{B}^{\mathrm{peak}}$ &  &  $14.5 \pm 0.0440$~mag\\
$M_{B}^{\mathrm{peak}}$ &  &  $-18.5 \pm 0.0440$~mag\\
\hline
\end{tabular}
\end{center}
\label{table:Observations}
\tablecomments{Extinction corrections have only been applied to the presented apparent magnitudes, not the absolute magnitudes.}
\end{table}

\begin{figure*}
\centering
\subfigure[]{\includegraphics[width=0.49\textwidth]{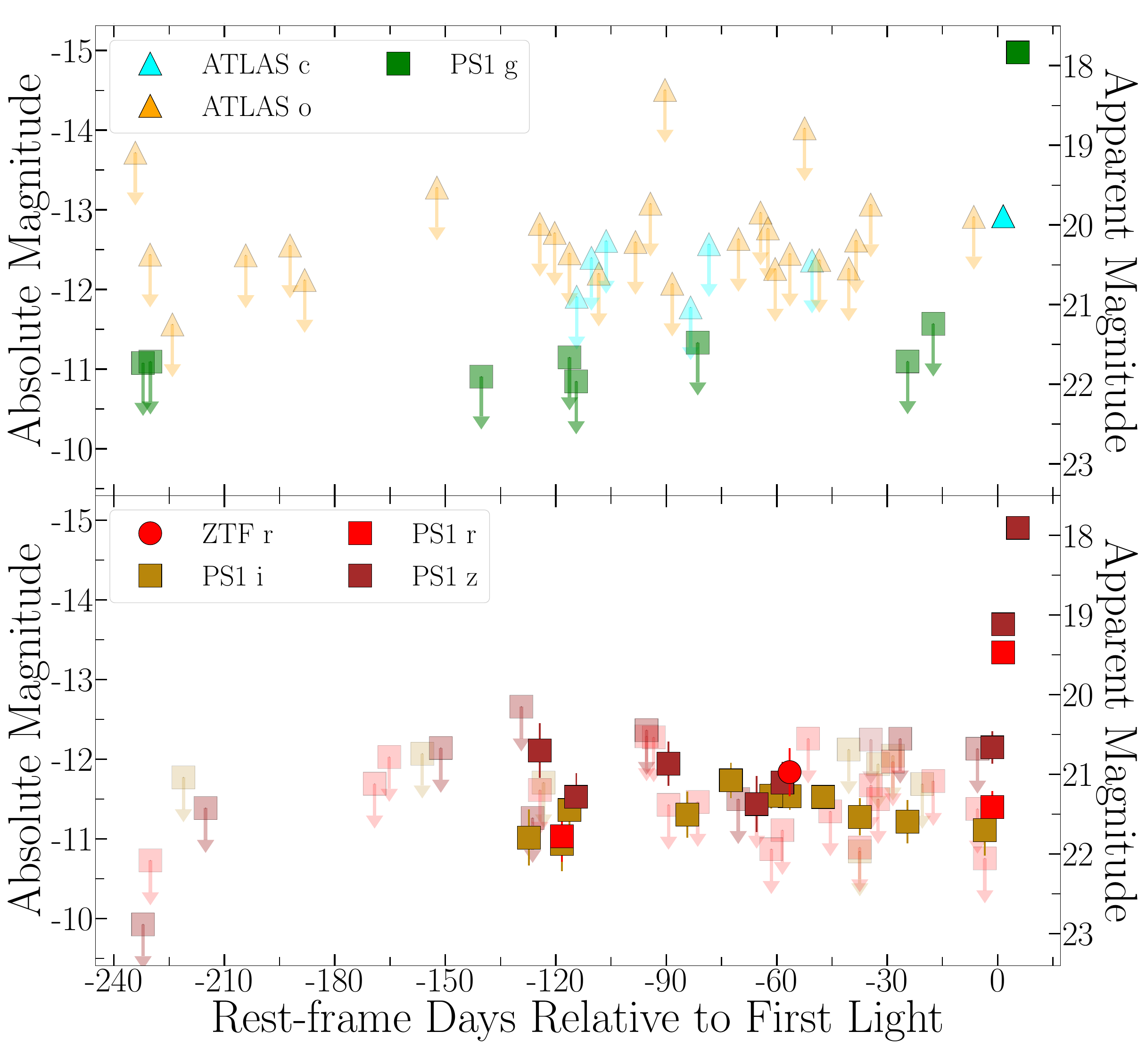}}
\subfigure[]{\includegraphics[width=0.49\textwidth]{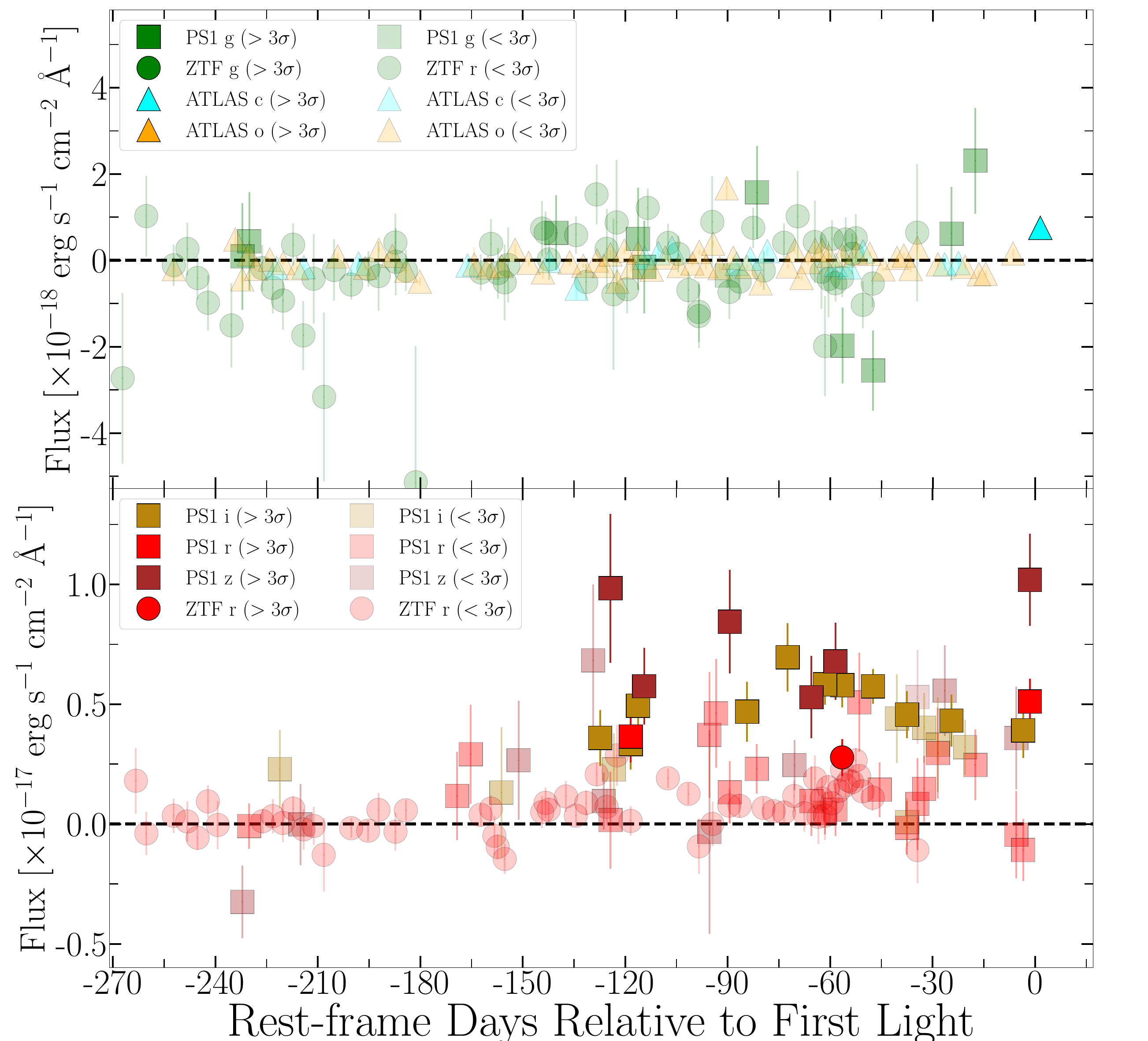}}
\caption{(a)/(b) Pre-explosion $c/o-$band ATLAS (triangles), $r-$band ZTF (circles) and $riz-$band PS1 (squares) light curves; magnitudes presented to the left, apparent fluxes presented to the right. 3$\sigma$ PS1 $riz-$band detections shown in bottom panel for $\sim130$~days before first light. \label{fig:preSN_LCphot} }
\end{figure*}

Based on the above data reduction, we find evidence for a statistically significant (>3$\sigma$) pre-explosion flux excess at the SN location ($m \approx 20.7 - 21.9$~mag) in $riz$-bands from MJD 58971.42 -- 59097.24 ($\delta t = -127.3 ~ {\rm to} ~ -1.49$~days before first light). However, we find no evidence for similar pre-explosion emission in the YSE $g-$band images from $\delta t = -232.1 ~ {\rm to} ~ -17.49$~days before first light. We present the pre-explosion $griz$-band stacked PS1 images in Figure \ref{fig:preSN_stack} over the phase range of $\delta t = -169.7 \ {\rm to} \ -3.7$~days before first light (MJD 58929-59095). The multi-band, pre-explosion PS1 light curve is displayed in Figure \ref{fig:preSN_LCphot}. Furthermore, there is no evidence for significant flux in earlier pre-explosion PS1 3$\pi$ survey imaging of the SN site from 28 Feb. 2011 to 21 Feb. 2014 ($\delta t = -3478 ~ {\rm to} ~ -2389$~days before first light). For PS1 $griyz$-bands, we derived 3$\sigma$ upper limits over this pre-explosion phase range of $>22.24$, $>22.28$, $>22.02$, $>21.50$, and $>21.75$~mag, respectively.


\subsection{Additional Pre-Explosion Observations}
\label{sec:preSN_ZTF_ATLAS}

Pre-explosion imaging of \sn{} was also acquired by the Zwicky Transient Facility (ZTF; \citealt{bellm19,graham19}) and ATLAS \citep[][]{2018PASP..130f4505T}. ZTF $g/r-$band photometry was obtained through the ZTF forced-photometry service \citep{Masci19} and covers a phase range of $\delta t = -900.4 \ {\rm to} -34.5$~days before first light. We follow the procedure outlined in the ZTF forced-photometry manual to apply a signal-to-noise threshold (SNT) of 3 to the data i.e., all photometry with SNR > 3 are considered >3$\sigma$ detections. After the SNT is applied, we find evidence for tentative pre-explosion ZTF $r-$band flux ($m \approx 21.2$~mag) ranging from $\delta t = -128.4 \ {\rm to} \ -51.50$~days since first light. To further test the validity of these ``detections,'' we downloaded the public difference image pre-explosion data from the Infrared Processing and Analysis Center (IPAC)\footnote{https://irsa.ipac.caltech.edu/applications/ztf} and performed the same random background aperture analysis on the images as discussed in \S\ref{subsec:YSE}. We find evidence for $>3\sigma$ emission in only one epoch of $r-$band ZTF data at a phase $\delta t = -56.5$~days prior to first light. This ZTF $r-$band detection is consistent with the PS1 detections and is presented in the pre-explosion light curve plot (Fig. \ref{fig:preSN_LCphot}a). Additionally, there is no evidence for detectable emission of pre-explosion flux in the ZTF $g-$band images ($m \geq 20.7$~mag).


Furthermore, we do not find evidence for significant emission in $c/o-$band ATLAS pre-explosion photometry during the phase range of $\delta t = -1714.1 \ {\rm to} -6.5$~days since first light. Similar to the YSE/PS1 pre-explosion image analysis described above, we model the background noise by placing random apertures near the explosion site and performing aperture photometry of these regions. The flux is then recorded in each these random background apertures for each pre-explosion epoch and used to create background light curves i.e., control light curves. To attempt and meature significant pre-SN flux detection at the location of SN 2020tlf, we apply several cuts on the total number of individual as well as averaged data in order to remove bad measurements. Our first cut uses the $\chi^2$ and uncertainty values of the PSF fitting to clean out bad data. We then obtain forced photometry of 8 control light curves located in a circular pattern around the location of the SN with a radius of 17\arcsec. The flux of these control light curves is expected be consistent with zero within the uncertainties, and any deviation from that would indicate that there are either unaccounted systematics or underestimated uncertainties.

We search for such deviations by calculating the 3$\sigma$ cut weighted mean of the set of control light curve measurements for a given epoch (for a more detailed discussion see Rest et al, in prep.). This weighted mean of these photometric measurements is expected to be consistent with zero and, if not, we flag and remove those epochs from the pre-SN light curve. This method allows us to identify potentially bad measurements in the SN light curve \emph{without} using the SN light curve itself. We then bin the \sn{} light curve by calculating a 3$\sigma$ cut weighted mean for each night (typically, ATLAS has 4 epochs per night), excluding the flagged measurements from the previous step. We find that this method successfully removes bad measurements that can mimic pre-SN emission (Rest et al., in prep.). We then calculate the rolling sum of the S/N with a Gaussian kernel of 30 days for the pre-SN and the control light curves and identify any significant flux excess in the rolling sum. The kernel size of 30 days is chosen to maximize the detection of pre-SN emission with similar time scales. We use the peaks in the control light curves as our empirical detection limit: since there is no transient in the control light curves (barring an extremely unlikely coincidence with a transient unrelated to pre-SN emission at the location of SN~2020tlf), any peaks in the control light curves are false positives. We choose as our conservative detection limit a rolling sum value of 20, and we find no evidence of pre-SN activity in \sn{} down to a magnitude limit of $m \gtrsim 20.3$~mag, which is consistent with PS1 and ZTF detections.

\section{Post-Explosion Observations} \label{sec:postSNobs}

\subsection{UV/Optical photometry}\label{SubSec:Phot}

\begin{figure*}
\centering
\includegraphics[width=\textwidth]{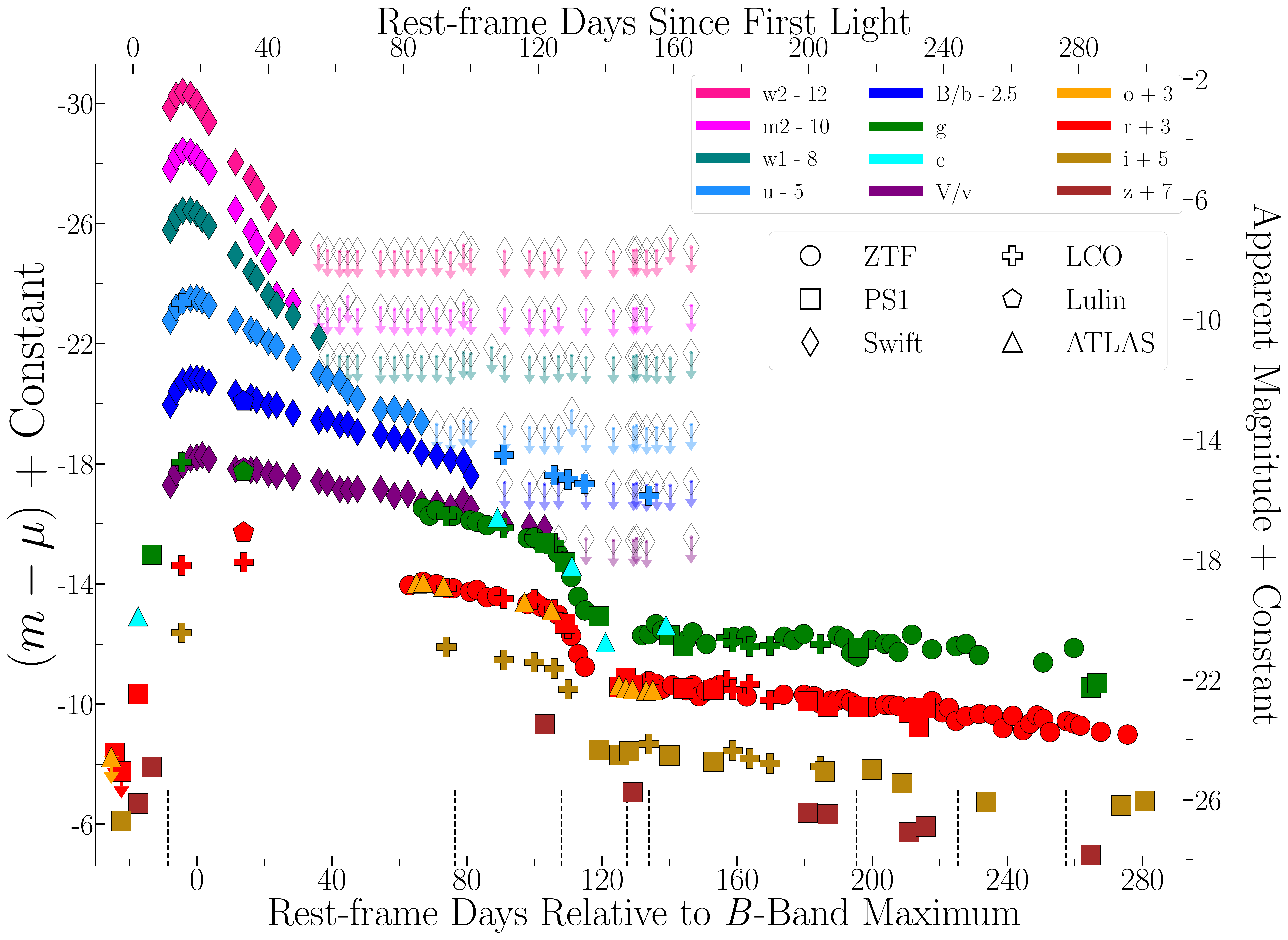}
\caption{UV/Optical/NIR light curve of \sn{} with respect to $B$-band maximum (bottom axis) and time since first light (top axis). Observed photometry presented in AB magnitude system and have not been corrected for any extinction. ATLAS data/3$\sigma$ upper limits are presented as triangles, PS1/YSE as squares, Las Cumbres Observatory (LCO) as plus signs, {\it Swift} as diamonds, ZTF as circles, Lulin observatory as pentagons. The epochs of our spectroscopic observations are marked by vertical black dashed lines. \label{fig:optical_LC} }
\end{figure*}

We started observing \sn{} with the Ultraviolet Optical Telescope (UVOT; \citealt{Roming05}) onboard the Neil Gehrels \emph{Swift} Observatory \citep{Gehrels04} on 9 Sept. 2020 until 18 Feb. 2021 ($\delta t=$ 11.0 -- 165.2 days since first light). We performed aperture photometry with a 5$\arcsec$ region with \texttt{uvotsource} within HEAsoft v6.26\footnote{We used the calibration database (CALDB) version 20201008.}, following the standard guidelines from \cite{Brown14}. In order to remove contamination from the host galaxy, we employed images acquired at $t\approx165$~days after first light, assuming that the SN contribution is negligible at this phase. This is supported by visual inspection in which we found no flux associated with \sn{}. We subtracted the measured count rate at the location of the SN from the count rates in the SN images following the prescriptions of \cite{Brown14}. We detect bright UV emission from the SN near optical peak (Figure \ref{fig:optical_LC}) until $t\approx60$ days after explosion. Subsequent non-detections in $w1, m2, w2$ bands indicate significant cooling of the photosphere and/or Fe-group line blanketing. 

Additional $griz$-band imaging of \sn{} was obtained through the Young Supernova Experiment (YSE) sky survey \citep{Jones2021} with the Pan-STARRS telescope \citep[PS1;][]{Kaiser2002} between 08 Sept. 2020 and 26 June 2021 ($\delta t= 1.5-292.3$~days since first light). The YSE photometric pipeline is based on {\tt photpipe} \citep{Rest+05}. Each image template was taken from stacked PS1 exposures, with most of the input data from the PS1 3$\pi$ survey. All images and templates are resampled and astrometrically aligned to match a skycell in the PS1 sky tessellation. An image zero-point is determined by comparing PSF photometry of the stars to updated stellar catalogs of PS1 observations \citep{Chambers2017}. The PS1 templates are convolved with a three-Gaussian kernel to match the PSF of the nightly images, and the convolved templates are subtracted from the nightly images with {\tt HOTPANTS} \citep{becker15}. Finally, a flux-weighted centroid is found for each SN position and PSF photometry is performed using ``forced photometry": the centroid of the PSF is forced to be at the SN position. The nightly zero-point is applied to the photometry to determine the brightness of the SN for that epoch.

\sn{} was observed with ATLAS ($\delta t = -9.40 - 157.8$~days since first light), a twin 0.5m telescope system installed on Haleakala and Mauna Loa in the Hawai'ian islands that robotically surveys the sky in cyan (\textit{c}) and orange (\textit{o}) filters \citep[][]{2018PASP..130f4505T}. The survey images are processed as described in \cite{2018PASP..130f4505T} and photometrically and astrometrically calibrated immediately \citep[using the RefCat2 catalogue;][]{2018ApJ...867..105T}. Template generation, image subtraction procedures and identification of transient objects are described
in \cite{Smith20}. Point-spread-function photometry is carried out on the difference images and all sources greater than 5$\sigma$ are recorded and all sources go through an automatic validation process that removes spurious objects \citep{Smith20}. Photometry on the difference images (both forced and non-forced) is from automated point-spread-function fitting as documented in \cite{2018PASP..130f4505T}. The photometry presented here are weighted averages of the nightly individual 30\,sec exposures, carried out with forced photometry at the position of \sn{}.

We observed \sn{} with the Las Cumbres Observatory Global Telescope Network 1-m telescopes and Las Cumbres Observatory imagers from 21 Sept 2020 to 29 March 2021 ($\delta t = 14.34-203.5$~days since first light) in $ugri$-bands.  We downloaded the calibrated BANZAI \citep{mccully18} frames from the Las Cumbres archive and re-aligned them using the command-line blind astrometry tool {\tt solve-field} \citep{lang10}.  Using the {\tt photpipe} imaging and photometry package \citep{Rest+05,Kilpatrick18}, we regridded each Las Cumbres Observatory frame with {\tt SWarp} \citep{swarp} to a common pixel scale of 0.389\arcsec\ centered on the location of \sn{}.  We then performed photometry on these frames with {\tt DoPhot} \citep{Schechter+93} and calibrated each frame using PS1 DR2 standard stars observed in the same field as \sn{} in $ugri$ bands \citep{Flewelling+16}.

Observations of \sn{} were obtained with the 1-m Lulin telescope located at Lulin Observatory on 09 Oct. 2020 ($\delta t = 32.71$~days since first light) in $BVgr$ bands.  The individual frames were corrected for bias and flat-fielded using calibration frames obtained on the same night and in the same instrumental configuration.  Within {\tt photpipe}, we solved for the astrometric solution in each frame using 2MASS astrometric standards \citep{2MASS} observed in the same field as \sn{}.  Finally, we performed photometry in each frame following the same procedures for Las Cumbres Observatory described above.

For both Las Cumbres Observatory and Lulin photometry, we re-processed the final light curve by calculating the mean astrometric position of \sn{} in all Las Cumbres Observatory and Lulin frames separately.  We then performed forced photometry using a custom version of {\tt DoPhot} at this position using the PSF parameters in each individual frame and solving only for the flux of \sn{} at the time.

The complete light curve of \sn{} is presented in Figure \ref{fig:optical_LC} and all photometric observations are listed in Appendix Table \ref{tbl:phot_table_s}. In addition to our observations, we include $g/r-$band photometry from the Zwicky Transient Facility (ZTF; \citealt{bellm19, graham19}) forced-photometry service \citep{Masci19}, which span from 27 Nov. 2020 to 28 June 2021 ($\delta t= 81.81-294.5$~days since first light).

The Milky Way (MW) $V$-band extinction and color excess along the SN line of site is $A_{V} = 0.043$~mag and \textit{E(B-V)} = 0.014~mag \citep{schlegel98, schlafly11}, respectively, which we correct for using a standard \cite{fitzpatrick99} reddening law (\textit{$R_V$} = 3.1). In addition to MW color excess, we estimate the contribution of galaxy extinction in the local SN environment. We use Equation 9 in \cite{Poznanski12} to convert the \ion{Na}{i} equivalent width (EW) of $0.10 \pm 0.010$~\AA \ in the first \sn{} spectrum to an intrinsic \textit{E(B-V)} and find a host galaxy extinction of $E(B-V)_{\textrm{host}} = 0.018 \pm 0.003$~mag, also corrected for using the \cite{fitzpatrick99} reddening law.

\begin{figure*}
\centering
\subfigure[]{\includegraphics[width=0.49\textwidth]{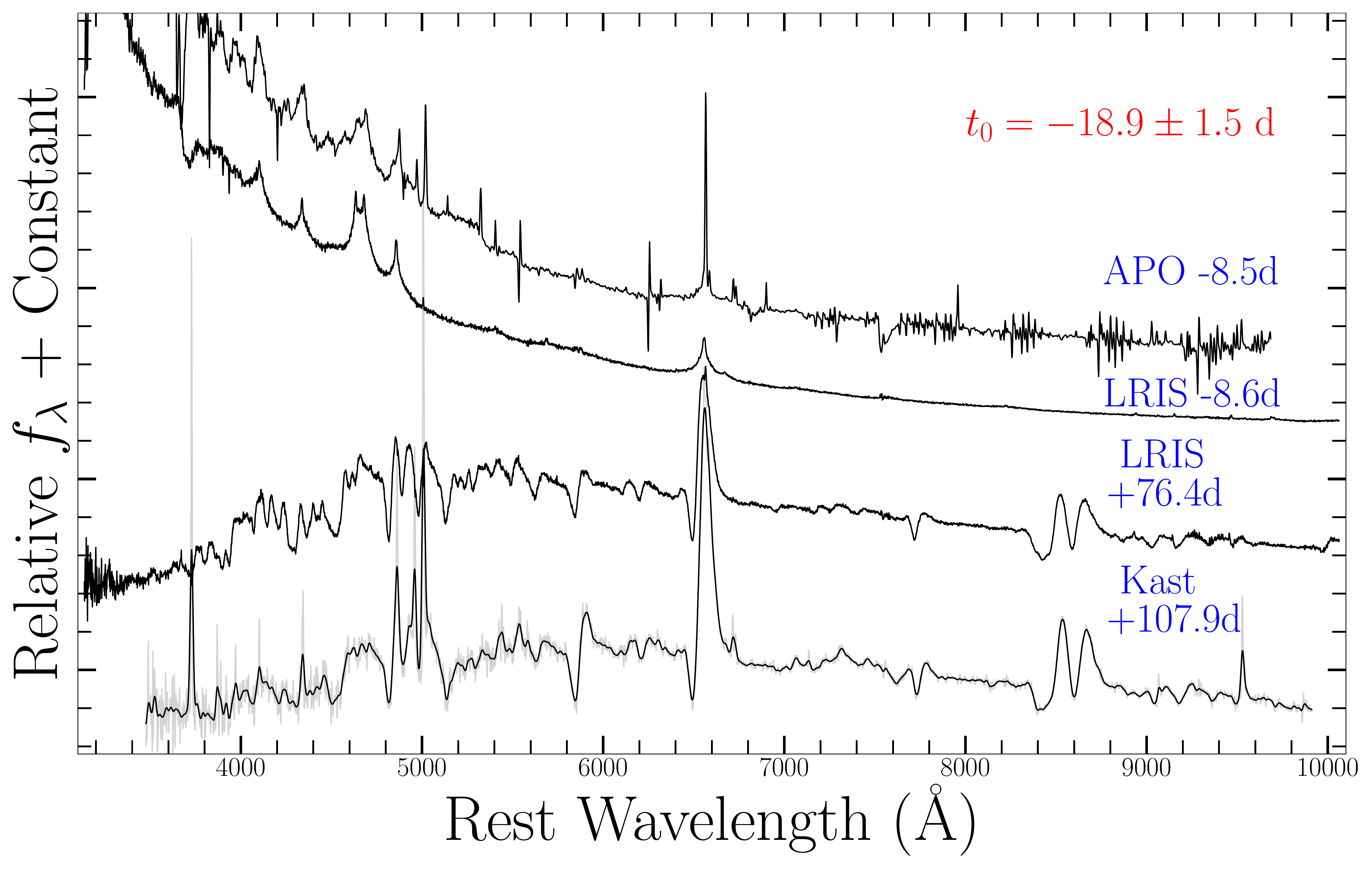}}
\subfigure[]{\includegraphics[width=0.49\textwidth]{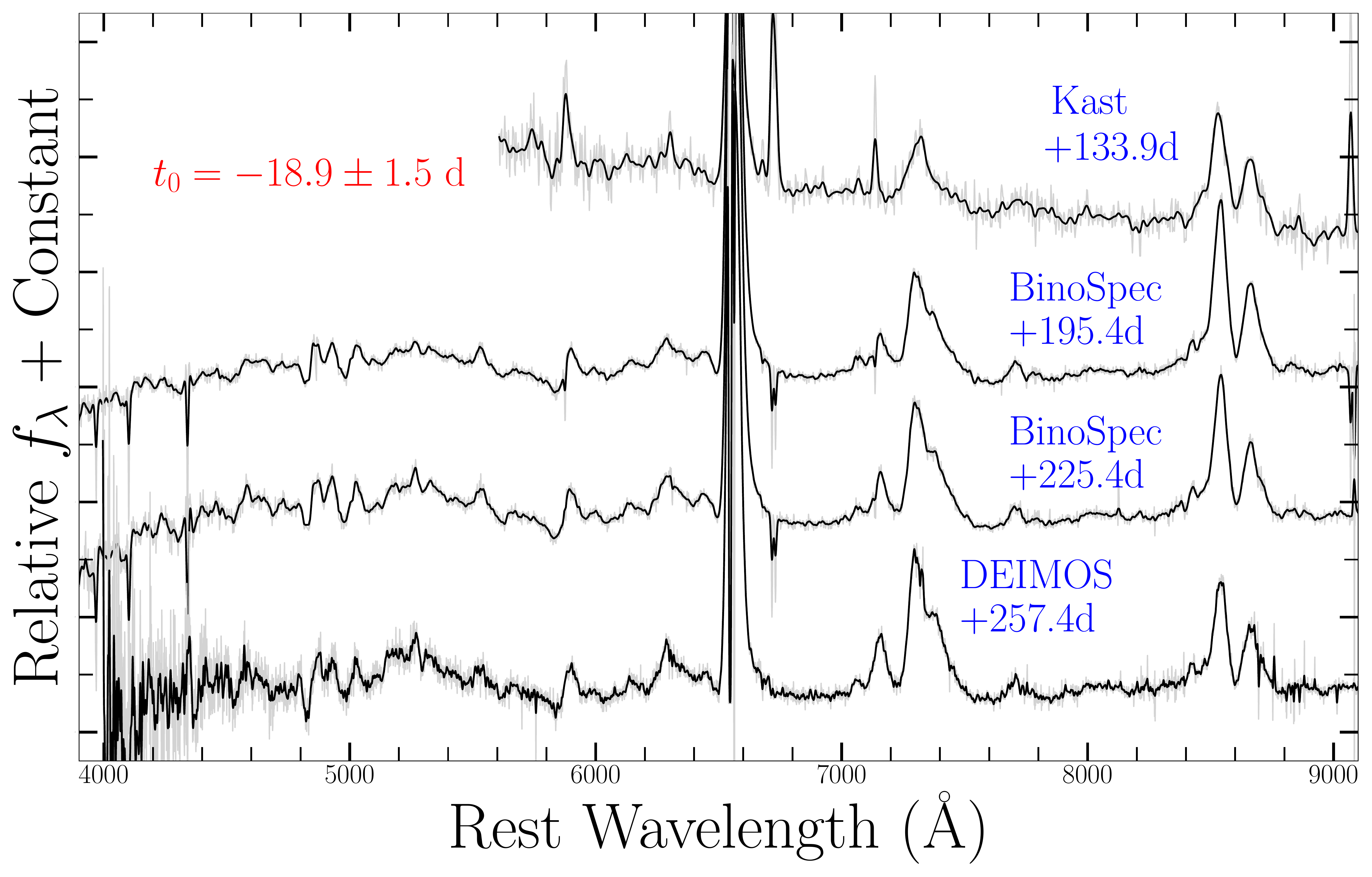}}
\caption{(a)/(b) Spectral observations of SN~2020tlf with phases (blue) marked with respect to $B$-band maximum. Time of first light relative to maximum listed in red. Unsmoothed spectra are shown in gray, and spectra shown in black have been smoothed with a Gaussian-filter. \label{fig:spectra} }
\end{figure*}

\subsection{Optical/NIR Spectroscopy} \label{SubSec:Spec}

In Figure \ref{fig:spectra}, we present the complete series of optical spectroscopic observations of \sn{} from -9 to +257~days relative to the $B$-band maximum ($\delta t = 10-270$ days relative to first light). A full log of spectroscopic observations is presented in Appendix Table \ref{tab:spec_table}. 

SN~2020tlf was observed with Shane/Kast \citep{KAST} and Keck/LRIS \citep{oke95} between -9 and +257~days relative to the $B-$band maximum. For all these spectroscopic observations, standard CCD processing and spectrum extraction were accomplished with \textsc{IRAF}\footnote{https://github.com/msiebert1/UCSC\_spectral\_pipeline}. The data were extracted using the optimal algorithm of \citet{1986PASP...98..609H}.  Low-order polynomial fits to calibration-lamp spectra were used to establish the wavelength scale and small adjustments derived from night-sky lines in the object frames were applied. We employed custom IDL routines to flux calibrate the data and remove telluric lines using the well-exposed continua of the spectrophotometric standard stars \citep{1988ApJ...324..411W, 2003PASP..115.1220F}. Details of these spectroscopic reduction techniques are described in \citet{2012MNRAS.425.1789S}.

Spectra of SN~2020tlf were also obtained with Keck NIRES and DEIMOS, as well as Binospec on MMT and the Dual Imaging Spectrograph (DIS) on the Astrophysical Research Consortium (ARC) 3.5-m telescope at Apache Point Observatory (APO). All of the spectra were reduced using standard techniques, which included correction for bias, overscan, and flat-field. Spectra of comparison lamps and standard stars acquired during the same night and with the same instrumental setting have been used for the wavelength and flux calibrations, respectively. When possible, we further removed the telluric bands using standard stars. Given the various instruments employed, the data-reduction steps described above have been applied using several instrument-specific routines. We used standard \textsc{IRAF} commands to extract all spectra.

\subsection{X-ray observations with Swift-XRT}\label{SubSec:XRT}

The X-Ray Telescope (XRT, \citealt{burrows05}) on board the \emph{Swift} spacecraft \citep{Gehrels04} started observing the field of \sn{} on 9 Sept. 2020 until 18 Feb. 2021 ($\delta t=$ 11.0 -- 165.2 days since first light) with a total exposure time of 35.2~ks, (Source IDs 11337 and 11339). We analyzed the data using HEAsoft v6.26 and followed the prescriptions detailed in \cite{margutti13}, applying standard filtering and screening using the latest CALDB files (version 2021008). We find no evidence for significant X-ray emission in any of the individual \emph{Swift}-XRT epochs, nor in merged images near optical/UV peak and at all observed phases. From the complete merged image, we extracted an X-ray spectrum using \texttt{XSELECT}\footnote{http://heasarc.nasa.gov/docs/software/lheasoft/ftools/xselect/} at the source location with a 35\arcsec\ source region (100\arcsec\ background region) and estimated the count-to-flux conversion by fitting an absorbed simple power-law spectral model with Galactic neutral H column density of $1.25\times 10^{20}$~cm$^{-2}$ \citep{Kalberla05} and spectral index $\Gamma = 2$ using \texttt{XSPEC} \citep{xspec}. Using a merged, 0.3-10~keV XRT image around UV peak ($\delta t=$ 11.0 -- 23.0 days since first light), we derive 3$\sigma$ upper limits on the count rate, unabsorbed flux and luminosity of $< 3.9 \times 10^{-3}$~ct s$^{-1}$, $< 1.7\times10^{-13}$~erg s$^{-1}$ cm$^{-2}$, and $< 2.6 \times 10^{40}$~erg s$^{-1}$, respectively. These limits assume no intrinsic absorption from material in the local SN environment e.g., $n_{\rm H, host} = 0$. This $n_{\rm H, host}$ value is chosen so as to provide the most conservative upper limit on X-ray emission despite the host reddening of $E(B-V)_{\rm host} = 0.018$~mag derived from optical spectra (\S\ref{SubSec:Phot}). 

\subsection{Radio observations with the VLA}\label{SubSec:VLA}

We acquired deep radio observations of \sn{} with the Karl G. Jansky Very Large Array (VLA) at $\delta t=146-320$ days since first light through project SD1096 (PI Margutti).  All observations have been obtained at 10 GHz (X-band) with 4.096 GHz bandwidth in standard phase referencing mode, with 3C286 as a bandpass and flux-density calibrator and QSO J1224+21 (in A and B configuration) and QSO J1254+114 (in D configuration) as complex gain calibrators. The data have been calibrated using the VLA pipeline in the Common  Astronomy  Software  Applications  package (CASA, \citealt{McMullin07}) v6.1.2 with additional flagging. \sn{} is not detected in our observations. We list the inferred upper-limits on the flux densities in Appendix Table \ref{Tab:radio}.

\begin{figure*}
\centering
\includegraphics[width=\textwidth]{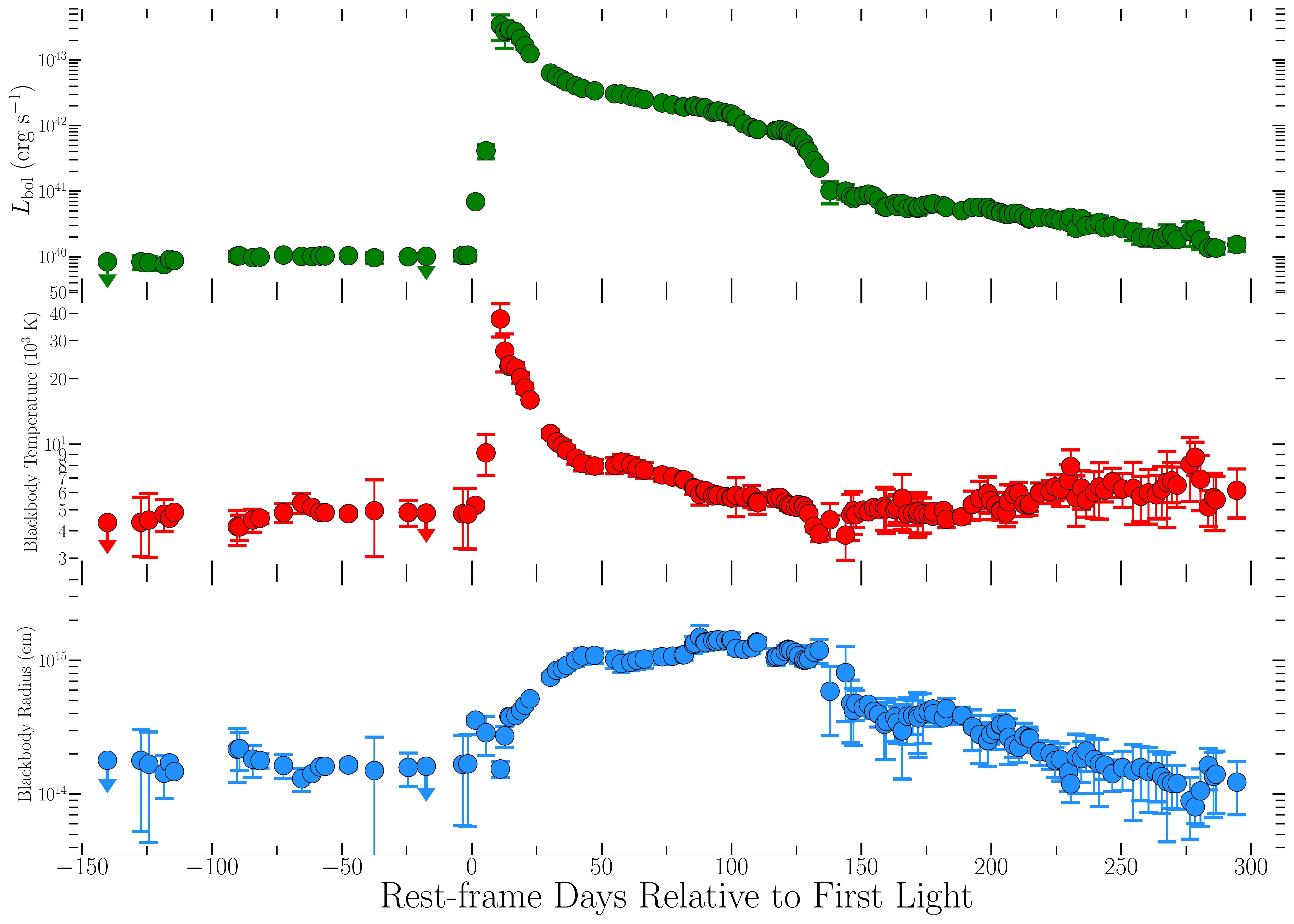}
\caption{Complete pre- and post-explosion bolometric light curve (top), blackbody temperatures (middle) and radii (bottom). Data shown is derived from SED blackbody modeling of all multi-color optical photometry. 
\label{fig:BB_LRT_all} }
\end{figure*}

\section{Host Galaxy Properties}\label{sec:host}


We determine an oxygen abundance 12 + log(O/H) in host galaxy NGC~5731 by using an SDSS spectroscopic observation taken on 14 April 2004. This spectrum was taken near the galactic core and therefore the metallicity at the explosion site could be slightly different. Using a combination of line flux ratios ([\ion{O}{iii}] / H$\beta$ and [\ion{N}{ii}]/H$\alpha$) into Equations 1 \& 3 of \cite{pettini04}, we determine a range of host metallicities of 12 + log(O/H) = $8.65 - 9.04$~dex ($0.99 - 1.04$~Z$_{\odot}$). Our derived metallicity range is higher than average SNe~II host metallicities of $\sim8.41 - 8.49$~dex \citep{anderson16}. However, the true metallicity at the SN explosion site could be lower than that estimated from the SDSS spectrum near the galactic core.


We utilize the same pre-explosion SDSS spectrum nearby the host galaxy center to determine a star formation rate. We calculate a total H$\alpha$ emission line luminosity of $L_{\textrm{H$\alpha$}} = 3.7 \times 10^{40}$~erg~s$^{-1}$. We then use Equation 2 from \cite{Kennicutt98} to estimate a star formation rate of SFR = $0.29 \ \Msun$ yr$^{-1}$ of the host galaxy. This star formation estimate is reflective of the star-forming characterization of host galaxy NGC~5731. The derived SFR is also consistent with with SFRs of other galaxies that hosted SNe~II that displayed photo-ionized emission features in their early spectra. For example, \cite{terreran21} find a SFR of 0.25--0.39\,$\Msun$ yr$^{-1}$ for the star-forming host of SN~2020pni. 

\section{Analysis} \label{sec:analysis}

\subsection{Photometric Properties}

The complete post-explosion, multi-band light curve of SN~2020tlf is presented in Figure \ref{fig:optical_LC} and pre-explosion $gcroiz-$band light curves are displayed in Figure \ref{fig:preSN_LCphot}. We define the time of first light as the average phase between the last photometric detection at the pre-SN flux threshold ($M \approx -12$~mag) and the first multi-color detections that rose above that flux threshold ($M \lesssim -12$~mag). This yields a time of first light of $t_{\rm exp} =  59098.7 \pm 1.5$, which is then used for reference through the analysis. We discuss potential uncertainties on this time when modeling the bolometric light curve (e.g., \S\ref{sec:modeling}). We fit a $3^{\rm rd}$-order polynomial to the \sn{} light curve to derive a peak absolute $B-$band magnitude of $M_B = -18.5 \pm 0.04$\,mag at MJD $59117.6\pm1.5$, where the uncertainty on peak magnitude is the $1\sigma$ error from the fit and the uncertainty on the peak phase is the same as the error on the time of first light. Using the adopted time of first light, this indicates a rise time of $t_r = 18.9 \pm 1.5$\,days with respect to $B$-band maximum.

As shown in Figure \ref{fig:flash_midcompare}(b), we compare the $r/V$-band light-curve evolution of \sn{} to popular SNe~II discovered within a few days of explosion, many of which showed photo-ionization features in the early-time spectra e.g., SNe 1998S \citep{Leonard00,fassia01,shivvers15}, 2013fs \citep{yaron17}, 2014G \citep{terreran16}, 2017ahn \citep{Tartaglia21}, and 2020pni \citep{terreran21}. Compared to these SNe, the peak $r/V$-band absolute magnitude of \sn{} is more luminous than that of SNe~2013ej, 2013fs, 2017ahn, and 2020pni, but less luminous than SNe~1998S and 2014G at peak. While the $r/V$-band rise time near maximum light is similar to SN~1998S, \sn{} was discovered at an even earlier phase with a fainter detection absolute magnitude of $\sim$-13.5~mag. The linear photometric evolution of \sn{} during its photospheric phase is comparable to most of these objects. However, \sn{} has the longest lasting plateau, extending out to $\sim$110~days after maximum light, suggesting a larger ejecta mass and/or larger stellar radius than other SNe~II with early-time signatures of CSM interaction.

\subsection{Bolometric Light Curve}

We construct a bolometric light curve by fitting the ZTF, PS1, Las Cumbres Observatory, ATLAS and \textit{Swift} photometry with a blackbody model that is dependent on radius and temperature. The extremely blue UV colors and early-time color evolution of \sn{} near maximum light impose non-negligible deviations from the standard \textit{Swift}-UVOT count-to-flux conversion factors. We account for this effect following the prescriptions by \cite{brown10}. Each spectral energy distribution (SED) was generated from the combination of multi-color UV/optical/NIR photometry in the $w2$, $m2$, $w1$, $u$, $b/B$, $v/V$, $g$, $c$, $o$, $r$, $i$, and $z$ bands (1500--10000\,\AA). In regions without complete color information, we extrapolated between light curve data points using a low-order polynomial spline. We present SN~2020tlf's pre- and post-explosion bolometric light curve in addition to its blackbody radius and temperature evolution in Figure \ref{fig:BB_LRT_all}. All uncertainties on blackbody radii and temperature were calculated using the co-variance matrix generated by the SED fits. At the time of first spectrum with photo-ionization emission features, the blackbody radius, temperature and luminosity is $R_{\rm BB} = (1.5 \pm 0.21) \times 10^{14}$\,cm, $T_{\rm BB} = (3.8 \pm 0.65) \times 10^4$\,K and $L_{\rm bol} = (3.4 \pm 1.4) \times 10^{43}$\,erg \,s$^{-1}$, respectively. This $R_{\rm BB}$ is technically the radius of thermalization ($\tau > 1$), which is much smaller than the photospheric radius ($\tau  = 1$; \citealt{dessart05}), and the assumption of a pure blackbody is not strictly accurate (see, e.g., \citealt{D15_2n}). Consequently, this can lead to the reported luminosities from blackbody fitting to be possble lower limits on the true bolometric luminosity of \sn{}.

\begin{figure}[t!]
\centering
\includegraphics[width=0.45\textwidth]{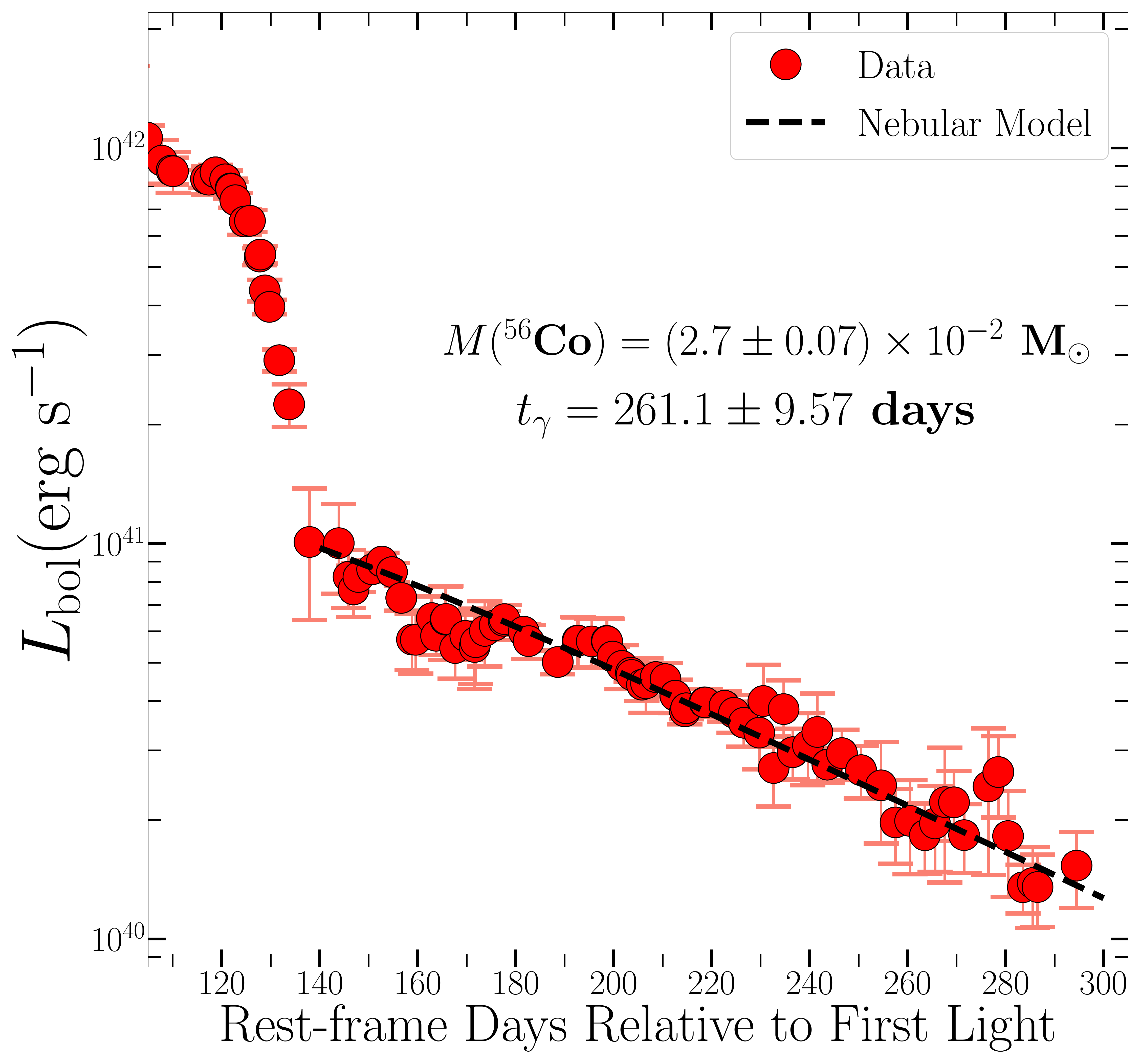}
\caption{Post-plateau bolometric light curve of \sn{} (red) with radioactive decay powered model shown in black for the energy released in ${}^{56}\textrm{Co}$ decay at late-times, following the decay of ${}^{56}\textrm{Ni}$ at early-times. The SN decline rate is consistent with a total ${}^{56}\textrm{Co}$ mass of $\sim0.03 \ \Msun$ and a $\gamma$-ray trapping timescale of $\sim$260~days. \label{fig:ni_model}}
\end{figure}

\begin{figure*}
\centering
\subfigure[]{\includegraphics[width=0.51\textwidth]{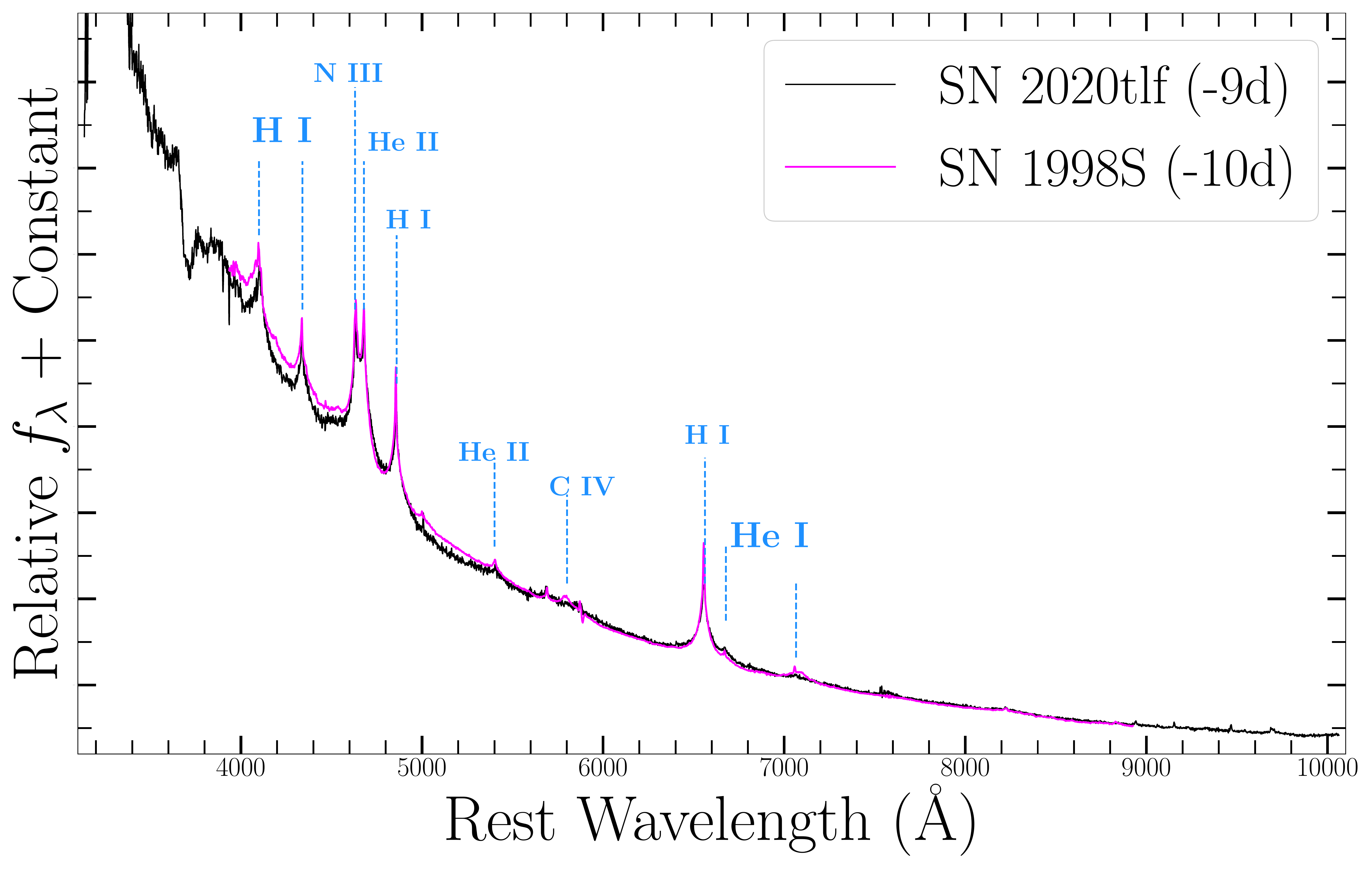}}
\subfigure[]{\includegraphics[width=0.22\textwidth]{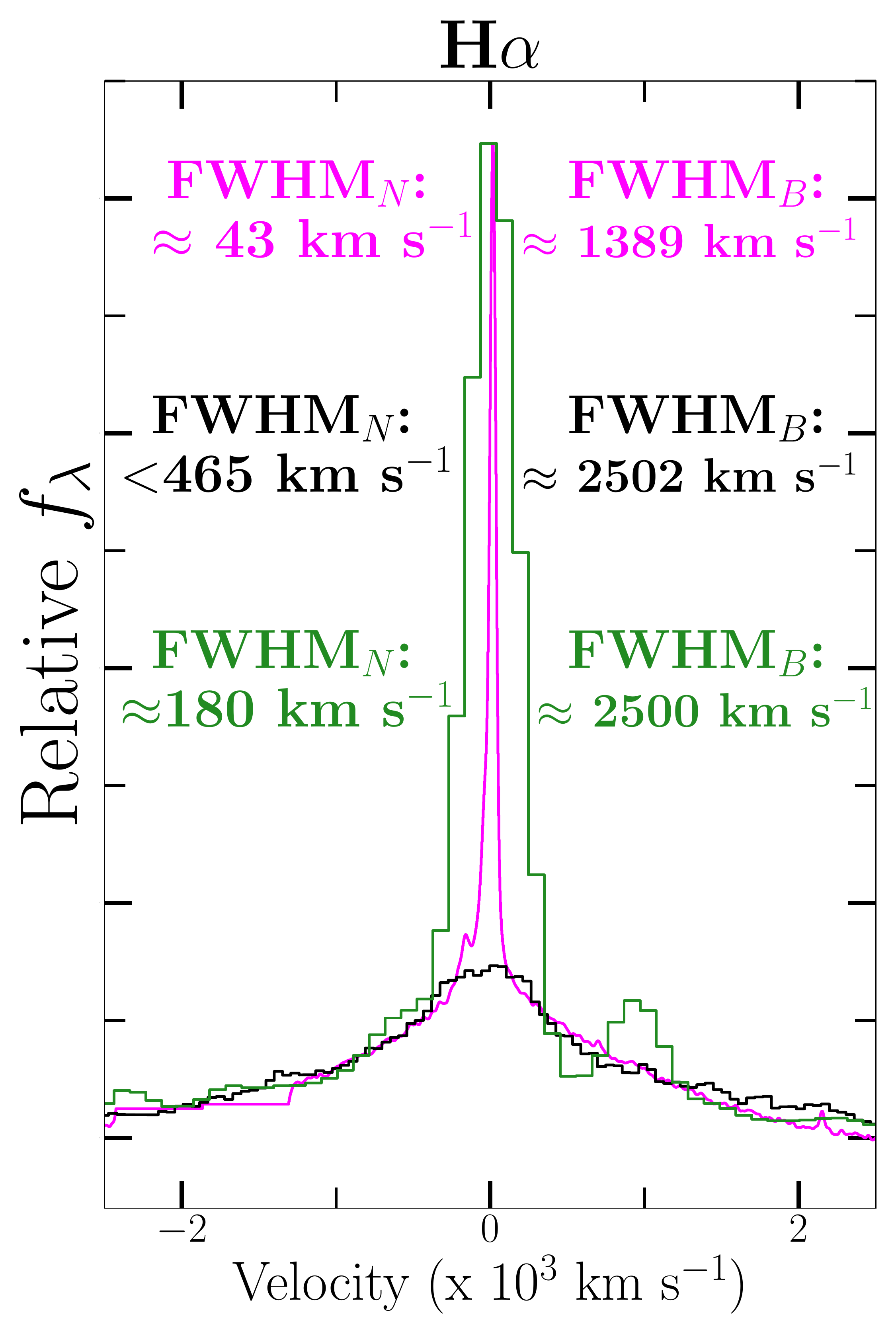}}
\subfigure[]{\includegraphics[width=0.22\textwidth]{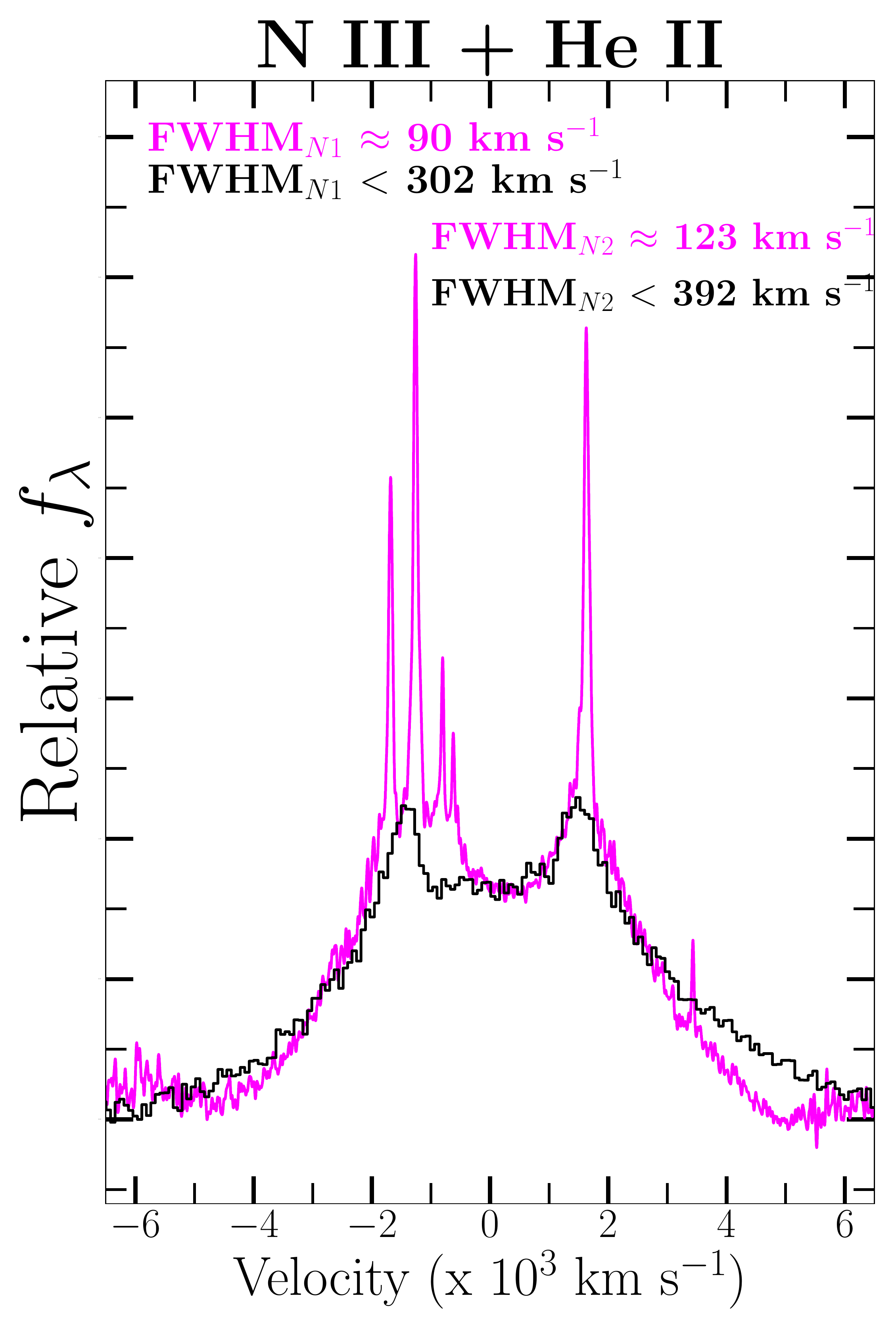}}
\caption{(a) Early-time spectra of SNe~1998S (magenta) and 2020tlf (black) with common narrow emission features labeled in blue; phases relative to $B-$band maximum. (b) H$\alpha$ emission profiles of SNe~1998S (magenta) and 2020tlf (Keck LRIS in black, APO DIS in green). Narrow component velocity is resolved in the high-resolution spectrum of SN~1998S \citep{shivvers15} to $v_w \approx 40 \ \kms$. (c) \ion{N}{iii} and \ion{He}{ii} emission profiles in SNe~1998S and 2020tlf spectra. \label{fig:flash_98S}}
\end{figure*}

As shown in Figure \ref{fig:ni_model}, we model the post-plateau ($t > 120$~days after maximum light) bolometric light evolution curve with energy injection from pure radioactive decay of newly synthesized ${}^{56}\textrm{Co}$. The complete analytic formalism behind this model is outlined in \citealt{valenti08}, \citealt{wheeler15}, and \citealt{jacobson-galan21}. From this modeling, we derive a total ${}^{56}\textrm{Co}$ mass of  $M_{\textrm{Co}}=(2.7 \pm 0.070) \times 10^{-2} \ \Msun$ and a $\gamma$-ray trapping timescale of $t_{\gamma} = 261.1 \pm 9.57$~days. The inferred ${}^{56}\textrm{Co}$ mass is lower than other SNe~II with early-time photo-ionization signatures e.g., SN~2014G ($\sim$0.06~$\Msun$; \citealt{terreran16}) or SN~1998S ($\sim$0.15~$\Msun$; \citealt{fassia01}). While the late-time light curve evolution is consistent with energy injection from the radioactive decay of ${}^{56}\textrm{Co}$, there are possibly small, but overall negligible, contributions from additional power sources at these phases such as CSM interaction. Furthermore, the nebular spectra of \sn{} (e.g., Fig. \ref{fig:spectra}b) show typical \ion{O}{i} and \ion{Ca}{ii} emission, which is compatible with \coVI \ decay being absorbed by the metal-rich inner ejecta rather than late-time power coming from the outer ejecta ramming into CSM, as observed in SN~1998S.

\subsection{Spectroscopic Properties}\label{subsec:spec_analysis}

The complete spectroscopic sequence of \sn{} from $-8.6$ to +257.4 days since maximum light is presented in Figure \ref{fig:spectra}. In the earliest spectrum, \sn{} shows narrow, symmetric emission features of \ion{H}{i}, \ion{He}{i}, \ion{He}{ii}, \ion{N}{iii} and \ion{C}{iii} (FWHM~$< 300~\kms$). As shown in Figure \ref{fig:flash_98S}, this spectrum is nearly identical to the early-time spectrum of SN~1998S at a phase of +3~days since first detection ($-10$~days from $B$-band peak; \citealt{fassia01}). However, the time of first light in SN~1998S is relatively uncertain given the last non-detection was 8~days prior to the first detection, indicating that the phase of this spectrum could be later than +3~days. Based on our adopted time of first light, \sn{} is at later phase of +10~days since first light ($-8.6$~days relative to $B$-band peak), despite the overall spectral similarity. This could indicate that the true time of first light for \sn{} is actually later than estimated, that first light emission from \sn{} was detected at earlier phases given the depth of PS1 compared to the instruments used to discover SN~1998S (plus the uncertainty on the time of first light for SN~1998S), or that the environment around each of the two SNe is different i.e., variations in the properties of the most local CSM or intrinsic extinction from the SN host galaxies. 

\begin{figure*}
\centering
\subfigure[]{\includegraphics[width=0.61\textwidth]{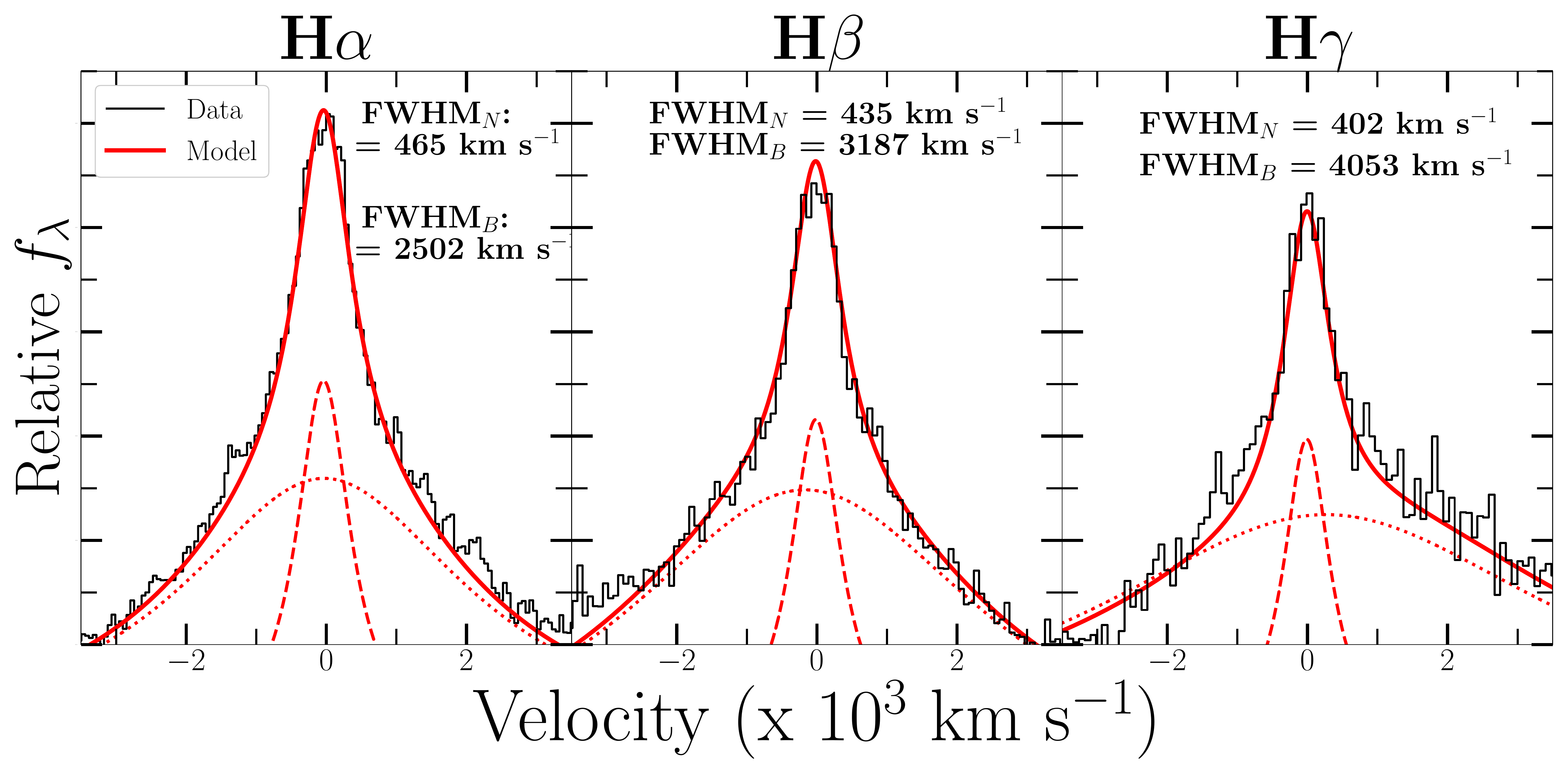}}
\subfigure[]{\includegraphics[width=0.35\textwidth]{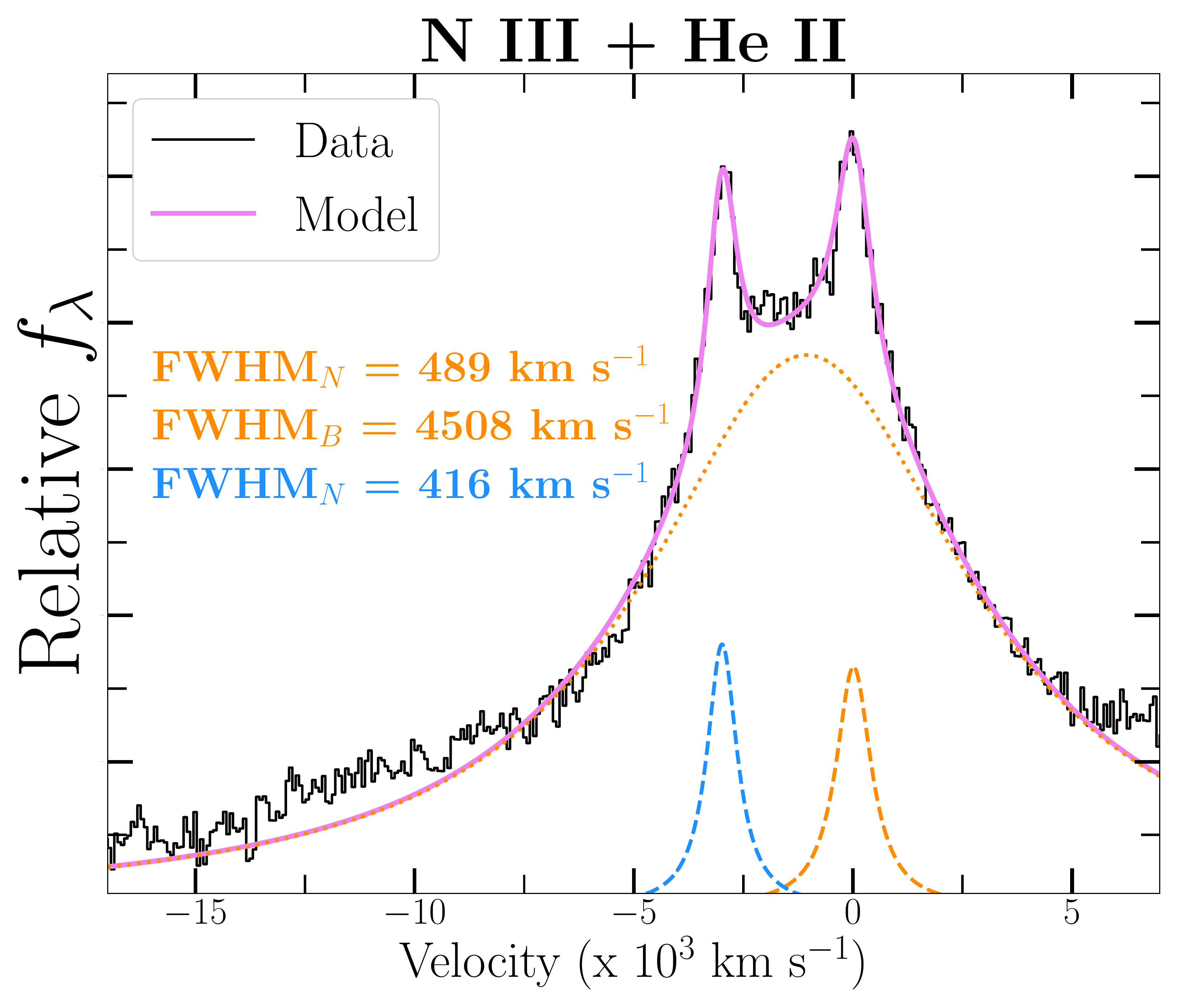}}
\caption{(a) Balmer emission features (black) from the LRIS photo-ionization spectrum with respect to multi-component Lorentzian models (red). The true velocities of the narrow component are unknown due to spectral resolution. (b) \ion{N}{iii} + \ion{He}{ii} feature (black) with complete Lorentzian emission model shown in violet. \ion{N}{iii} emission model presented in blue and \ion{He}{ii} model shown in orange.  \label{fig:flash_vels} }
\end{figure*}

In Figure \ref{fig:flash_98S}(b)/(c), we present velocity comparisons plots of \ion{H}{i} and \ion{N}{iii} + \ion{He}{ii} emission profiles for \sn{} and SN~1998S. The SN~1998S high-resolution spectrum is from \cite{shivvers15} and all line velocities can be resolved, unlike in the \sn{} LRIS spectrum. Nonetheless, while line velocities in the \sn{} LRIS spectrum can only be resolved to $\lesssim 300-400~\kms$ and $\lesssim 200~\kms$ in the APO DIS spectrum at the same phase, the overall similarity of the narrow features in \sn{} compared to SN~1998S indicates that the wind velocities of CSM around the \sn{} progenitor may be comparable to that of the CSM in SN~1998S. To test this, we convolve the high-resolution SN~1998S spectrum to the instrumental resolution of the \sn{} LRIS spectrum and find that the narrow Balmer series emission components in this spectrum, as well as those in the SN~1998S LRIS spectrum, can be modeled with a similar Lorentzian profile velocity ($\sim 300-400~\kms$) as observed in the \sn{} spectrum. Therefore, based on the SN~1998S spectra, it is possible that the H-rich CSM in \sn{} is moving at $\sim 50 ~ \kms$ (e.g., Fig. \ref{fig:flash_98S}b) and other CSM ions such as \ion{He}{ii} or \ion{N}{iii} (e.g., Fig. \ref{fig:flash_98S}c) are moving with wind velocities of $\sim$90-120~$\kms$. We present additional modeling of these photo-ionization line profiles in Figure \ref{fig:flash_vels} using combined Lorentzian profiles. Narrow components of each profile in the LRIS spectrum can only be resolved to FWHM $\lesssim 300 ~ \kms$ and FWHM $\lesssim 200 ~ \kms$ in the APO DIS spectrum, but the broad components of the profiles resulting from electron scattering (e.g., \citealt{Chugai01, Dessart09}) are fit using Lorentzian profiles with FWHM ~ $\sim 2000 - 3000 \kms$. Based on the comparison to SN~1998S and the Lorentzian profile fits, we conclude that the \sn{} progenitor likely had a wind velocity of $v_w \sim 50 - 200 ~ \kms$. For the \ion{N}{iii} + \ion{He}{ii} feature shown in Figure \ref{fig:flash_vels}(b), we explore the possibility of blueshifted, Doppler broadened \ion{He}{ii} from the SN ejecta being present in the line profile, in addition to the narrow \ion{He}{ii} and \ion{N}{iii} profiles derived from the wind. This specific combination of Lorentzian profiles is consistent with the overall profile shape as well as the flux excess on top of continuum emission, bluewards of the \ion{N}{iii} + \ion{He}{ii} feature. Doppler broadened \ion{He}{ii} from the SN ejecta has been proposed as an explanation for blue flux excesses in SNe~II-P that do not show spectral signatures of CSM interaction \citep{Dessart08}. 

\begin{figure*}[t]
\centering
\includegraphics[width=\textwidth]{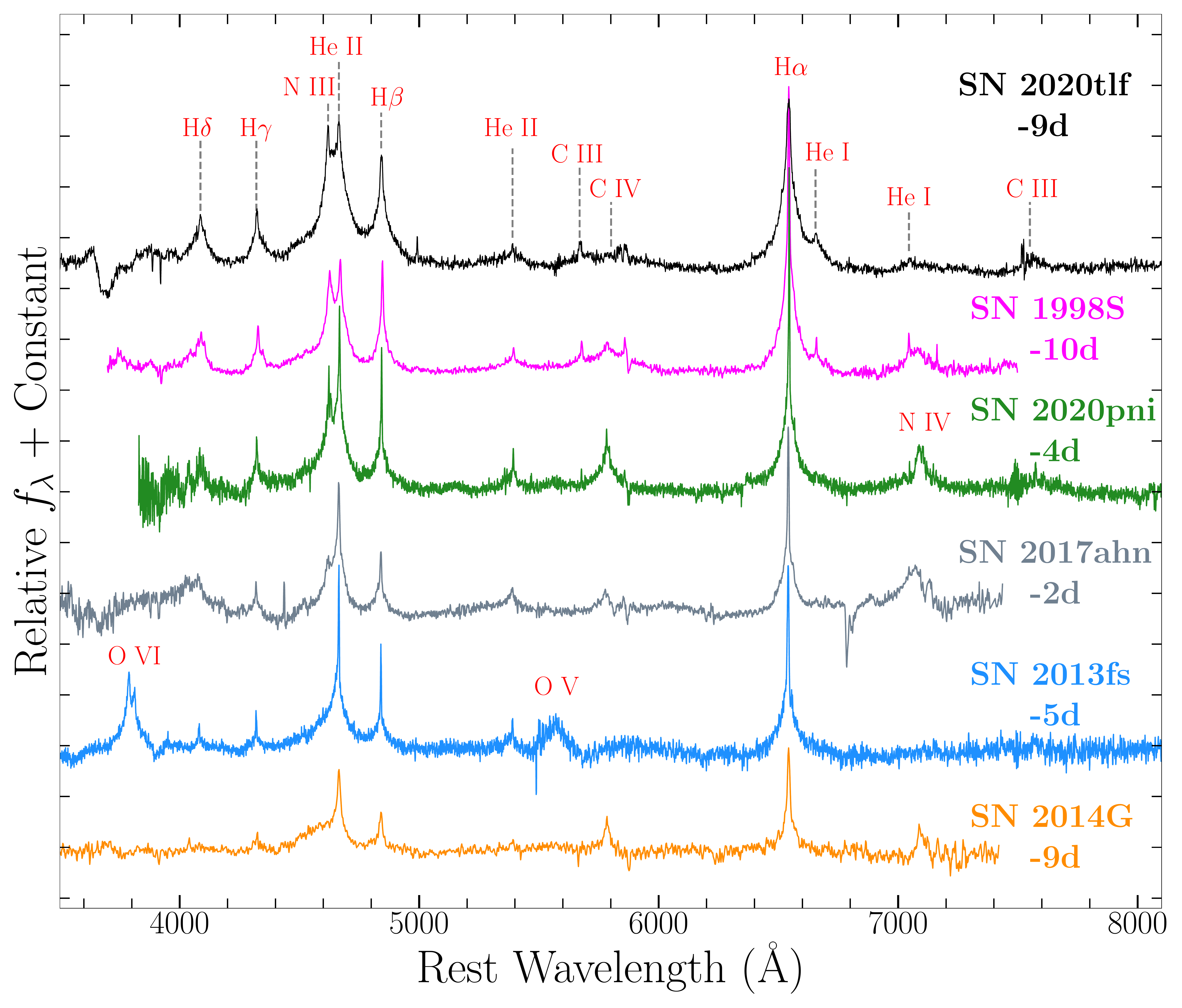}
\caption{Early-time, continuum-subtracted spectral comparison of SNe~II, 2020tlf (black), 1998S (magenta; \citealt{fassia01}), 2020pni (green; \citealt{terreran21}), 2017ahn (grey; \citealt{Tartaglia21}), 2013fs (blue; \citealt{yaron17}) and 2014G (orange; \citealt{terreran16}). Common photo-ionization ions labeled in red; phases relative to $B-$band maximum. Early-time spectrum of \sn{} shows nearly identical photo-ionization features to SNe~1998S and 2020pni, indicating a similar ionization temperature and CSM extent. \label{fig:flash_compare} }
\end{figure*}

In Figure \ref{fig:flash_compare}, we compare the continuum-subtracted spectrum of \sn{} to other well-studied events with photo-ionization features such as SNe~1998S, 2014G, 2013fs, 2017ahn, and 2020pni \citep{fassia01, terreran16, yaron17,Tartaglia21, terreran21}. The \ion{H}{i}, \ion{He}{ii}, and \ion{N}{iii} emission lines present in the early-time spectrum of \sn{} are similar to those found in most other objects. \sn{} differs slightly from SN~2013fs in that it does not contain high-ionization lines such as \ion{O}{iv--vi}, which indicates a more extended CSM and thus lower ionization temperature for \sn{} \citep{dessart17}. SNe~2013fs and 2014G also do not have the detectable \ion{N}{iii} unlike \sn{}, SN~1998S, 2020pni and 2017ahn, which have clear \ion{N}{iii} emission in the double-peaked \ion{N}{iii} + \ion{He}{ii} feature. Furthermore, \sn{} does not have significant \ion{C}{iv} or \ion{N}{iv} emission like most other objects, with the exception of SN~2013fs. 

\begin{figure*}
\centering
\subfigure[]{\includegraphics[width=0.49\textwidth, height=2.94in]{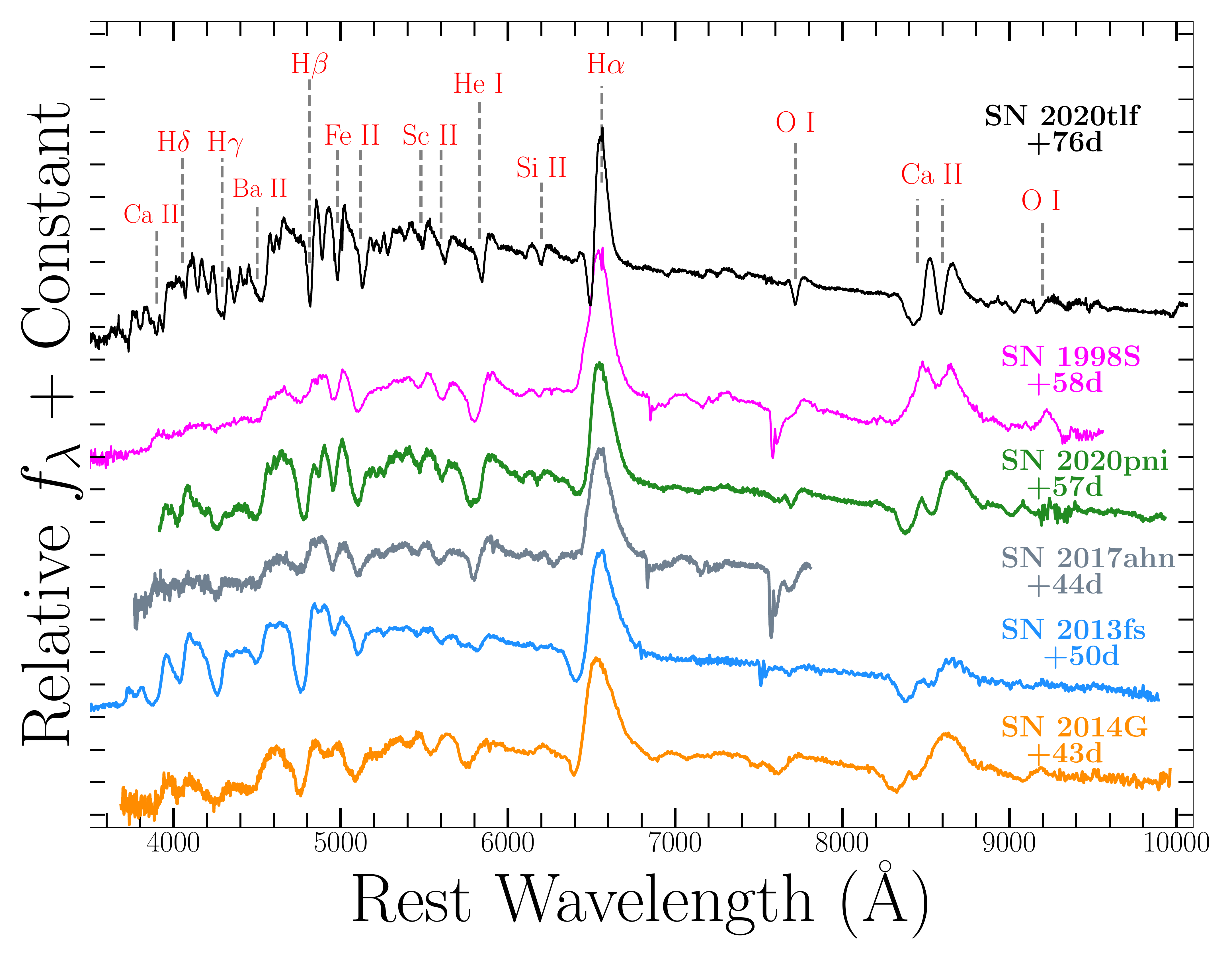}}
\subfigure[]{\includegraphics[width=0.49\textwidth, height=2.9in]{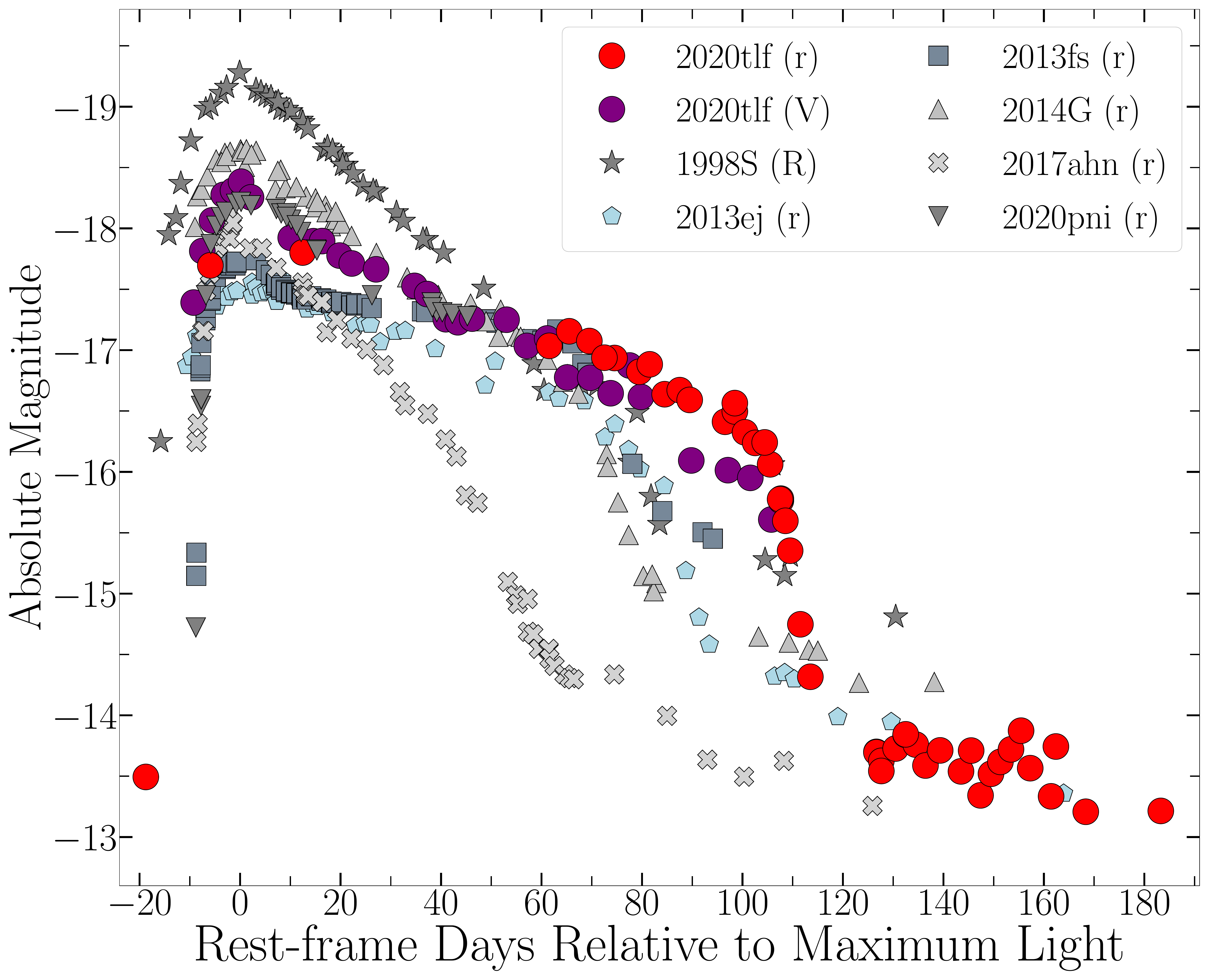}}
\caption{(a) Spectral comparison of SN~2020tlf (black) and other SNe~II discovered with photo-ionization spectra at approximately the same phase relative to $B-$band maximum \citep{fassia01, terreran16, yaron17, Tartaglia21, terreran21}. Common ions are marked by grey lines. (b) Early-time $r/V-$band light curve comparison of \sn{} (red/purple circles) and known SNe~II with photo-ionization spectral features (shades of gray) plus SN~2013ej (light blue), which was a SN~II discovered  very young but without spectroscopic evidence of CSM photo-ionization. \label{fig:flash_midcompare} }
\end{figure*}

\begin{figure}[t!]
\centering
\includegraphics[width=0.47\textwidth]{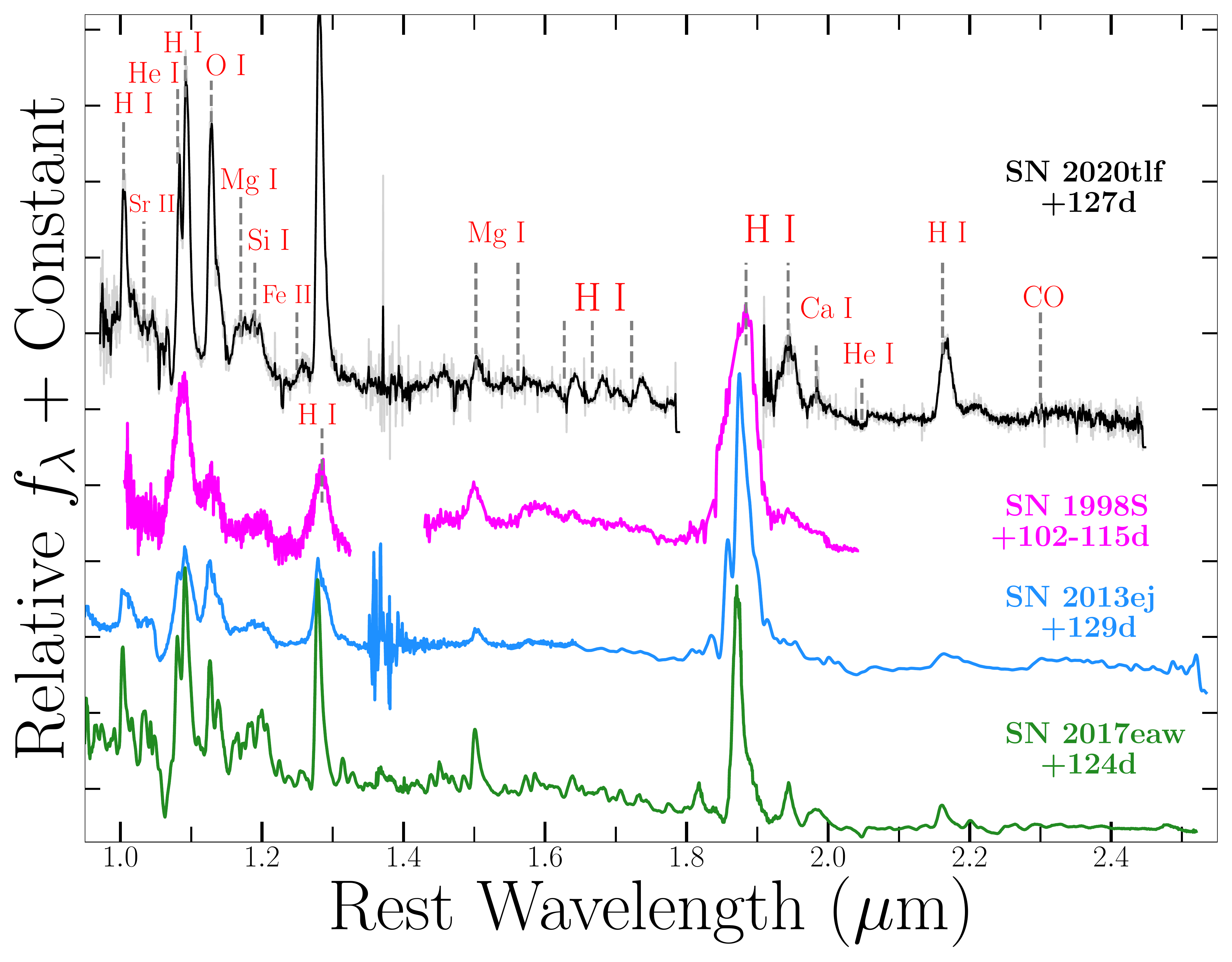}
\caption{Infrared spectral comparison of SNe~2020tlf, 1998S and 2013ej. Common ions marked by grey lines; phases relative to $B-$band maximum. \sn{} has identical IR spectral features to SNe~II, 1998S and 2013ej but overall lower photospheric velocities based on the line profiles. Line profile widths are smaller in \sn{} than in other SNe II, which is compatible with a larger $R_{\star}$ and lower $E_{k}/M_{ej}$.  \label{fig:ir_compare}}
\end{figure}

We also compare the mid-time spectra ($\delta t = +40 - 60$~days since peak) of this sample to the second spectrum of SN~2020tlf at +76~days since $B-$band peak, which was obtained once the SN was visible to ground-based observatories (Figure \ref{fig:flash_midcompare}a). At this phase, the SN is in its recombination phase, with strong signatures of line blanketing by metals in the H-rich ejecta and a red spectrum.  Overall, \sn{} has similar ions to other events e.g., strong Balmer series, Fe-group, \ion{O}{i} and \ion{Ca}{ii} profiles. However, absorption profiles in \sn{} are noticeably narrower than other objects, which could be due to the later phase and/or larger $R_{\star}$ or lower $E_k/M_{ej}$. The \sn{} spectrum is still photospheric at +76~days (+95~days since explosion) and contains a bluer continuum with weaker line blanketing compared to SNe II at similar epochs. This could indicate persistent energy injection from a more extended envelope or additional CSM interaction powering the SN at this phase. Additionally, we compare the IR spectrum of \sn{} at +127~days post-peak to IR spectra of SNe~1998S, 2013ej, 2017eaw at a similar phase in Figure \ref{fig:ir_compare}. All four SNe show similar ions at this phase such as prominent H emission, Fe-group elements and \ion{Mg}{i}. Additionally, the IR spectrum of \sn{} appears to show evidence for CO emission, similar to that confirmed in SN~2017eaw by \cite{Rho18}.

We present the late-time spectra of \sn{} in Figure \ref{fig:spectra}(b) over a phase range of $\delta t = 153 - 277$~days since first light. At these phases, \sn{} displays strong emission lines such as H$\alpha$, [\ion{O}{i}]\,$\lambda\lambda$\,$6300,\,6364$ [\ion{Ca}{ii}]\,$\lambda\lambda$\,$7291,\,7323$ emission. The SN appears to not be fully nebular by the +277~days post-explosion as it still shows H$\alpha$ and Fe-group element absorption profiles. However, some of these line transitions are optically thick and can exhibit a P-Cygni profile during the nebular phase when the continuum optical depth is low. 

To constrain the zero age main sequence (ZAMS) mass of the \sn{} progenitor, we compare the late-time spectra to nebular-phase radiative transfer models that have, in other SN~II studies, shown that the [\ion{O}{i}] emission profile is a direct tracer of progenitor mass. In Figure \ref{fig:neb_models}(a), we compare the nebular-phase models from \cite{Jerkstrand14} for 12-19~$\Msun$ progenitors to \sn{} at +250~days post-explosion. We find that at this phase, the 12~$\Msun$ model best reproduces the nebular transitions observed in \sn{}. We also compare the +277~day spectrum of \sn{} to the nebular models from \cite{Dessart21} that are generated from 9.5-15~$\Msun$ progenitors at +350~days post explosion and find that the 10~$\Msun$ model is the most consistent with the data. We therefore conclude that the progenitor of \sn{} had a ZAMS of $\sim$~10-12~$\Msun$. The estimated \sn{} progenitor mass is comparable to that derived from nebular emission in sample studies of SNe~II-P ($\sim$~12-15~$\Msun$; \citealt{Silverman17}), but lower than that of other SNe~II with photo-ionization spectra e.g., SN~2014G had an estimated progenitor ZAMS mass of 15-19~$\Msun$ \citep{terreran16}. We note that we cannot completely rule out the possibility that the progenitor of \sn{} was a low-mass ($\sim 9~\Msun$) super-asymptotic giant branch star, as proposed to be the progenitors of electron-capture SN candidates (e.g., see \citealt{Hiramatsu21}). However, based on the observed bolometric light curve evolution and total synthesized \niVI \ mass, it is unlikely that \sn \ was an electron-capture SN from such a progenitor star. 

\subsection{Precursor Emission}\label{subsec:precursor}

\sn{} is the first SN with typical SN~II-P/L-like spectral and light curve behavior that has a confirmed detection of precursor flux. Precursor emission was also identified $\sim 60$~years prior to SN~II, iPTF14hls in archival imaging \citep{Arcavi17}. However, while the spectral evolution of iPTF14hls resembles a normal SN~II, the extremely long-lasting and time variable light curve evolution indicated that this event, as well as its progenitor star, were very different than standard SN~II explosions. The pre-explosion light curve, presented in Figure \ref{fig:preSN_LCphot}(a), shows >3$\sigma$ detections in PS1 $riz$-bands starting from $\delta t = -130$~days and persisting with a consistent flux until first SN light. The lack of precursor detections in bluer bands such as PS1/ZTF $g-$band or ATLAS $c-$band suggests a moderately cool emission or an extended, low-temperature emitting surface of whatever physical mechanism caused this pre-explosion flux. We construct a pre-explosion bolometric light curve, as well as temperatures and radii, by modeling the SED containing 3$\sigma$ $riz-$band detections and $g-$band upper limits with a blackbody model, same as that used in \S 5.2. We show the pre-explosion bolometric light curve, blackbody temperatures and radii in Figure \ref{fig:preSN_LC}(b). It should be noted that the pre-SN bolometric light curve relies on only 3 optical/NIR bands and thus contributions from undetected parts of the blueward (or IR) ends of SED could cause variations from what is observed. Furthermore, the presence of spectral emission lines during the precursor (e.g., H$\alpha$) could lead to increased flux in $r-$band, for example, relative to other bands. We find that the precursor has a bolometric luminosity of $\sim 10^{40}$~erg s$^{-1}$ ($\sim 2\times 10^6 ~ \Lsun$), and has an average blackbody temperature and radius of $\sim 5000$~K and $\sim 10^{14}$~cm ($\sim1500 ~ \Rsun$), respectively. For reference, we also plot the predicted luminosity, surface temperature and radius evolution of a $15~\Msun$ RSG progenitor undergoing wave-driven mass loss as presented in \cite{fuller17}. This model has a consistent emitting radius to the \sn{} precursor emission, but has significantly lower luminosities and temperatures at phases where pre-explosion emission is detected. 

The pre-SN activity prior to \sn{} is considerably fainter than other SNe with confirmed precursor emission. In Figure \ref{fig:preSN_compare}, we compare the multi-color \sn{} pre-explosion detections to popular SNe~IIn, 2009ip, 2010mc and 2016bhu, all of which had confirmed precursor emission prior to explosion. As shown in the plot, the \sn{} precursor only reaches $\sim -11.5$~mag in all bands, while the plotted SNe~IIn precursors have absolute $r$-band magnitudes ranging from $-13$~mag to $-15.5$~mag. Precursor emission from the \sn{} progenitor system is also fainter than the average absolute magnitude of -13~mag found in the sample of ZTF-observed SNe~IIn with pre-explosion outbursts presented by \cite{Strotjohann21}. However, as shown in Figure \ref{fig:preSN_compare}, because the limiting magnitude of ZTF (<20.5~mag; \citealt{bellm19, graham19}) is $\sim$1~mag shallower than YSE (<21.5~mag; \citealt{Jones2021}), pre-explosion emission in SNe~II-like events would not have been detected at the flux level of the precursor of \sn{}. Nevertheless, searches for pre-SN emission from SN~II progenitors at closer distances (e.g., $\lesssim 50$~Mpc) in transient survey archival data (e.g., ZTF, ATLAS, YSE, etc) will allow us to determine whether more 20tlf-like precursor events are possible.

Integrating the pre-explosion bolometric light curve yields a total radiated energy of $\sim10^{47}$~erg over the $\sim$~130~day precursor event. Coincidentally, this derived radiated energy is approximately the binding energy of a H-rich envelope in a typical RSG \citep{dessart10}. We explore potential power sources for the precursor emission in the form of CSM interaction-powered and wind-driven emission. For the former, the precursor emission would result from interaction between material ejected in a progenitor outburst and CSM from a previous outburst and/or steady-state wind-driven mass loss, causing a fraction of the kinetic energy to be converted in radiative energy. In this process, the relation between radiated and kinetic energy, as well as CSM properties, goes as:
\begin{equation}
    E_{\rm rad} = \frac{\epsilon}{2} M_{\rm pre} v^2_{\rm pre}
\end{equation}
where $\epsilon$ is the fraction of converted kinetic energy, $M_{\rm pre}$ is mass ejected in the precursor and $v_{\rm pre}$ is the velocity of that material. For the observed precursor radiated energy of $E_{\rm rad} \approx 10^{47}$~erg, efficiency $\epsilon = 1$, and velocities discussed in \S\ref{subsec:spec_analysis} (e.g., $v_w = 50 - 200 ~ \kms$), the total mass ejected in the precursor is $M_{\rm pre} \approx 4.3 - 0.27 ~ \Msun$, respectively. However, if CSM interaction is the mechanism for precursor emission, the conversion efficiency is definitely much less than 100\% \citep{smith10} and therefore the derived $M_{\rm pre}$ is at least $\gtrsim \Msun$ for the largest $v_{\rm pre}$ that is consistent with observations. Furthermore, it should be noted that a material ejected in a precursor that then collides with pre-existing CSM may lead to to formation of a semi-static CSM shell of constant density (i.e., $s = 0$), which is different than the wind-like density CSM that is typically invoked to model events with photo-ionization spectra (e.g., see \S \ref{sec:modeling}).

If the precursor emission from the \sn{} progenitor was instead from a super-Eddington, continuum-driven wind, we follow the mass loss prescription outlined in \cite{shaviv01a} that goes as: 
\begin{equation}
    E_{\rm rad} \approx \frac{1}{W} M_{\rm CSM} c_s c 
\end{equation}
where $W$ is an empirical factor found to be $\sim 5$, $c_s$ is the speed of sound at the base of the optically thick wind (e.g., $\sim 60 ~ \kms$; \citealt{shaviv01b}), and $c$ is the speed of light. For $E_{\rm rad} \approx 10^{47}$~erg, we derive a total amount of material lost in a potential super-Eddington wind to be $M_{\rm CSM} \approx 2\times 10^{-3}~\Msun$. However, it should be noted that this formalism is designed for SN~IIn progenitors such as LBVs. Furthermore, a super-Eddington wind is likely unphysical for a 10-12~$\Msun$ progenitor mass range as derived from the nebular spectra of SN~2020tlf. 

Another possible mechanism to explain the pre-SN activity in \sn{} is stellar interaction between the primary RSG progenitor and a smaller binary companion star. This can manifest as a ``common envelope'' phase in the progenitor's evolution \citep{Sana12}, which can result in the merging of primary and binary companions, the result of which is a slightly luminous, short-lived transient \citep{Kochanek14}. While this scenario has been invoked as an explanation for Luminous Red Novae (LRN) or Intermediate Luminosity Optical Transients (ILOT), the resulting luminosity produced by this physical mechanism appears to be too faint ($\sim 10^{2-4}~\Lsun$; \citealt{Pejcha17}) to match the pre-explosion luminsoty in \sn{} ($\sim 10^{6}~\Lsun$). Therefore, it is more likely that an eruption from the primary progenitor alone is the most likely cause of the pre-SN activity observed in \sn{}.

\section{Light Curve and Spectral Modeling} \label{sec:modeling}

\begin{figure}[t!]
\centering
\includegraphics[width=0.45\textwidth]{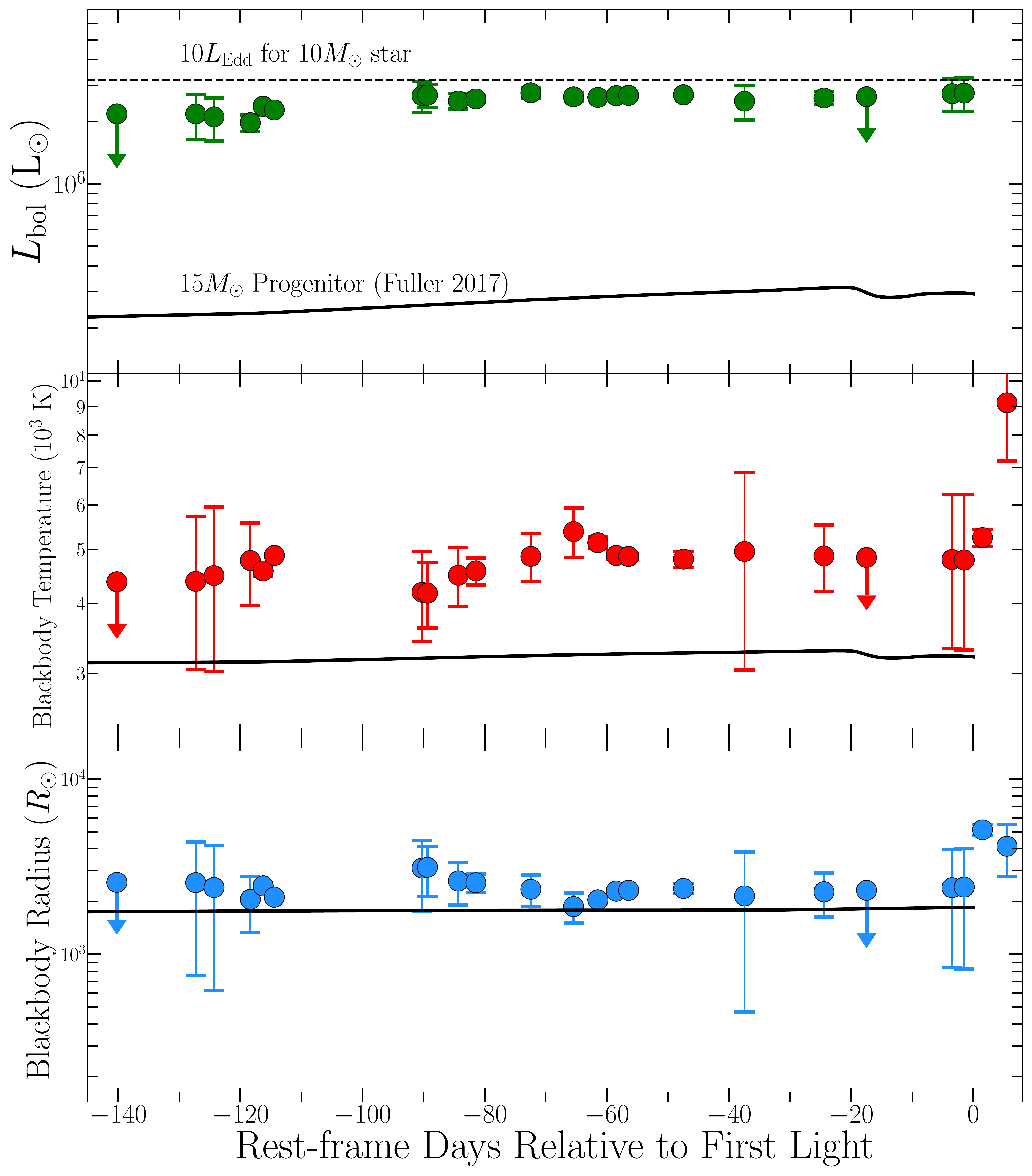}\caption{Pre-explosion bolometric light curve (top), blackbody temperatures (middle) and radii (bottom) from SED modeling of multi-band photometry associated with $\geq 3 \sigma$ flux excesses. Shown in black is a progenitor model from \cite{fuller17} of a 15~$\Msun$ RSG undergoing wave-driven mass loss. \label{fig:preSN_LC}}
\end{figure}

\begin{figure}[t!]
\centering
\includegraphics[width=0.47\textwidth]{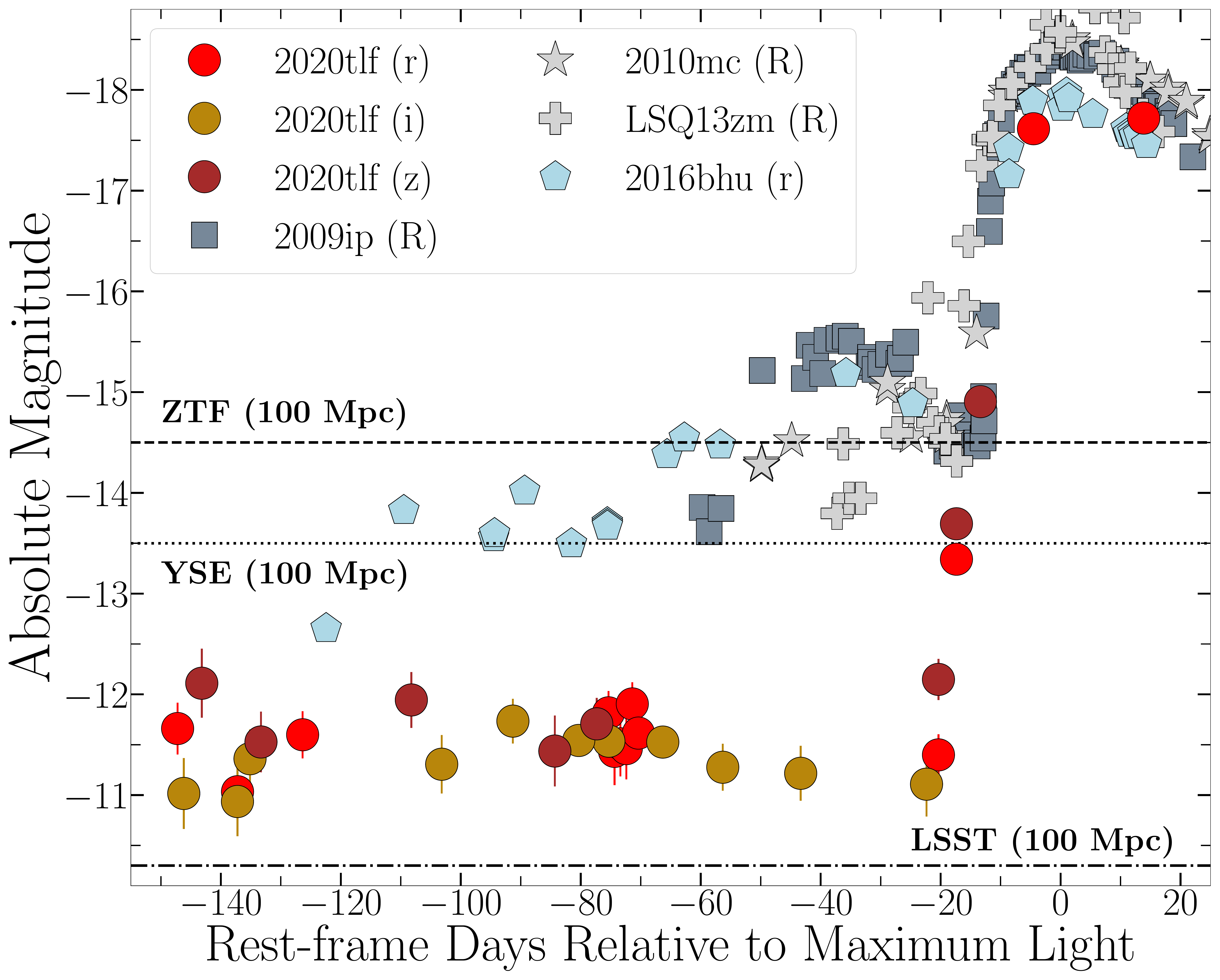}
\caption{Light curve comparison of SN~2020tlf (circles) and SNe~IIn with confirmed precursor emission. SN~2009ip $R-$band shown as squares, SN~2010mc $R-$band shown as stars, LSQ13zm $R-$band as plus signs, and SN~2016bhu $r-$band shown as pentagons. Limiting magnitudes at $D < 100$~Mpc for ZTF, YSE and LSST surveys shown as black lines. These limits represent detection magnitudes for single epoch, pre-SN observations whose detection is dependent on relatively deep template imaging that can then be applied in difference imaging. \label{fig:preSN_compare}}
\end{figure}

\begin{figure*}
\centering
\includegraphics[width=\textwidth]{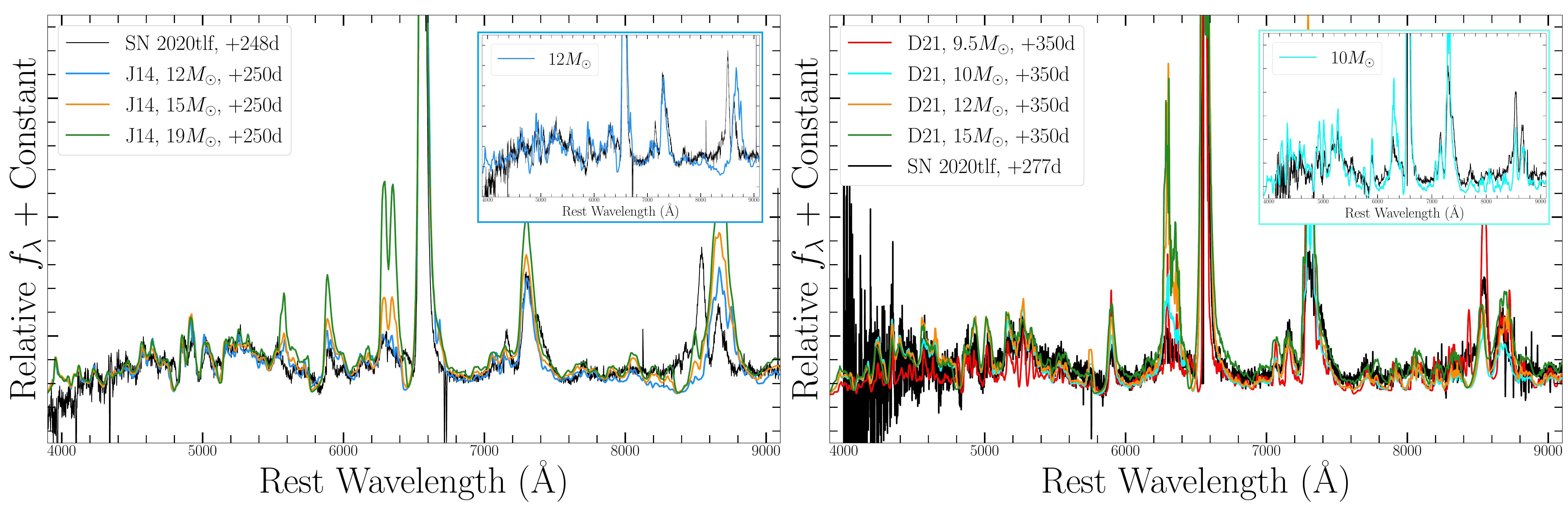}
\caption{\textit{Left:} Nebular spectrum of SN~2020tlf at +248~days post-explosion (black) compared to nebular spectral models at a similar phase from \cite{Jerkstrand14} for varying progenitor ZAMS masses: $12~\Msun$ (blue), $15~\Msun$ (orange) and $19~\Msun$ (green). The $12~\Msun$ ZAMS mass model, shown in upper right panel, is the best match to the nebular \sn{} spectrum at +248~days. \textit{Right:} Nebular models from \cite{Dessart21} for $9.5~\Msun$ (red), $10~\Msun$ (cyan), $12~\Msun$ (orange) and $15~\Msun$ (green) progenitor ZAMS masses with respect to \sn{} at +277~days post-explosion. Here, the $10~\Msun$ ZAMS mass model, shown in upper right panel, is the best match to the nebular \sn{} spectrum. \label{fig:neb_models} }
\end{figure*}

We performed non-LTE, radiative transfer modeling of the complete light curve and spectral evolution of \sn{} in order to derive properties of the progenitor and its CSM. Our modeling approach was similar to that presented in \citet{dessart17}, both in terms of initial conditions for the ejecta and CSM, the simulations of the interaction with the radiation-hydrodynamics code \heracles\ \citep{gonzalez_heracles_07,vaytet_mg_11,D15_2n}, and the post-processing with the non-LTE radiative-transfer code \cmfgen. For the progenitor star, we considered three models of RSGs produced by three different choices of mixing length parameter $\alpha_{\rm MLT}$. A greater $\alpha_{\rm MLT}$ boosts the convective energy transport in the H-rich envelope and produces a more ``compact" progenitor. This choice is generally required to match the color evolution of standard (i.e., non-interacting) Type II SNe (see discussion in \citealt{DH11_sn2p}; \citealt{d13_sn2p}) since more extended RSGs yield SNe II-P that both recombine and turn red too late in their evolution. The progenitor with increased radius may be more compatible with the pre-SN properties of \sn{} given the evidence for an inflated progenitor star prior to explosion (e.g., Fig. \ref{fig:preSN_LC}). 

In practice, we employed model m15mlt3 ($R_{\star}= 501~\Rsun$), m15 ($R_{\star}=768~\Rsun$), and m15mlt1 ($R_{\star} = 1107~\Rsun$) from \citet{d13_sn2p}. Taking these models at a time of a few 1000\,s before shock breakout, we stitch a cold, dense, and extended material from the progenitor photosphere out to some large radius. For simplicity, this material corresponds to a constant velocity wind ($v_{w}=50$\,$\kms$), a temperature of 2000\,K, and a composition set to the surface mixture of the progenitor \citep{davies_dessart_19}. We note that only a wind-like density profile (e.g., $s = 2$) is considered in our simultionas and not a shell-like profile of constant density (e.g., $s=0$). The former has proved to be the most realistic CSM structure for modeling similar events \citep{shivvers15, dessart17, terreran21} and the latter could be considered in future modeling. Nonetheless, we choose to adopt a CSM with a non-homogeneous density profile given that the most local CSM around massive stars appears to have complex CSM structure i.e., not constant density or shell-like. 

We consider wind mass loss rates of 0.01 and 0.03\,\msunyr\ from the progenitor surface out to a distance of order 10$^{15}$\,cm, beyond which the wind density is forced to smoothly decrease to 10$^{-6}$\,\msunyr\ at 6 or $8 \times 10^{15}$\,cm. These specific mass loss rates were chosen because simulations with these values, combined with a range of CSM extents, are most consistent with the observed SN properties e.g., early-time light curve evolution, peak luminosity and spectral features. A higher/lower $\dot{M}$ value outside of our adopted range is likely more inconsistent with our observations given the dependence of mass loss with increasing/decreasing the light curve rise time and peak luminosity, for example \citep{dessart17, Moriya17}. The dense part of the CSM is limited in extent to reflect the temporary boost in luminosity observed in \sn{}. That is, by increasing (decreasing) the radius that bounds the dense part of the CSM, one can lengthen (shorten) the duration over which the luminosity is boosted as a result of the change in diffusion time through the CSM and the amount of shock/ejecta energy trapped by the CSM. 

The interaction configurations described above are used as initial conditions for the multi-group radiation-hydrodynamics simulations with the code \heracles. For simplicity, we assume spherical symmetry and perform all simulations in 1-D; an asymmetric explosion could cause variations in the observed light curve and/or spectral evolution such as an extended SBO or slower evolving early-time light curve evolution. We use eight groups that cover from the ultraviolet to the far infrared: one group for the entire Lyman continuum, two groups for the Balmer continuum, two for the Paschen continuum, and three groups for the Brackett continuum and beyond. We also compute gray variants for some of the calculations: these tend to yield a shorter and brighter initial luminosity peak because the gray opacity underestimates the true opacity of a cold CSM crossed by high-energy radiation (see \citealt{D15_2n} for discussion). The difference between multi-group and gray transport is, however, modest because of the relative small CSM mass and extent. We adopt a simple equation of state that treats the gas as ideal with adiabatic index of $\gamma=$\,5/3. 

From the \heracles\ simulations, we extract the total luminosity crossing the outer grid radius as a function of time (the time origin for our light curves is usually set when the total luminosity recorded first exceeds 10$^{41}$\,\ergs). We also extract the hydrodynamical quantities (radius, velocity, density, and temperature) at selected epochs to post-process with the non-monotonic velocity solver in the non-LTE code \cmfgen\ (e.g., see \citealt{D15_2n}) and compute the emergent spectrum from the ultraviolet to the infrared. This approach captures the relative contributions from the fast ejecta, the dense shell at the interface between the ejecta and the CSM, the unshocked ionized CSM, as well as the outer cooler unshocked CSM. One limitation with this version of \cmfgen\ is the use of the Sobolev approximation (line transfer is therefore simplistic and line blanketing is underestimated) and the necessity to fix the temperature, which results from the hydrodynamics solution and the influence of the shock. This temperature from \heracles\ is not very accurate since the radiation hydrodynamics code treats the gas in a simplistic manner (the kinetic equations are not solved for). The composition adopted in our \cmfgen\ calculations at early times is homogeneous and corresponds to mass fractions of $X_{\rm He}=$\,0.34, $X_{\rm C}=1.28 \times 10^{-3}$, $X_{\rm N}=3.29 \times 10^{-3}$, $X_{\rm O}=4.67 \times 10^{-3}$ (and other metals at their solar metallicity value; $X_{\rm H}=$\,1-$X_{\rm all}$), which are the values predicted for a 15~$\Msun$ star \citep{davies_dessart_19}. The model atoms used in \cmfgen\ differ for early and late post-explosion times. At early times, we include \ion{H}{i}, \ion{He}{i/ii}, \ion{C}{i--iv}, \ion{N}{i--iv}, \ion{O}{ii--vi}, \ion{Mg}{ii}, \ion{Si}{ii}, \ion{S}{ii}, \ion{Ca}{ii}, \ion{Cr}{ii--iii}, \ion{Fe}{i--iv}, \ion{Co}{ii--iii}, and \ion{Ni}{ii--iii}. At later times, we drop the high ionization stages and add the atoms or ions \ion{Na}{i}, \ion{Mg}{i}, \ion{Si}{i}, \ion{S}{i}, \ion{Ca}{i}, \ion{Sc}{i--iii}, \ion{Ti}{ii--iii}. 

\begin{figure}[t!]
\centering
\includegraphics[width=0.45\textwidth]{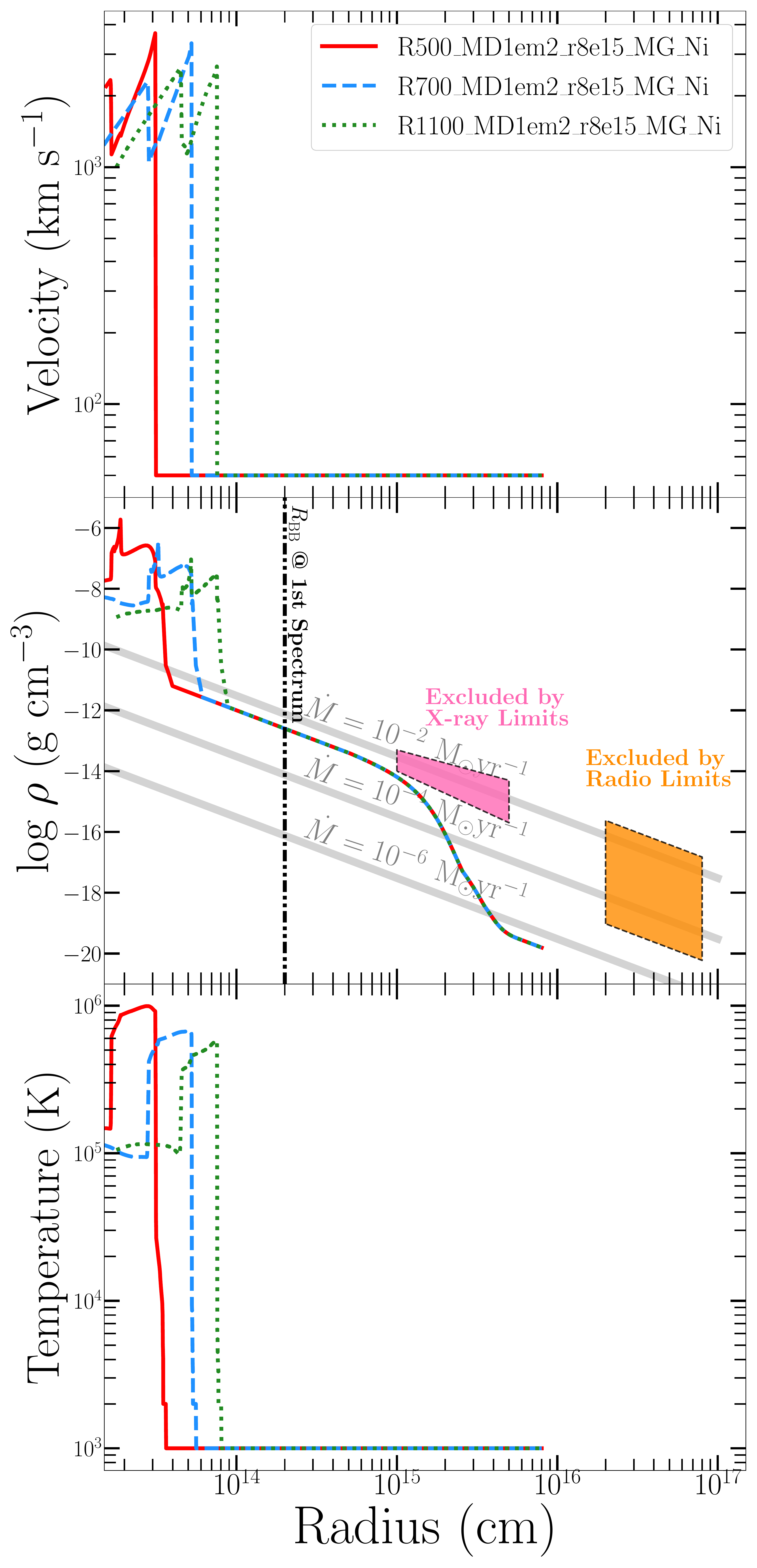}
\caption{Initial stellar structure and circumstellar environment at a few 1000s before shock breakout for the three most consistent models: $500~\Rsun$ (solid red line), $700~\Rsun$ (dashed blue line) and $1100~\Rsun$ (dotted green line). In middle panel, lines of constant mass loss are shown for $v_w = 50~ \kms$. Regions of density parameter space excluded by X-ray and radio limits shown in pink and orange, respectively. Blackbody radius as derived from first spectrum shown as black line. \label{fig:VDT}}
\end{figure}

\begin{figure}[t!]
\centering
\includegraphics[width=0.45\textwidth]{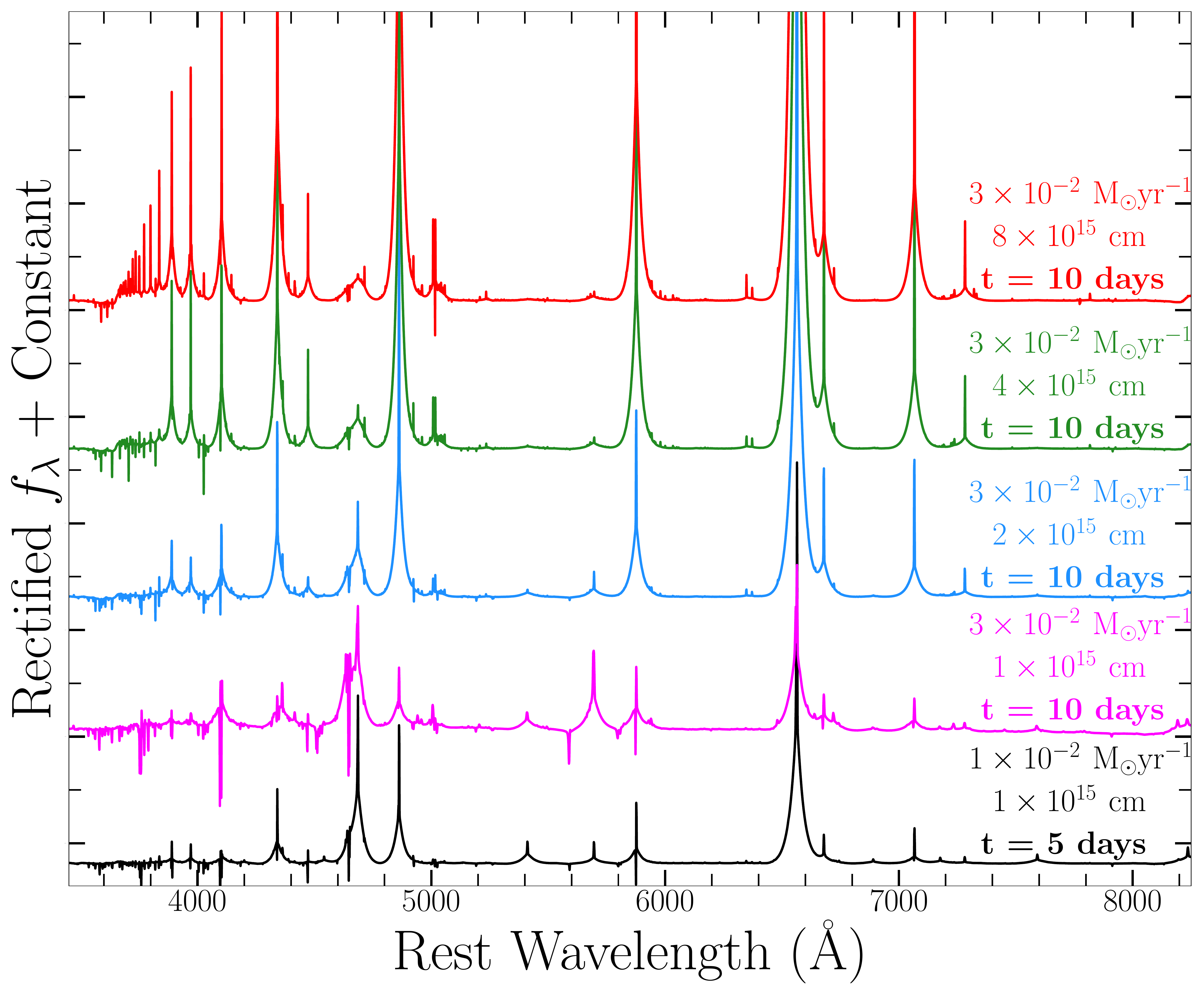}
\caption{Comparison of early-time \cmfgen\ model spectra for varying CSM radius extent and wind mass loss rates. Spectra shown in red ($r_{\rm CSM} = 8 \times 10^{15}$~cm), green ($r_{\rm CSM} = 4 \times 10^{15}$~cm), blue ($r_{\rm CSM} = 2 \times 10^{15}$~cm), and magenta ($r_{\rm CSM} = 1 \times 10^{15}$~cm) include a mass loss rate of $\dot{M} = 0.03~\Msun$~yr$^{-1}$ and phase of +10~day. Model shown in black includes $\dot{M} = 0.01~\Msun$~yr$^{-1}$ and $r_{\rm CSM} = 1 \times 10^{15}$~cm at a phase of +5~day.  \label{fig:csm_extent}}
\end{figure}

All model characteristics are presented in Table \ref{tbl:m_table} and the CSM structure of most consistent models are plotted in Figure \ref{fig:VDT}. Furthermore, in Figure \ref{fig:csm_extent}, we show how early-time \cmfgen\ spectral models are influenced by both the extent of the CSM and the progenitor mass loss rate. We show that for $\dot{M} = 0.03~\Msun$~yr$^{-1}$ at a phase of +10~days since explosion, models with more extended CSM radii (e.g., $4 - 8 \times 10^{15}$~cm) have wider, more prominent emission profiles from CSM interaction than models with less extended CSM (e.g., $1-2 \times 10^{15}$~cm). We also show that for a model with $\dot{M} = 0.01~\Msun$~yr$^{-1}$ and CSM radius of $10^{15}$~cm, narrow emission lines are less prominent and shorter lived than other models with larger CSM radii and mass loss rates. Furthermore, the more compact the CSM, the higher the ionization, which influences the spectral features present because a smaller optically thick volume leads to a higher radiation temperature and consequently a higher gas temperature.

\begin{figure*}
\centering
\subfigure[]{\includegraphics[width=0.49\textwidth, height=2.5in]{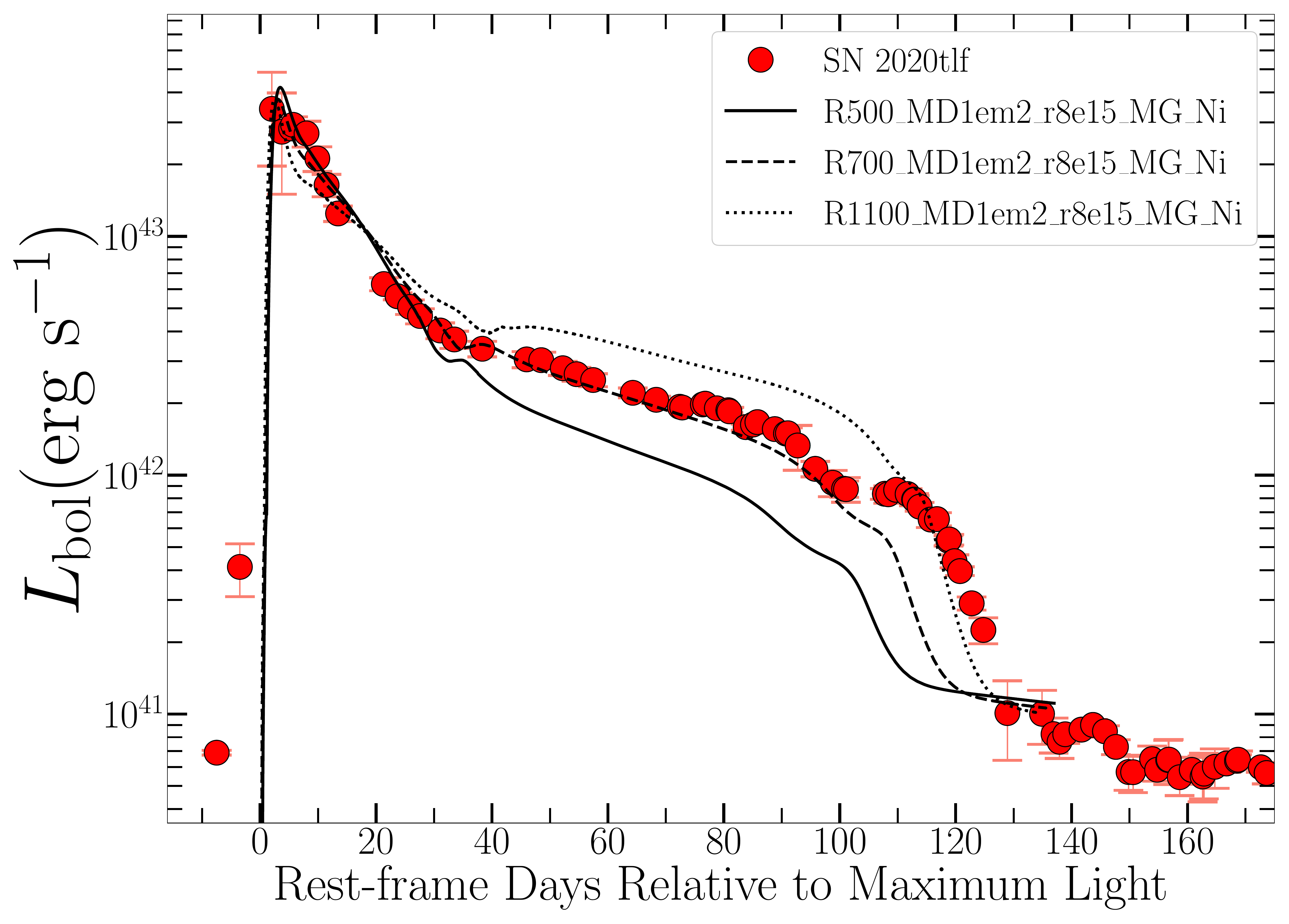}}
\subfigure[]{\includegraphics[width=0.47\textwidth, height=2.5in]{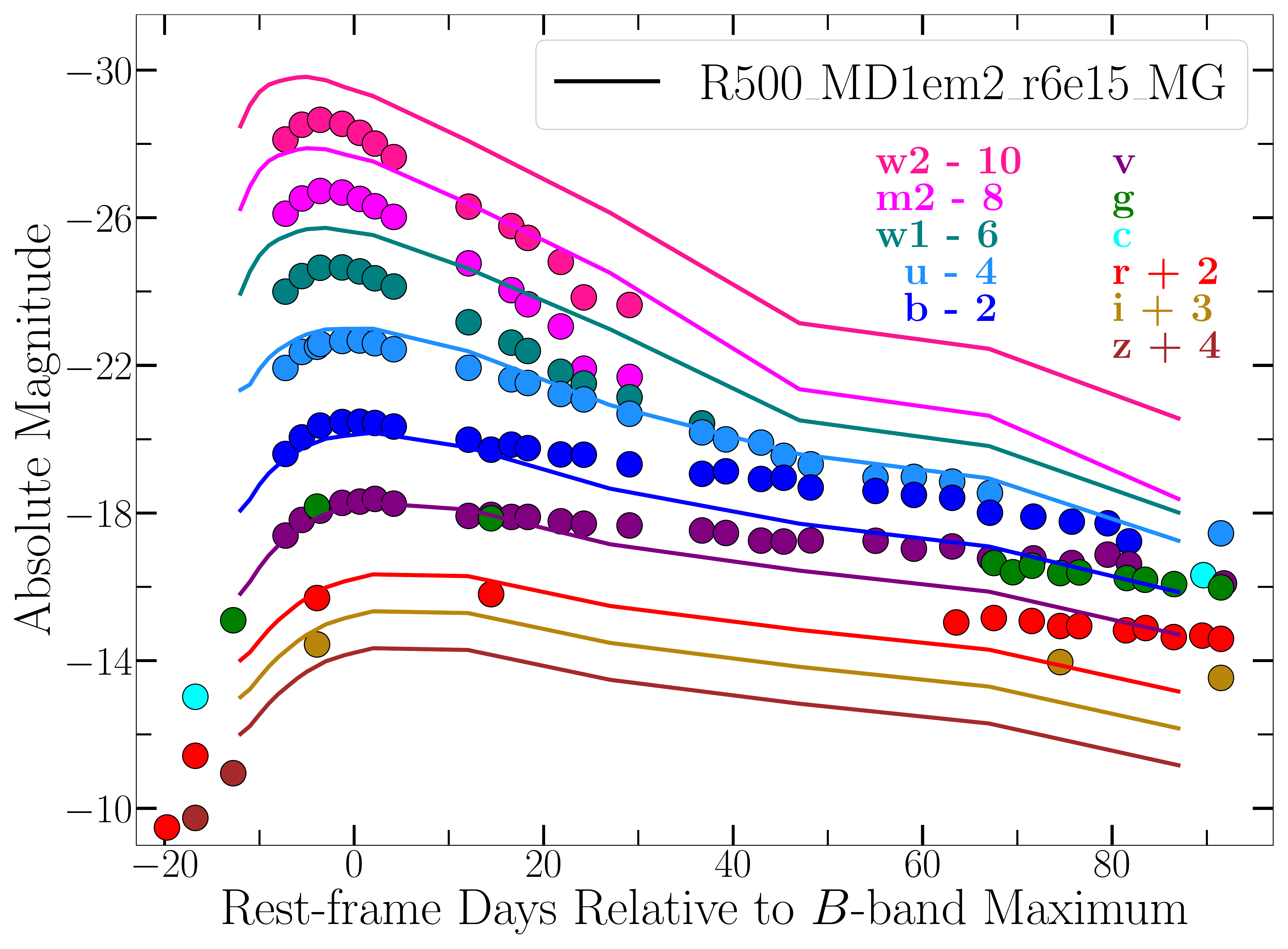}}
\caption{(a) Bolometric light curve models shown in black for CSM that extends to $r = 8\times 10^{15}$~cm around $15~\Msun$ progenitor ($\dot M = 0.01$~$\Msun$ yr$^{-1}$) of varying envelope radii: $501~\Rsun$ (solid line), $768~\Rsun$ (dashed line) and $1107~\Rsun$ (dotted line). Despite the imperfect match to the complete bolometric evolution, the most extended progenitor model ($R_{\star} = 1107~\Rsun$) is the only simulation that can reproduce the elongated light curve plateau observed in \sn{}. (b) Multi-band, early-time light curve model for extended CSM ($r = 6\times 10^{15}$~cm) and mass loss rate of $\dot M = 0.01$~$\Msun$ yr$^{-1}$. Models do not extend in time to the phases of the earliest \sn{} photometry given the low luminosity of multi-band SN detections show above (e.g., $\delta < -15$~days relative to maximum).  \label{fig:Dmodel_LC} }
\end{figure*}

In Figure \ref{fig:Dmodel_LC}, we present the most consistent bolometric \heracles\ models and multi-band \cmfgen\ models with respect to \sn{} observations. We find that an extended progenitor radius of $\sim$~1100~$\Rsun$ (dotted line in Fig. \ref{fig:Dmodel_LC}a) is the most consistent with the long-lived and very luminous plateau phase in \sn{}. Additionally, the early light curve of \sn{}, which is strongly influenced by the interaction of the ejecta with the CSM, is best modeled by a mass loss rate of $\dot M = 10^{-2}$~$\Msun$ yr$^{-1}$ ($v_{w} = 50~\kms$) and a dense CSM that extends out to a radius of $r_{\rm CSM} = 10^{15}$~cm -- the influence of the more tenuous CSM beyond that radius is modest and eventually naught (i.e., at $>$\,40~days). As shown in Figure \ref{fig:Dmodel_LC}(b), the light curve model matches the multi-band early-time photometry in most optical/NIR bands, but it over-predicts the UV peak in \emph{Swift} filters by $\sim 1$~mag. There are many possible reasons for this inconsistency given the simplicity of our assumptions. For example, one possible cause is that there is additional host extinction near the explosion site that was not able to be measured through typical reddening estimates (e.g., see \S\ref{SubSec:Phot}). Additionally, while the model light curves are consistent with the peak bolometric luminosity and decline rate, they cannot reproduce the long rise-time observed in \sn{} following the pre-SN activity. However, model first light is defined when the simulation bolometric light curve rises above $10^{41}$~erg s$^{-1}$ and thus the two bolometric light curve points in Figure \ref{fig:Dmodel_LC}(a) would not be reproduced by the models given their low luminosities. Nonetheless, it is worth noting that the model light curves predict a faster rise ($t_{\rm exp} \approx 59108$~MJD) than our estimate based on the earliest photometry ($t_{\rm exp} \approx 59098.7$~MJD). If the former is the true time of explosion, the earliest detections may represent additional precursor activity or SBO emission from an asymmetric explosion or CSM. 

\begin{figure*}
\centering
\subfigure[]{\includegraphics[width=0.49\textwidth]{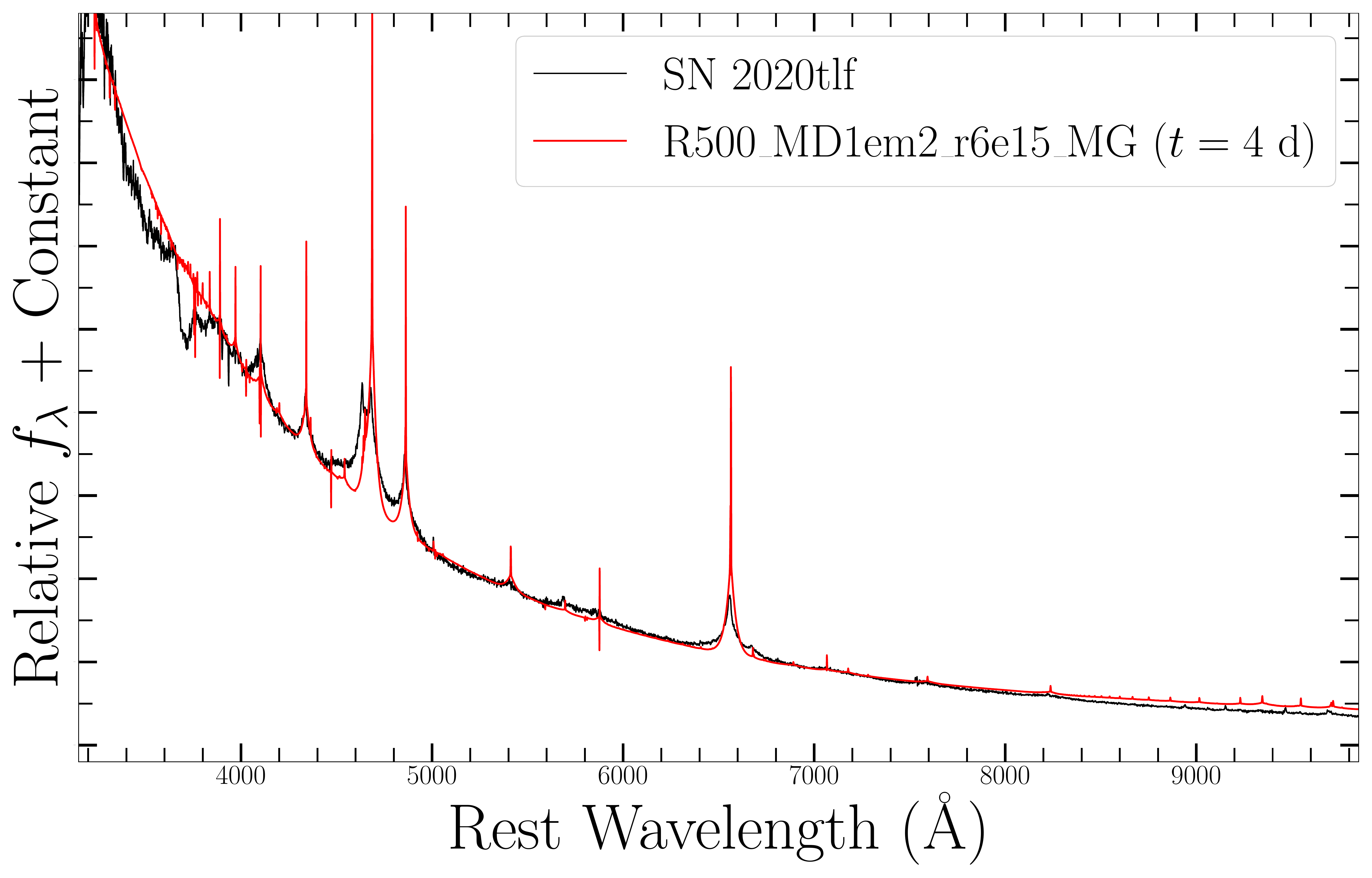}}
\subfigure[]{\includegraphics[width=0.49\textwidth]{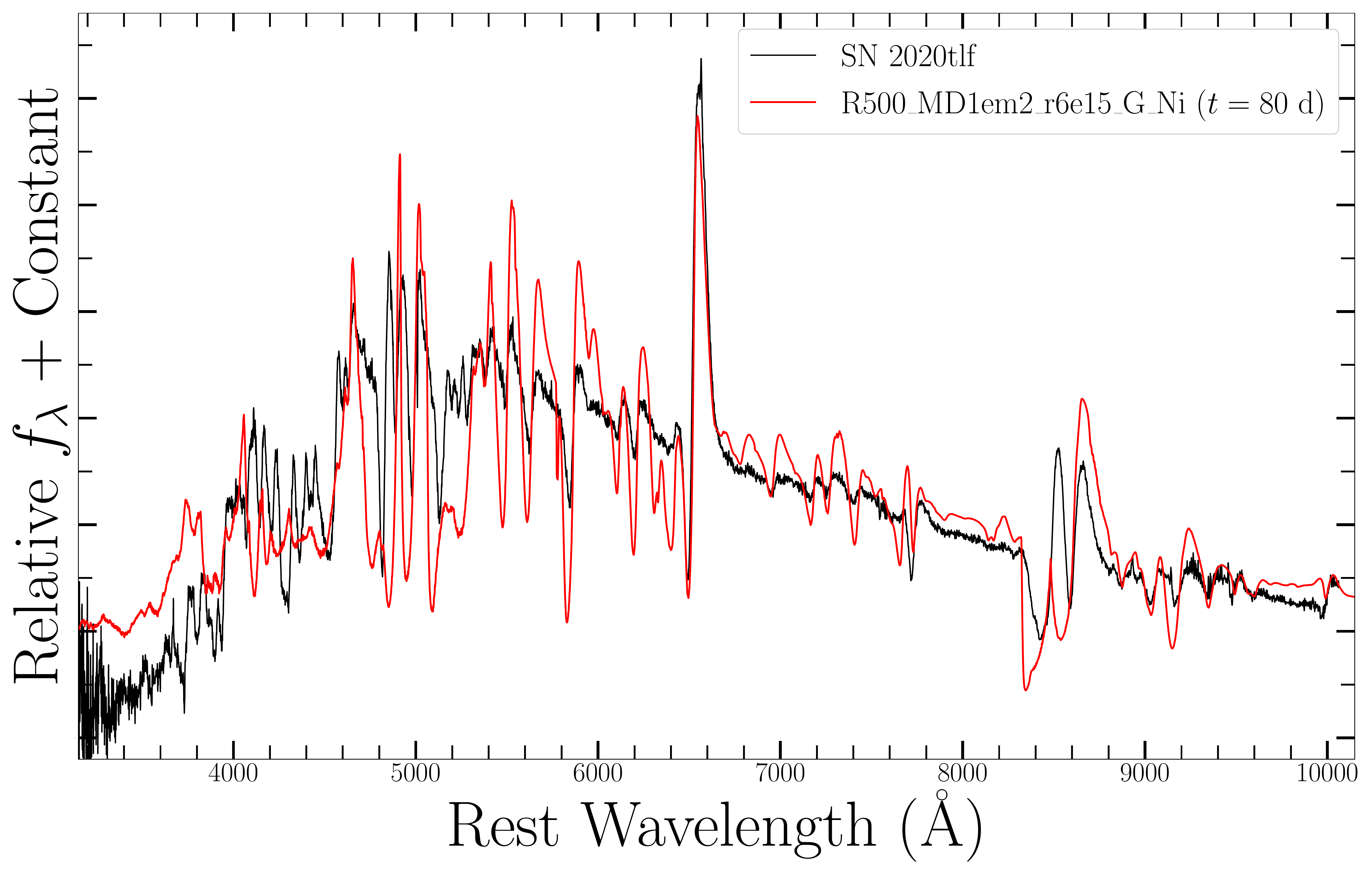}}
\caption{(a) Early-time, LRIS photo-ionization spectrum of SN~2020tlf (black) compared to \cmfgen\ CSM interaction model (red) at +4~days after model first light. Model CSM that extends to $r = 6\times 10^{15}$~cm around $15~\Msun$ progenitor ($\dot M = 0.01$~$\Msun$ yr$^{-1}$). (b) Mid-time \cmfgen\ model spectrum at +80~days after model first light with gray variant solver and with ${}^{56}\textrm{Ni}$ included. \label{fig:Dmodel_spec} }
\end{figure*}

In Figure \ref{fig:Dmodel_spec}(a), we present the most consistent \cmfgen\ model with respect to the first spectrum of \sn{}. The model spectrum is a consistent match to the widths and strengths of emission features such as \ion{H}{i}, \ion{He}{i--ii}, and \ion{C}{iii--iv}, as well as the continuum shape and temperature. Despite the presence of N in the model CSM composition, the most consistent model cannot perfectly reproduce the \ion{N}{iii} emission feature on the bluewards side of the \ion{N}{iii} + \ion{He}{ii} feature. Alternative \cmfgen\ model procedures that include a static wind structure (e.g., see \citealt{shivvers15, Boian20, terreran21}) reproduce this \ion{N}{iii} line but employ a strong N enrichment, incompatible with the 10-12~$\Msun$ progenitor mass inferred for \sn{} (e.g., see \S\ref{subsec:spec_analysis}). Furthermore, the most consistent early-time spectral model is for a phase of +4~days after explosion, therefore indicating a time of first light of MJD 59105, which is between the estimates derived from either early-time photometry or light curve modeling. We also present a late-time \cmfgen\ model at +80~days with respect to the +95~day spectrum in Figure \ref{fig:Dmodel_spec}(b). This model accurately matches most features and line profiles, as well as the boosted continuum at blueward wavelengths that could be the result of persistent CSM interaction. 

The modeling of \sn{}'s light curve and early-time spectrum suggests similar CSM properties and progenitor mass loss to other SNe~II with \cmfgen\ modeling of early-time spectra. Compared to the sample of \cmfgen-modeled interacting SNe~II presented by \cite{Boian20} and expanded by \cite{terreran21}, the \sn{} progenitor mass loss rate of $10^{-2}$~$\Msun$ yr$^{-1}$ is consistent but slightly greater than that of some events with early photo-ionization signatures such as SNe~1998S, 2017ahn, 2013fs and 2020pni ($\dot M \approx 5-8 \times 10^{-3}$~$\Msun$ yr$^{-1}$, $v_w = 40 - 200~\kms$), and is lower than SNe~2013fr, 2014G, and 2018zd ($\dot M \approx 0.04 - 0.2$~$\Msun$ yr$^{-1}$, $v_w = 500 - 800~\kms$). The mass loss derived for \sn{} is also very similar to SN~IIn 2010mc ($v_w = 300~\kms$) that also had confirmed precursor emission but whose narrow emission lines persisted for all of the SN evolution. In terms of the disappearance of narrow emission features in these events, \sn{} cannot be constrained as well as other SNe~II with higher cadence early-time spectral coverage, but does have a lower limit on this timescale of $t \geq 10.3$~days since first light. Compared to the SN sample presented in Figure 14 of \cite{terreran21}, the time of narrow line disappearance in \sn{} is most likely greater than all other presented events besides SN~1998S, whose narrow features persisted until $\sim30$~days since first light. This indicates a much more extended CSM in the case of SNe~1998S and, to a lesser degree, 2020tlf, than other events where the observed narrow features persisted for $\lesssim 12$~days since first light. 

\section{CSM Constraints from X-ray/Radio Emission} \label{sec:xray_radio}

The shock interaction with a dense CSM is a well-known source of X-ray emission (e.g, \citealt{Chevalier06}). To constrain the parameter space of CSM densities that are consistent with the lack of evidence for X-ray emission at the location of SN2020tlf ($\delta t=$ 11.0 -- 23.0 days since first light; \S\ref{SubSec:XRT}), we start by generating a grid of intrinsic $n_{\rm H,host}$ values. We then assumed an absorbed bremsstrahlung spectrum with $T=20$~keV, in analogy to other strongly interacting SNe (e.g. 2014C, \citealt{Margutti17}) with different levels of $n_{\rm H,host}$ and converted the upper limit on the observed count-rate into an upper limit on the observed flux $F_x$ using \texttt{XSPEC}. The resulting luminosity limits are derived as  $L_x=4\pi D^2 F_x$. We then compare the grid of $L_X$ upper limits to the X-ray luminosities from the analytic formalism presented in \cite{Chevalier06} for  free-free emission from reverse-shocked CSM:
\begin{equation}
    L_{\rm ff} = 3\times 10^{35} \frac{(n-3)(n-4)^2}{4(n-2)} \beta^{1/2} \zeta_2^{-1} A_{\star}^2 t_{10}^{-1} \ {\rm erg \ s^{-1}}
\end{equation}
where $n$ is the index of the progenitor outer density profile $\rho(r)\propto r^{-n}$, $\beta$ is the ratio of electron to equilibrium temperatures (e.g., $T_e /T_{\rm eq}$), $\zeta$ is a chemical composition parameter and $\zeta = 1$ for H-rich material, $A_{\star}$ is a mass loss parameter calibration such that $A_{\star} = 1$ for $\dot{M} = 10^{-5}$~$\Msun$ yr$^{-1}$ and $v_w = 1000~\kms$, and $t_{10} = (t_{\rm exp} / 10$~days). For this model, we use $n = 15$ as expected for extended progenitor stars, $\beta = 1$ (equilibrium), and $t_{10} = 2$ (at maximum light) \citep{Chevalier06}. For a given $n_H$, allowed model X-ray model luminosities must be less than the flux limit derived from the stacked XRT image and the specific $n_H$ value must be less than that derived from the model $A_{\star}$ value e.g., $n_H = 1000 \cdot A_{\star} / (4 \pi R v_w^2m_p)$ for $R = (1-5)\times 10^{15}$~cm and $v_w =  50~\kms$. All X-ray luminosities that satisfy these conditions are used to find the resulting $A_{\star}$ values that are then converted into a range of $\dot{M}$ that are permitted by the observed luminosity limit. We then find an allowed range of progenitor mass loss rates of $\dot{M} < 0.001$~$\Msun$ yr$^{-1}$ or $\dot{M} > 0.02-0.08$~$\Msun$ yr$^{-1}$, for $v_w = 50~\kms$. Furthermore, we convert these mass loss limits into limits on the CSM density at radius $r \approx (1-5)\times 10^{15}$~cm (positions of shock at peak, traveling at $\sim 0.03-0.1c$) and present them in Figure \ref{fig:VDT}.

We interpret the radio upper limits of \S\ref{SubSec:VLA} ($\delta t=146-320$ days since first light) in the context of synchrotron emission from electrons accelerated to relativistic speeds at the explosion's forward shock, as the SN shock expands into the medium. We adopt the synchrotron self-absorption (SSA) formalism by \cite{Chevalier98} and we self-consistently account for free-free absorption (FFA) following \cite{Weiler02}. For the calculation of the free-free optical depth $\tau_{\rm ff}(\nu)$, we adopt a wind-like density profile $\rho_{\rm{csm}}\propto r^{-2}$ in front of the shock, and we conservatively assume a gas temperature $T=10^4\,\rm{K}$ (higher gas temperatures  would lead to tighter density constraints). The resulting SSA+FFA synchrotron spectral energy distribution depends on the radius of the emitting region, the magnetic field, the environment density and on the shock microphysical parameters $\epsilon_B$ and $\epsilon_e$ (i.e.~the fraction of post-shock energy density in magnetic fields and relativistic electrons, respectively). Additional details on these calculations can be found in the Appendix of \cite{terreran21}.

We find that for a typical shock velocity of $\sim 0.1c$ \citep{Chevalier06} and microphysical parameters $\epsilon_B = 0.1$ and $\epsilon_e = 0.1$, the lack of detectable radio emission is consistent with either a low-density medium with density corresponding to $\dot M<1.3\times 10^{-5}\,\rm{M_{\sun}\,yr^{-1}}$, or a higher density medium with $\dot M>0.032\,\rm{M_{\sun}\,yr^{-1}}$ that would absorb the emission (e.g., $\rho = \dot{M} R_{\rm CSM} v_w^{-1} V^{-1})$. However, this high density limit is excluded based on the optical photometry and spectroscopy. These $\dot M$ values are for a wind velocity $v_w=50\,\rm{km\,s^{-1}}$ and CSM radii of $r_w = (2-8)\times 10^{16}$~cm. We present these limits as excluded regions of the \sn{} CSM density parameter space in Figure \ref{fig:VDT}. These derived mass loss rates suggest a confined, dense CSM around the \sn{} progenitor star from enhanced mass loss in the final months-to-year before explosion, as well as more diffuse, lower density material extending out to large radii, suggestive of a steady-state RSG wind. The $\dot{M}$ values inferred from radio and X-ray observations are also consistent with other photo-ionization events with multi-wavelength observations e.g., SNe~2013fs \citep{yaron17} and 2020pni \citep{terreran21}.   

\section{Discussion} \label{sec:discussion}

\subsection{A Physical Progenitor Model}\label{subsec:visual}

\begin{figure*}[t]
\centering
\includegraphics[width=\textwidth]{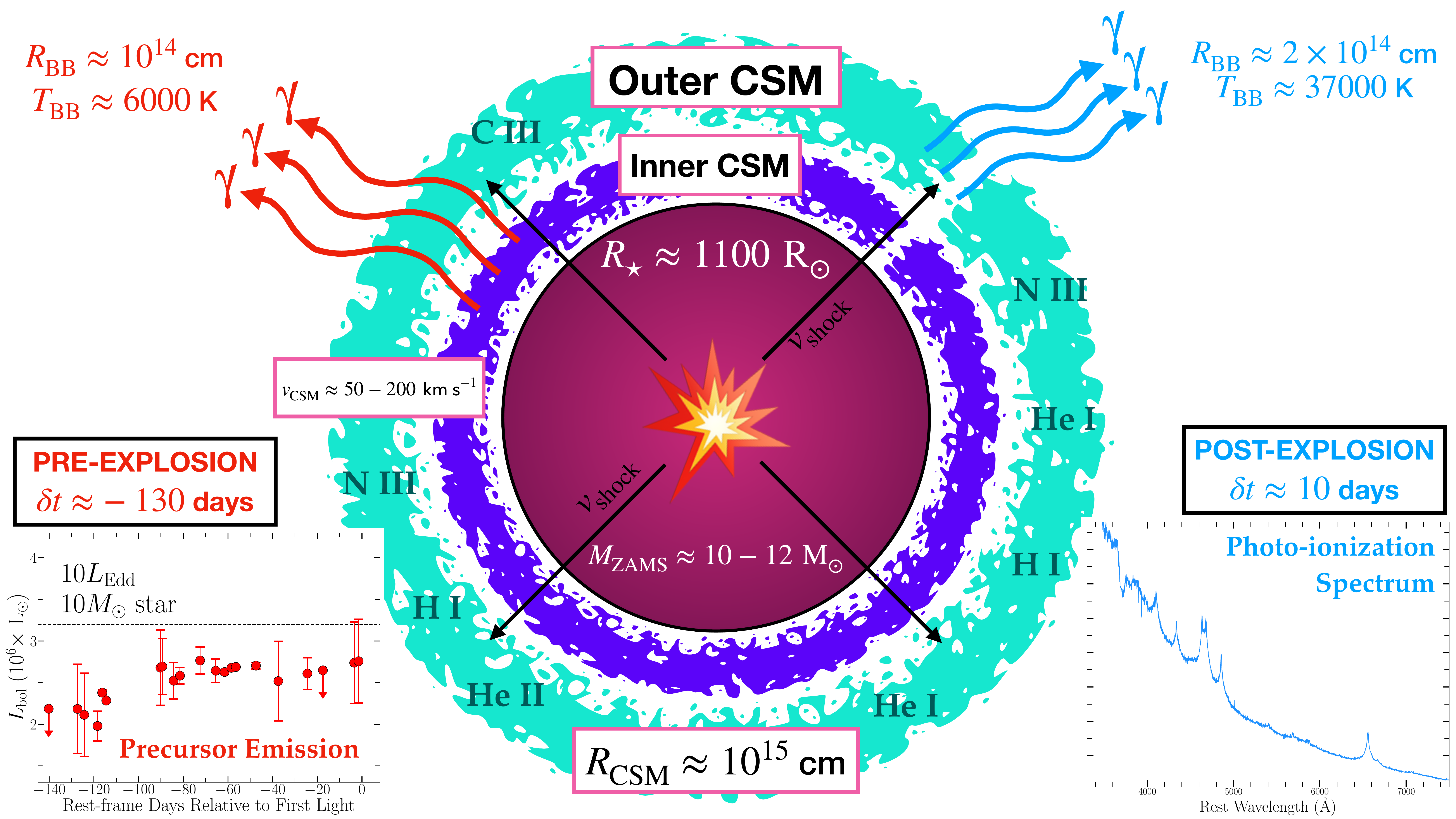}
\caption{Visual representation  of SN~2020tlf's progenitor system at the time of explosion (\S\ref{subsec:visual}). Here, the SN shock breaks out from an extended H-rich envelope of a 10-12~$\Msun$ RSG progenitor star and collides with dense CSM ($r \sim 10^{15}$~cm, $v_w \approx 50-200~\kms$), inducing photo-ionized spectral lines observed in the earliest SN spectrum (shown in blue). Precursor emission was detected for $\sim 130$~days prior to explosion (shown in red) due to the ejection of stellar material. For slower wind velocities ($v-w \lesssim 50~\kms$), outer CSM (cyan circle) represents the material ejected prior to the precursor ejection of inner CSM (dark blue circle). However, material driven off in the pre-SN activity could be the same material as is visible in the photo-ionization spectrum for wind velocities of $v_w \approx 50-200~\kms$.   \label{fig:visual} }
\end{figure*}

Pre- and post-explosion panchromatic observations have provided an unprecedented picture of the \sn{} progenitor system. In Figure \ref{fig:visual}, we attempt to combine inferences made from observation and modeling to create a visualization of the explosion and surrounding progenitor environment. Our model is a snapshot of the SN at the time of first light and contains physical scales and parameters such as distance, velocity and composition estimates. The illustration also includes progenitor properties derived from precursor emission in the $\sim$130~days leading up to SBO.

As discussed in \S\ref{sec:modeling}, \cmfgen\ modeling of the \sn{} light curve and photo-ionization spectrum indicate that the 10-12~$\Msun$ (ZAMS e.g., see \S \ref{subsec:spec_analysis}) progenitor star had radius of $\sim$1100~$\Rsun$ and was losing mass at an enhanced rate of $\dot M = 10^{-2}$~$\Msun$ yr$^{-1}$ in the final months before explosion, leading to the creation of dense CSM (shown in sea foam green; Fig. \ref{fig:visual}) at distances $r \lesssim 10^{15}$~cm; lower density CSM extended out to $r \approx 8 \times 10^{15}$~cm. These models suggest that the \sn{} progenitor star had a total CSM mass of $\sim 0.05-0.07~\Msun$ in the local environment at the time of explosion. At the time of the photo-ionization spectrum ($\delta t \approx 10$~days post-explosion), \sn{} had a blackbody temperature $T\approx 3.7 \times 10^{4}$~K at the thermalization depth and an emitting radius of $\sim 2 \times 10^{14}$~cm (shown in light blue; Fig. \ref{fig:visual}). The identification of narrow emission lines from photo-ionized material in the earliest spectrum confirms that the CSM was comprised of high-ionization species such as \ion{He}{ii}, \ion{N}{iii} and \ion{C}{iii--iv}, as well as lower ionization species such as \ion{H}{i} and \ion{He}{i}. As observed in the photo-ionization spectrum, the wind velocity of the CSM is likely $v_w \approx 50 - 200$~$\kms$. 

Prior to explosion, the \sn{} progenitor star produced detectable precursor emission for $\sim$130~days prior to SBO. The observed emission is relatively constant leading up to explosion ($\sim 10^{40}$~erg s$^{-1}$), with an average emission radius and temperature of $\sim 10^{14}$~cm and $\sim$5000~K, respectively (shown in red; Fig. \ref{fig:visual}). Because the blackbody radius rate of change during the pre-SN activity is $\sim 1000~\Rsun$ over a timescale of $\sim 30$~days, it is likely that the observed pre-SN emission is not derived from the stellar surface; the Kelvin-Helmholtz timescale for a $\sim10~\Msun$ progenitor to change in radius at this rate is $\tau_{th} \gtrsim 200$~days. As discussed in \S \ref{subsec:precursor}, this precursor emission could have resulted from the ejection, and subsequent CSM interaction, of $>0.3$~$\Msun$ of stellar material that was most local to the progenitor star (shown in dark blue; Fig. \ref{fig:visual}). However, this estimated mass of precursor material is larger than the CSM mass of $\sim 0.05-0.07~\Msun$ in the most consistent \cmfgen\ models. There is also a possibility that the precursor emission arose from a super-Eddington wind that drove off $> 10^{-3}$~$\Msun$. However, this mass loss mechanism may be unphysical for the low mass progenitor of \sn{}.

An open question in understanding the pre-explosion activity of the \sn{} progenitor star is whether material ejected in the detected precursor is the same CSM responsible for the photo-ionization spectrum at $\sim 10$~days post-explosion. The validity of this conclusion is dependent on what wind velocity we adopt in the range of possible CSM velocities ($\sim 50 - 200~\kms$) derived in \S \ref{subsec:spec_analysis}. If the precursor material was ejected with a velocity of $v_w \approx 50-200~\kms$, that specific CSM could reach radii of $r \approx (0.6 - 2.4) \times 10^{14}$~cm in the $\sim$~140~days before the photo-ionization spectrum was obtained. However, if the material was driven off from the surface of a progenitor star with an extended radius of $\sim$1100~$\Rsun$, the distance reached by this material in $\sim$~140~days increases to $r \approx (1.4 - 3.2) \times 10^{14}$~cm. These distances are consistent with the blackbody radius of $\sim 2 \times 10^{14}$~cm at the time of the photo-ionization spectrum. Therefore, unless the wind velocities are $< 50 ~ \kms$, it is feasible that the material driven off to cause the precursor emission is the same CSM material that was photo-ionized by the SN shock wave, resulting in the narrow emission lines present in the early-time spectrum.

\subsection{Progenitor Mass Loss Mechanisms}

The detection of precursor emission, combined with the presence of dense CSM (e.g., see \S \ref{sec:modeling}) around the $\sim$10-12~$\Msun$ progenitor of \sn{} necessitates a physical mechanism for enhanced mass loss and luminosity, together with a likely structural change to the stellar envelope (inflation), in the final year to months before core collapse. As shown in \S \ref{subsec:precursor}, powering the precursor emission would require $>0.3~\Msun$ of material through CSM interaction and $>10^{-3}~\Msun$ of material via a super-Eddington wind, the latter of which is much smaller than the CSM mass derived from light curve and spectral modeling (e.g., \S \ref{sec:modeling}). However, a super-Eddington wind is most likely unphysical given the small progenitor ZAMS mass derived from the nebular spectra; it will also lead to larger CSM densities than those derived from modeling (\S\ref{sec:modeling}). Therefore, in the final $\sim$year of stellar evolution, a physical mechanism is needed to produce enhanced mass loss (e.g., 0.01~$\Msun$ yr$^{-1}$ derived from modeling) and detectable precursor flux. 

As discussed initially in \S \ref{subsec:precursor}, wave-driven mass loss is one process that occurs in late stage stellar evolution that could lead to the ejection of material from the progenitor surface, also resulting in detectable pre-explosion emission. The excitation of gravity waves by oxygen or neon burning in the final years before SN can allow for the injection of energy (e.g., $\sim10^{46-48}$~erg) into the outer stellar layers, resulting in an inflated envelope and/or eruptive mass loss episodes \citep{meakin07, arnett09, Quataert12, Shiode14, fuller17, Wu21}. While this mass loss mechanism is a potential explanation for the precursor activity in \sn{}, there are currently no wave-driven models that can match the observed pre-explosion activity. As shown in Figure \ref{fig:preSN_LC}(b), the model for a 15$\Msun$ RSG undergoing wave-driven mass loss by \cite{fuller17} does not reproduce the bolometric luminosities of the \sn{} precursor, but is consistent in radius in the final $\sim$130~days before core-collapse. In an updated study of wave-driven models, \cite{Wu21} show that pronounced pre-SN outbursts could occur in progenitor stars of similar mass to that of \sn{} (e.g., $<14~\Msun$). However, the timescales of these mass loss episodes are inconsistent with relatively constant emission observed in the \sn{} precursor in the final $\sim 130$~days before explosion. 

A related, promising explanation for enhanced mass loss is the sudden deposition of energy into the internal layers of a massive star outlined by \cite{dessart10}. Agnostic to the mechanism for energy injection, these models show that a release of energy ($E_{\rm dep}$) that is on the order of the binding energy of the stellar envelope ($E_{\rm bind}$) will create a shock front that will propagate outwards, causing a partial ejection of the stellar envelope. As shown in Figures 8 \& 9 in \cite{dessart10} for an 11~$\Msun$ progenitor, energy injection of $E_{\rm dep} \sim E_{\rm bind}$ will produce a detectable pre-SN outburst that is continuous for hundreds of days and matches the observable in the \sn{} precursor e.g., $L \approx 10^{6}~\Lsun$, $T \approx 5000$~K and $R \approx 1500~\Rsun$. Possible causes for such energy release could be gravity waves from neon/oxygen burning or even a silicon-flash in the final 100-200~days before explosion. For the latter, \cite{woosley15} show that low mass progenitors (9-11~$\Msun$) can produce precursor emission in the final $\sim$~year before explosion as a result of silicon deflagration in their cores. Specifically, the 10.0C progenitor model listed in Table 3 of \cite{woosley15} has consistent pre-SN properties to that observed in the \sn{} precursor e.g., $L \approx 10^{40}$~erg s$^{-1}$, $R \approx 10^{14}$~cm. Overall, the simulations from both of these studies are promising scenarios to explain the enhanced mass loss observed in \sn{}.

\subsection{Pre-Explosion Variability in SN II Progenitors}

\sn{} represents the first instance of a SN~II where significant variability has been detected in the RSG progenitor star prior to explosion. These observations reveal a clear disjuncture from the findings by other studies that examined the pre-SN activity of SN~II progenitors in the final years before core-collapse. For example, the progenitor behavior prior to SN~II-P, 2017eaw has been studied extensively using pre-explosion UV/optical/IR imaging in the final decades before explosion \citep{kilpatrick17eaw, Rui19, Tinyanont19, vandyk19}. However, the $\sim 11-13~\Msun$ RSG progenitor of SN~2017eaw only reached a luminosity of $\sim 4.7~\Lsun$ prior to explosion \citep{kilpatrick17eaw}, with IR variability estimated to be at most $\Delta \nu L_\nu \approx 5000~\Lsun$ \citep{Tinyanont19}; both of these progenitor luminosity estimates being orders of magnitude lower than the precursor recorded prior to \sn{}. Similar quiescent behavior is also observed in sample studies on the long-term variability of SN~II progenitors by \cite{Johnson18} as well as the single object study of SN~II-P, ASASSN-16fq by \cite{Kochanek17}. Based on the findings of the former, the \sn{} progenitor lies in the $< 37\%$ of RSGs that exhibit extended outbursts after O ignition i.e., $\sim 1000-100$~days before explosion, depending on the progenitor mass. Furthermore, \cite{Kochanek17} and \cite{Johnson18} both find that these SN~II progenitors show very little variability (e.g., $\Delta \nu L_\nu \lesssim 3000~\Lsun$) for years-to-days before core-collapse. Interesting, none of these SNe~II showed spectroscopic evidence of interaction with CSM shed by the progenitor during episodes of enhanced mass loss, as detected directly in the earliest spectrum of \sn{}. This may indicate that only RSG progenitors with CSM that is dense enough to be detectable in early-time spectra of young SNe~II are also able to produce luminous precursor emission of $\sim10^6~\Lsun$, as observed prior to \sn{}.

\section{Conclusions} \label{sec:conclusion}

In this paper we have presented pre- and post-explosion (-130 to +300 days) panchromatic observations of the nearby SN~II, 2020tlf located in the star-forming SAcd-type galaxy NGC~5731 at $d\approx36.8$~Mpc. Our observations and modeling cover the electromagnetic spectrum from the X-rays to the radio band, specifically high cadence coverage in UV/optical/NIR. Future studies (e.g., ``Final Moments II--'') will focus on samples of 20tlf-like events in order to constrain the late-stage evolution of RSG progenitors through pre-SN emission and ``flash'' spectroscopy. Below we summarize the primary observational findings that make \sn{} one of the most intriguing SNe~II to date: 

\begin{itemize}

\item \sn{} is the first normal SN~II-P/L with confirmed precursor emission for $\sim130$~days prior to first light. Pre-explosion activity was detected in $riz$-band YSE/PS1 filters, which showed an average pre-SN bolometric luminosity, blackbody radius and temperature of $\sim 10^{40}$~erg s$^{-1}$ ($\sim 2 \times 10^{6}~\Lsun$), $\sim 10^{14}$~cm ($\sim 1500 ~ \Rsun$), and $\sim 5000$~K, respectively. 

\item The early-time optical spectrum of \sn{} is nearly identical to the earliest spectra of SN~1998S and includes most of the same narrow, IIn-like emission features. Following classification, \sn{} evolved into a normal SN~II-P/L with an extended and luminous plateau light curve phase and strong P-Cygni H$\alpha$ emission in its spectra. 

\item Early-time spectroscopic observations of \sn{} revealed prominent narrow emission lines from the photo-ionization of dense CSM shed in enhanced mass loss episodes in the final months before explosion. 

\item The nebular spectrum of \sn{} is compatible with a 10-12~$\Msun$ ZAMS mass RSG star. The weak [\ion{O}{i}] $\lambda$6300 line flux robustly rejects a higher mass progenitor.

\item Early-time ($\delta t < 10$~days)  \emph{Swift}-XRT non-detections in \sn{} suggest complete absorption of thermal bremsstrahlung X-ray emission by the most local CSM. At larger radii of $r \approx (1-5) \times 10^{15}$~cm, X-ray limits indicate a low density medium ($\rho \lesssim (4 - 0.2) \times 10^{-15}$~g cm$^{-3}$, respectively) incapable of producing detectable X-ray emission. For more distant CSM at $r = (2-8) \times 10^{16}$~cm, radio non-detections reveal a limit on the progenitor mass loss rate of $\dot M<1.3 \times 10^{-5}\,\rm{M_{\sun}\,yr^{-1}}$. 

\item Light curve and spectral modeling with \cmfgen\ supports an extended progenitor star at the time of explosion with radius $R_{\star} \approx 1100~\Rsun$, a mass loss rate of $\dot{M} = 0.01~\Msun$~yr$^{-1}$ ($v_{w} = 50~\kms$) resulting in dense CSM confined within $r < 10^{15}$~cm. Because of the pre-SN activity, this large progenitor radius may reflect a phase of inflation or expansion prior to core-collapse, concomitant with the phase of enhanced mass loss.

\item Given the progenitor mass range derived from nebular spectra, it is likely that the enhanced mass loss and precursor emission is the result of instabilities deeply rooted in the stellar interior, most likely associated with the final nuclear burning stages. Energy deposition from either gravity waves generated in neon/oxygen burning stages or a silicon flash in the progenitor's final $\sim$130~days could have ejected stellar material that was then detected in both pre-explosion flux and the early-time SN spectrum. 

\end{itemize}

Based on the novel detection of precursor flux prior to \sn{}, pre-SN emission should be common in SNe~II-P/L and has eluded detection until now simply because it is very faint (i.e., below the detection level of most surveys). This statement is supported by the relatively common presence of bright UV emission that dominates the energy release in SNe~IIP at early times. As Figure \ref{fig:preSN_compare} shows, LSST, with its improved sensitivity, is uniquely equipped to test our hypothesis and detect pre-SN emission at the level of the pre-\sn{} outburst in newly discovered SNe~IIP at $D \lesssim 200$~Mpc. 

\section{Acknowledgements} \label{Sec:ack}

Research at Northwestern University and CIERA is conducted on the stolen land of the Council of Three Fires, the Ojibwe, Potawatomi, and Odawa people, as well as the Menominee, Miami and Ho-Chunk nations. Research at UC Berkeley is conducted on the territory of Huichin, the ancestral and unceded land of the Chochenyo speaking Ohlone people, the successors of the sovereign Verona Band of Alameda County. Keck I/II, ATLAS, and PS1 observations were conducted on the stolen land of the k\={a}naka `\={o}iwi people. We stand in solidarity with the Pu'uhonua o Pu'uhuluhulu Maunakea in their effort to preserve these sacred spaces for native Hawai`ians. MMT observations were conducted on the stolen land of the Tohono O'odham and Hia-Ced O'odham nations; the Ak-Chin Indian Community, and Hohokam people. ZTF observations were conducted on the stolen land of the Pauma and Cupe\~{n}o tribes; the Kumeyaay Nation and the Pay\'{o}mkawichum (Luise\~{n}o) people. Shane 3-m observations were conducted on the stolen land of the Ohlone (Costanoans), Tamyen and Muwekma Ohlone tribes. VLA observations were conducted on the stolen land of the Chiricahua and Mescalero Apache tribes, and the Pueblo people.

The Young Supernova Experiment and its research infrastructure is supported by the European Research Council under the European Union's Horizon 2020 research and innovation programme (ERC Grant Agreement No.\ 101002652, PI K.\ Mandel), the Heising-Simons Foundation (2018-0913, PI R.\ Foley; 2018-0911, PI R.\ Margutti), NASA (NNG17PX03C, PI R.\ Foley), NSF (AST-1720756, AST-1815935, PI R.\ Foley; AST-1909796, AST-1944985, PI R.\ Margutti), the David \& Lucille Packard Foundation (PI R.\ Foley), VILLUM FONDEN (project number 16599, PI J.\ Hjorth), and the Center for AstroPhysical Surveys (CAPS) at the National Center for Supercomputing Applications (NCSA) and the University of Illinois Urbana-Champaign.

We thank Jim Fuller and Samantha Wu for stimulating discussion and RSG models. W.J-G is supported by the National Science Foundation Graduate Research Fellowship Program under Grant No.~DGE-1842165 and the IDEAS Fellowship Program at Northwestern University. W.J-G acknowledges support through NASA grants in support of {\it Hubble Space Telescope} programs GO-16075 and GO-16500. This research was supported in part by the National Science Foundation under Grant No. NSF PHY-1748958.  The Margutti team at UC Berkeley and Northwestern is supported in part by the National Science Foundation under Grant No. AST-1909796 and AST-1944985, by NASA through Award Number 80NSSC20K1575 and by the Heising-Simons Foundation under grant \# 2018-0911 (PI: Margutti). Raffaella Margutti is a CIFAR Azrieli Global Scholar in the Gravity \& the Extreme Universe Program 2019, and a Sloan Fellow in Physics, 2019. 

This work was granted access to the HPC resources of CINES under the allocation 2019 -- A0070410554 and 2020 -- A0090410554 made by GENCI, France. 

MRD acknowledges support from the NSERC through grant RGPIN-2019-06186, the Canada Research Chairs Program, the Canadian Institute for Advanced Research (CIFAR), and the Dunlap Institute at the University of Toronto.

D. A. Coulter acknowledges support from the National Science Foundation Graduate Research Fellowship under Grant DGE1339067. Q.W. acknowledges financial support provided by the STScI Director's Discretionary Fund. M.~R.~S. is supported by the National Science Foundation Graduate Research Fellowship Program Under Grant No. 1842400. A.G. is supported by the National Science Foundation Graduate Research Fellowship Program under Grant No.~DGE–1746047. A.G. also acknowledges funding from the Center for Astrophysical Surveys Fellowship at UIUC/NCSA and the Illinois Distinguished Fellowship. D.O.J is supported by NASA through the NASA Hubble Fellowship grant HF2-51462.001 awarded by the Space Telescope Science Institute, which is operated by the Association of Universities for Research in Astronomy, Inc., for NASA, under contract NAS5-26555. This work was supported by a VILLUM FONDEN Young Investigator Grant to C.G. (project number 25501).

Parts of this research were supported by the Australian Research Council Centre of Excellence for All Sky Astrophysics in 3 Dimensions (ASTRO 3D), through project number CE170100013. This work was supported by a VILLUM FONDEN Investigator grant to J.H. (project number 16599).

The ZTF forced-photometry service was funded under the Heising-Simons Foundation grant
\#12540303 (PI: Graham).

IRAF is distributed by NOAO, which is operated by AURA, Inc., under cooperative agreement with the National Science Foundation (NSF).

The UCSC team is supported in part by NASA grant 80NSSC20K0953, NSF grant AST-1815935, the Gordon \& Betty Moore Foundation, the Heising-Simons Foundation, and by a fellowship from the David and Lucile Packard Foundation to R.J.F.

Some of the data presented herein were obtained at the W. M. Keck Observatory, which is operated as a scientific partnership among the California Institute of Technology, the University of California, and NASA. The Observatory was made possible by the generous financial support of the W. M. Keck Foundation. The authors wish to recognize and acknowledge the very significant cultural role and reverence that the summit of Maunakea has always had within the indigenous Hawaiian community. We are most fortunate to have the opportunity to conduct observations from this mountain. We recognize the destructive history of colonialism endured by native Hawaiians as we strive to hear the voice of those whose sacred land we continue to utilize for scientific gain.

A major upgrade of the Kast spectrograph on the Shane 3~m telescope at Lick Observatory was made possible through generous gifts from the Heising-Simons Foundation as well as William and Marina Kast. Research at Lick Observatory is partially supported by a generous gift from Google.

The National Radio Astronomy Observatory is a facility of the National Science Foundation operated under cooperative agreement by Associated Universities, Inc.

This work makes use of observations from the Las Cumbres Observatory global telescope network following the approved NOIRLab programs 2020B-0250 and 2021A-0239. Las Cumbres Observatory telescope time was granted by NOIRLab through the Mid-Scale Innovations Program (MSIP). MSIP is funded by NSF.

W. M. Keck Observatory 
access was supported by Northwestern University and the Center for Interdisciplinary Exploration and Research in Astrophysics (CIERA).

Based in part on observations obtained with the Samuel Oschin 48-inch Telescope at the Palomar Observatory as part of the Zwicky Transient Facility project. ZTF is supported by the NSF under grant AST-1440341 and a collaboration including Caltech, IPAC, the Weizmann Institute for Science, the Oskar Klein Center at Stockholm University, the University of Maryland, the University of Washington, Deutsches Elektronen-Synchrotron and Humboldt University, Los Alamos National Laboratories, the TANGO Consortium of Taiwan, the University of Wisconsin at Milwaukee, and the Lawrence Berkeley National Laboratory. Operations are conducted by the Caltech Optical Observatories (COO), the Infrared Processing and Analysis Center (IPAC), and the University of Washington (UW).

This work has made use of data from the Asteroid Terrestrial-impact Last Alert System (ATLAS) project. The Asteroid Terrestrial-impact Last Alert System (ATLAS) project is primarily funded to search for near earth asteroids through NASA grants NN12AR55G, 80NSSC18K0284, and 80NSSC18K1575; byproducts of the NEO search include images and catalogs from the survey area. This work was partially funded by Kepler/K2 grant J1944/80NSSC19K0112 and HST GO-15889, and STFC grants ST/T000198/1 and ST/S006109/1. The ATLAS science products have been made possible through the contributions of the University of Hawaii Institute for Astronomy, the Queen’s University Belfast, the Space Telescope Science Institute, the South African Astronomical Observatory, and The Millennium Institute of Astrophysics (MAS), Chile.

The Pan-STARRS1 Surveys (PS1) and the PS1 public science archive have been made possible through contributions by the Institute for Astronomy, the University of Hawaii, the Pan-STARRS Project Office, the Max-Planck Society and its participating institutes, the Max Planck Institute for Astronomy, Heidelberg and the Max Planck Institute for Extraterrestrial Physics, Garching, The Johns Hopkins University, Durham University, the University of Edinburgh, the Queen's University Belfast, the Harvard-Smithsonian Center for Astrophysics, the Las Cumbres Observatory Global Telescope Network Incorporated, the National Central University of Taiwan, STScI, NASA under grant NNX08AR22G issued through the Planetary Science Division of the NASA Science Mission Directorate, NSF grant AST-1238877, the University of Maryland, Eotvos Lorand University (ELTE), the Los Alamos National Laboratory, and the Gordon and Betty Moore Foundation.

This publication has made use of data collected at Lulin Observatory, partly supported by MoST grant 108-2112-M-008-001.

\facilities{Neil Gehrels \emph{Swift} Observatory, VLA, Zwicky Transient Facility, ATLAS, YSE/PS1, Las Cumbres Observatory, Lulin Observatory, Shane (Kast), MMT (Binospec), Keck I/II (LRIS/DEIMOS/NIRES)}

\software{IRAF (Tody 1986, Tody 1993),  photpipe \citep{Rest+05}, DoPhot \citep{Schechter+93}, HOTPANTS \citep{becker15}, HEAsoft (v6.22; HEASARC 2014), CMFGEN \citep{Hillier96}, HERACLES \citep{gonzalez_heracles_07, vaytet_mg_11}, SWarp \citep{swarp}, CASA (v6.1.2; \citealt{McMullin07}), solve-field \citep{lang10}}

\bibliographystyle{aasjournal} 
\bibliography{references} 


\clearpage
\appendix

\renewcommand\thetable{A\arabic{table}} 
\setcounter{table}{0}

\begin{deluxetable*}{cccccc}[h!]
\tablecaption{Optical Spectroscopy of SN~2020tlf \label{tab:spec_table}}
\tablecolumns{5}
\tablewidth{0.45\textwidth}
\tablehead{
\colhead{UT Date} & \colhead{MJD} &
\colhead{Phase\tablenotemark{a}} &
\colhead{Telescope} & \colhead{Instrument} & \colhead{Wavelength Range}\\
\colhead{} & \colhead{} & \colhead{(days)} & \colhead{} & \colhead{} & \colhead{(\AA)}
}
\startdata
2020-09-17 & 59109.0 & $-8.6$ & Keck I & LRIS & 3200--10800 \\
2020-09-17  & 59109.1 & $-8.5$ &  APO 3.5m & DIS &  3650--9830 \\
2020-12-11 & 59194.0 & $+76.4$ & Keck I & LRIS & 3200--10800 \\
2021-01-11 & 59225.5 & $+107.9$ & Shane & Kast & 4000--9200 \\
2021-01-31 & 59245.0 & $+127.4$ & Keck II & NIRES & 9500--24500 \\
2021-02-06 & 59251.5 & $+133.9$ & Shane & Kast & 4000--9200 \\
2021-04-09 & 59313.0 & $+195.4$ & MMT & Binospec & 4000--9200 \\
2021-05-09 & 59343.0 & $+225.4$ & MMT & Binospec & 4000--9200 \\
2021-06-10 & 59375.0 & $+257.4$ & Keck II & DEIMOS & 3400--10200 \\
\enddata
\tablenotetext{a}{Relative to $B$-band maximum (MJD 59117.6)}
\end{deluxetable*}

\begin{table*}
\centering
    \caption{VLA radio observations of SN~2020tlf (Project SD1096, PI Margutti).}
    \label{Tab:radio}
    \begin{tabular}{ccccc}
    \hline
    \hline
  Start Date & Phase\footnote{Relative to $B-$band maximum (MJD 59117.6)} & Frequency & Bandwidth & Flux Density\footnote{\label{Tab_radio_1}Upper-limits are quoted at $3\sigma$.} \\
(UT) & (days) & (GHz) & (GHz) & ($\mu$Jy/beam)\\
    \hline
21-Feb-19 12:11:48UT & +146.4 & 10 & 4.096  & $\leq12$\\
21-May-12 03:27:16UT & +228.7 & 10 & 4.096 &  $\leq72$\footnote{There was significant contribution from the host, as the VLA was in D-configuration. The quoted upper-limit is flux density in a synthesized beam centered at the optical position of \sn{} plus 3 times the RMS.}\\
21-Aug-12 02:11:59UT & +320.7 & 10 & 4.096 & $\leq42$\footnote{Briggs weighting with a robust parameter of -2 was used to minimize the host contribution.}\\
\hline
\end{tabular}
\end{table*}

\begin{deluxetable*}{cccccccc}
\tablecaption{\cmfgen\ Models \label{tbl:m_table}}
\tablecolumns{8}
\tablewidth{\textwidth}
\tablehead{
\colhead{Model Name} & \colhead{ZAMS Mass} & \colhead{Radius} & \colhead{$\dot M$} & \colhead{$M_{\rm CSM}$} & \colhead{$r_{\rm CSM}$} & \colhead{$M({}^{56}\textrm{Ni})$} & \colhead{Gray/Multi-Group}\\
\colhead{} & \colhead{($\Msun$)} & \colhead{($\Rsun$)} & \colhead{($\Msun$ yr$^{-1}$)} & \colhead{($\Msun$)} & \colhead{(cm)} & \colhead{($\Msun$)} & \colhead{}
}
\startdata
 R500\_MD1em2\_r6e15\_MG & $15$ & $501$ & $10^{-2}$ & $0.052$ & $6 \times 10^{15}$ & -- & Multi-Group \\
R500\_MD1em2\_r6e15\_MG\_Ni & $15$ & $501$ & $10^{-2}$ & $0.052$ & $6 \times 10^{15}$ & 0.02 & Multi-Group \\
R500\_MD1em2\_r6e15\_G\_Ni & $15$ & $501$ & $10^{-2}$ & $0.052$ & $6 \times 10^{15}$ & 0.02 & Gray \\
R500\_MD1em2\_r8e15\_MG\_Ni & $15$ & $501$ & $10^{-2}$ & $0.073$ & $8 \times 10^{15}$ & 0.02 & Multi-Group \\
R700\_MD1em2\_r8e15\_MG\_Ni & $15$ & $768$ & $10^{-2}$ & $0.073$ & $8 \times 10^{15}$ & 0.02 & Multi-Group \\
R1100\_MD1em2\_r8e15\_MG\_Ni & $15$ & $1107$ & $10^{-2}$ & $0.073$ & $8 \times 10^{15}$ & 0.02 & Multi-Group \\
\enddata
\tablecomments{Most consistent model to observations includes mass loss of $\dot{M} = 0.01~\Msun$~yr$^{-1}$ and progenitor radius of $R_{\star} = 1107~\Rsun$. Distinction between multi-group and gray variant solvers is discussed in \S\ref{sec:modeling}. }
\end{deluxetable*}

\begin{deluxetable}{cccccc}[h!]
\tablecaption{Optical Photometry of SN~2020tlf \label{tbl:phot_table_s}}
\tablecolumns{6}
\tablewidth{0.45\textwidth}
\tablehead{
\colhead{MJD} &
\colhead{Phase\tablenotemark{a}} &
\colhead{Filter} & \colhead{Magnitude} & \colhead{Uncertainty} & \colhead{Instrument}
}
\startdata
59104.24 & -13.36 & $g$ & 17.84 & 0.01 & PS1 \\
59220.66 & +103.06 & $g$ & 17.45 & 0.01 & PS1 \\
59226.63 & +109.03 & $g$ & 18.08 & 0.02 & PS1 \\
59236.66 & +119.06 & $g$ & 19.88 & 0.09 & PS1 \\
59257.57 & +139.97 & $g$ & 20.52 & 0.10 & PS1 \\
59261.63 & +144.03 & $g$ & 20.87 & 0.13 & PS1 \\
59313.53 & +195.93 & $g$ & 20.94 & 0.15 & PS1 \\
58980.38 & -137.22 & $r$ & 21.78 & 0.32 & PS1 \\
59097.24 & -20.36 & $r$ & 21.41 & 0.20 & PS1 \\
59100.24 & -17.36 & $r$ & 19.47 & 0.04 & PS1 \\
59226.63 & +109.03 & $r$ & 17.13 & 0.01 & PS1 \\
59242.64 & +125.04 & $r$ & 19.24 & 0.10 & PS1 \\
59244.62 & +127.02 & $r$ & 18.93 & 0.08 & PS1 \\
59245.66 & +128.06 & $r$ & 19.23 & 0.06 & PS1 \\
59246.64 & +129.04 & $r$ & 19.17 & 0.06 & PS1 \\
59261.63 & +144.03 & $r$ & 19.29 & 0.03 & PS1 \\
59270.58 & +152.98 & $r$ & 19.34 & 0.09 & PS1 \\
59298.51 & +180.91 & $r$ & 19.72 & 0.07 & PS1 \\
59304.51 & +186.91 & $r$ & 19.90 & 0.11 & PS1 \\
59313.53 & +195.93 & $r$ & 19.91 & 0.08 & PS1 \\
59328.46 & +210.86 & $r$ & 20.09 & 0.19 & PS1 \\
59331.40 & +213.80 & $r$ & 20.57 & 0.25 & PS1 \\
59333.43 & +215.83 & $r$ & 19.94 & 0.07 & PS1 \\
58971.42 & -146.18 & $i$ & 21.79 & 0.35 & PS1 \\
58980.37 & -137.23 & $i$ & 21.87 & 0.34 & PS1 \\
58982.46 & -135.14 & $i$ & 21.45 & 0.29 & PS1 \\
59014.42 & -103.18 & $i$ & 21.50 & 0.29 & PS1 \\
59026.29 & -91.31 & $i$ & 21.08 & 0.22 & PS1 \\
59037.27 & -80.33 & $i$ & 21.27 & 0.16 & PS1 \\
59042.29 & -75.31 & $i$ & 21.27 & 0.17 & PS1 \\
59051.28 & -66.32 & $i$ & 21.28 & 0.14 & PS1 \\
59061.28 & -56.32 & $i$ & 21.53 & 0.23 & PS1 \\
59074.27 & -43.33 & $i$ & 21.59 & 0.27 & PS1 \\
59095.25 & -22.35 & $i$ & 21.70 & 0.32 & PS1 \\
59236.66 & +119.06 & $i$ & 19.34 & 0.06 & PS1 \\
59242.64 & +125.04 & $i$ & 19.51 & 0.11 & PS1 \\
59245.66 & +128.06 & $i$ & 19.38 & 0.06 & PS1 \\
59257.57 & +139.97 & $i$ & 19.52 & 0.04 & PS1 \\
59270.58 & +152.98 & $i$ & 19.73 & 0.11 & PS1 \\
59303.53 & +185.93 & $i$ & 20.05 & 0.13 & PS1 \\
59317.49 & +199.89 & $i$ & 19.99 & 0.09 & PS1 \\
59326.40 & +208.80 & $i$ & 20.44 & 0.21 & PS1 \\
59351.37 & +233.77 & $i$ & 21.07 & 0.29 & PS1 \\
59391.30 & +273.70 & $i$ & 21.19 & 0.19 & PS1 \\
59398.30 & +280.70 & $i$ & 21.04 & 0.29 & PS1 \\
58974.42 & -143.18 & $z$ & 20.70 & 0.34 & PS1 \\
58984.29 & -133.31 & $z$ & 21.28 & 0.30 & PS1 \\
59009.36 & -108.24 & $z$ & 20.87 & 0.28 & PS1 \\
59033.28 & -84.32 & $z$ & 21.37 & 0.35 & PS1 \\
59040.27 & -77.33 & $z$ & 21.10 & 0.26 & PS1 \\
\enddata
\tablenotetext{a}{Relative to $B$-band maximum (MJD 59117.6)}
\end{deluxetable}

\begin{deluxetable}{cccccc}[h!]
\tablecaption{Optical Photometry of SN~2020tlf \label{tbl:phot_table}}
\tablecolumns{6}
\tablewidth{0.45\textwidth}
\tablehead{
\colhead{MJD} &
\colhead{Phase\tablenotemark{a}} &
\colhead{Filter} & \colhead{Magnitude} & \colhead{Uncertainty} & \colhead{Instrument}
}
\startdata
59097.24 & -20.36 & $z$ & 20.66 & 0.20 & PS1 \\
59100.24 & -17.36 & $z$ & 19.12 & 0.06 & PS1 \\
59104.24 & -13.36 & $z$ & 17.91 & 0.02 & PS1 \\
59220.66 & +103.06 & $z$ & 16.48 & 0.01 & PS1 \\
59246.65 & +129.05 & $z$ & 18.75 & 0.05 & PS1 \\
59298.50 & +180.90 & $z$ & 19.43 & 0.14 & PS1 \\
59304.51 & +186.91 & $z$ & 19.48 & 0.07 & PS1 \\
59328.46 & +210.86 & $z$ & 20.07 & 0.20 & PS1 \\
59333.43 & +215.83 & $z$ & 19.89 & 0.10 & PS1 \\
59382.27 & +264.67 & $z$ & 20.83 & 0.18 & PS1 \\
59184.53 & +66.93 & $g$ & 16.28 & 0.01 & ZTF \\
59186.53 & +68.93 & $g$ & 16.53 & 0.02 & ZTF \\
59188.55 & +70.95 & $g$ & 16.36 & 0.01 & ZTF \\
59193.55 & +75.95 & $g$ & 16.54 & 0.01 & ZTF \\
59198.55 & +80.95 & $g$ & 16.69 & 0.01 & ZTF \\
59200.48 & +82.88 & $g$ & 16.74 & 0.03 & ZTF \\
59203.55 & +85.95 & $g$ & 16.86 & 0.01 & ZTF \\
59215.57 & +97.97 & $g$ & 17.29 & 0.02 & ZTF \\
59217.55 & +99.95 & $g$ & 17.26 & 0.02 & ZTF \\
59219.45 & +101.85 & $g$ & 17.42 & 0.02 & ZTF \\
59221.52 & +103.92 & $g$ & 17.49 & 0.01 & ZTF \\
59224.53 & +106.93 & $g$ & 17.79 & 0.02 & ZTF \\
59226.53 & +108.93 & $g$ & 18.05 & 0.02 & ZTF \\
59228.51 & +110.91 & $g$ & 18.58 & 0.03 & ZTF \\
59230.50 & +112.90 & $g$ & 19.24 & 0.06 & ZTF \\
59232.53 & +114.93 & $g$ & 19.69 & 0.06 & ZTF \\
59249.45 & +131.85 & $g$ & 20.52 & 0.26 & ZTF \\
59251.44 & +133.84 & $g$ & 20.49 & 0.16 & ZTF \\
59253.53 & +135.93 & $g$ & 20.15 & 0.09 & ZTF \\
59255.39 & +137.79 & $g$ & 20.36 & 0.15 & ZTF \\
59262.51 & +144.91 & $g$ & 20.68 & 0.18 & ZTF \\
59264.41 & +146.81 & $g$ & 20.42 & 0.13 & ZTF \\
59268.50 & +150.90 & $g$ & 20.81 & 0.16 & ZTF \\
59276.45 & +158.85 & $g$ & 20.58 & 0.25 & ZTF \\
59280.39 & +162.79 & $g$ & 20.54 & 0.19 & ZTF \\
59291.50 & +173.90 & $g$ & 20.56 & 0.16 & ZTF \\
59294.26 & +176.66 & $g$ & 20.68 & 0.34 & ZTF \\
59297.31 & +179.71 & $g$ & 20.48 & 0.32 & ZTF \\
59307.33 & +189.73 & $g$ & 20.53 & 0.13 & ZTF \\
59309.33 & +191.73 & $g$ & 20.63 & 0.16 & ZTF \\
59311.37 & +193.77 & $g$ & 21.10 & 0.20 & ZTF \\
59313.29 & +195.69 & $g$ & 21.23 & 0.35 & ZTF \\
59317.29 & +199.69 & $g$ & 20.67 & 0.14 & ZTF \\
59321.37 & +203.77 & $g$ & 20.80 & 0.17 & ZTF \\
59323.30 & +205.70 & $g$ & 20.80 & 0.22 & ZTF \\
59325.37 & +207.77 & $g$ & 21.08 & 0.32 & ZTF \\
59329.35 & +211.75 & $g$ & 20.50 & 0.31 & ZTF \\
59335.29 & +217.69 & $g$ & 20.98 & 0.25 & ZTF \\
59342.36 & +224.76 & $g$ & 20.89 & 0.17 & ZTF \\
59345.37 & +227.77 & $g$ & 20.81 & 0.19 & ZTF \\
\enddata
\tablenotetext{a}{Relative to $B$-band maximum (MJD 59117.6)}
\end{deluxetable}

\begin{deluxetable}{cccccc}[h!]
\tablecaption{Optical Photometry of SN~2020tlf \label{tbl:phot_table}}
\tablecolumns{6}
\tablewidth{0.45\textwidth}
\tablehead{
\colhead{MJD} &
\colhead{Phase\tablenotemark{a}} &
\colhead{Filter} & \colhead{Magnitude} & \colhead{Uncertainty} & \colhead{Instrument}
}
\startdata
59349.24 & +231.64 & $g$ & 21.18 & 0.23 & ZTF \\
59368.20 & +250.60 & $g$ & 21.42 & 0.31 & ZTF \\
59377.32 & +259.72 & $g$ & 20.94 & 0.20 & ZTF \\
58970.39 & -147.21 & $r$ & 21.15 & 0.26 & ZTF \\
58991.23 & -126.37 & $r$ & 21.21 & 0.23 & ZTF \\
59042.23 & -75.37 & $r$ & 20.99 & 0.22 & ZTF \\
59043.24 & -74.36 & $r$ & 21.38 & 0.33 & ZTF \\
59044.22 & -73.38 & $r$ & 21.30 & 0.32 & ZTF \\
59045.20 & -72.40 & $r$ & 21.35 & 0.31 & ZTF \\
59046.20 & -71.40 & $r$ & 20.91 & 0.21 & ZTF \\
59047.22 & -70.38 & $r$ & 21.19 & 0.26 & ZTF \\
59180.55 & +62.95 & $r$ & 15.86 & 0.01 & ZTF \\
59184.52 & +66.92 & $r$ & 15.74 & 0.01 & ZTF \\
59188.52 & +70.92 & $r$ & 15.82 & 0.01 & ZTF \\
59193.53 & +75.93 & $r$ & 15.96 & 0.01 & ZTF \\
59198.46 & +80.86 & $r$ & 16.07 & 0.01 & ZTF \\
59200.54 & +82.94 & $r$ & 16.01 & 0.01 & ZTF \\
59203.51 & +85.91 & $r$ & 16.25 & 0.01 & ZTF \\
59206.52 & +88.92 & $r$ & 16.21 & 0.02 & ZTF \\
59215.53 & +97.93 & $r$ & 16.48 & 0.01 & ZTF \\
59217.49 & +99.89 & $r$ & 16.40 & 0.01 & ZTF \\
59219.51 & +101.91 & $r$ & 16.57 & 0.01 & ZTF \\
59221.55 & +103.95 & $r$ & 16.65 & 0.05 & ZTF \\
59224.51 & +106.91 & $r$ & 16.83 & 0.01 & ZTF \\
59226.49 & +108.89 & $r$ & 17.12 & 0.01 & ZTF \\
59228.49 & +110.89 & $r$ & 17.54 & 0.01 & ZTF \\
59230.53 & +112.93 & $r$ & 18.14 & 0.02 & ZTF \\
59232.51 & +114.91 & $r$ & 18.57 & 0.03 & ZTF \\
59249.49 & +131.89 & $r$ & 19.16 & 0.06 & ZTF \\
59251.47 & +133.87 & $r$ & 19.05 & 0.05 & ZTF \\
59253.50 & +135.90 & $r$ & 19.13 & 0.05 & ZTF \\
59255.43 & +137.83 & $r$ & 19.30 & 0.05 & ZTF \\
59258.35 & +140.75 & $r$ & 19.18 & 0.06 & ZTF \\
59262.50 & +144.90 & $r$ & 19.35 & 0.06 & ZTF \\
59264.52 & +146.92 & $r$ & 19.18 & 0.05 & ZTF \\
59266.39 & +148.79 & $r$ & 19.55 & 0.13 & ZTF \\
59268.47 & +150.87 & $r$ & 19.37 & 0.06 & ZTF \\
59270.34 & +152.74 & $r$ & 19.27 & 0.11 & ZTF \\
59272.45 & +154.85 & $r$ & 19.17 & 0.12 & ZTF \\
59280.43 & +162.83 & $r$ & 19.56 & 0.10 & ZTF \\
59291.37 & +173.77 & $r$ & 19.50 & 0.07 & ZTF \\
59297.41 & +179.81 & $r$ & 19.50 & 0.09 & ZTF \\
59300.41 & +182.81 & $r$ & 19.54 & 0.17 & ZTF \\
59302.40 & +184.80 & $r$ & 19.81 & 0.16 & ZTF \\
59305.35 & +187.75 & $r$ & 19.69 & 0.10 & ZTF \\
59307.35 & +189.75 & $r$ & 19.69 & 0.07 & ZTF \\
59309.40 & +191.80 & $r$ & 19.64 & 0.07 & ZTF \\
59311.33 & +193.73 & $r$ & 19.74 & 0.07 & ZTF \\
59313.35 & +195.75 & $r$ & 19.87 & 0.10 & ZTF \\
59317.41 & +199.81 & $r$ & 19.91 & 0.07 & ZTF \\
\enddata
\tablenotetext{a}{Relative to $B$-band maximum (MJD 59117.6)}
\end{deluxetable}

\begin{deluxetable}{cccccc}[h!]
\tablecaption{Optical Photometry of SN~2020tlf \label{tbl:phot_table}}
\tablecolumns{6}
\tablewidth{0.45\textwidth}
\tablehead{
\colhead{MJD} &
\colhead{Phase\tablenotemark{a}} &
\colhead{Filter} & \colhead{Magnitude} & \colhead{Uncertainty} & \colhead{Instrument}
}
\startdata
59321.41 & +203.81 & $r$ & 19.83 & 0.07 & ZTF \\
59323.35 & +205.75 & $r$ & 19.84 & 0.09 & ZTF \\
59325.31 & +207.71 & $r$ & 19.87 & 0.09 & ZTF \\
59329.29 & +211.69 & $r$ & 19.90 & 0.16 & ZTF \\
59335.33 & +217.73 & $r$ & 19.71 & 0.20 & ZTF \\
59338.39 & +220.79 & $r$ & 20.09 & 0.11 & ZTF \\
59340.31 & +222.71 & $r$ & 19.94 & 0.08 & ZTF \\
59342.41 & +224.81 & $r$ & 20.37 & 0.12 & ZTF \\
59345.32 & +227.72 & $r$ & 20.22 & 0.11 & ZTF \\
59349.30 & +231.70 & $r$ & 20.14 & 0.10 & ZTF \\
59353.32 & +235.72 & $r$ & 20.17 & 0.12 & ZTF \\
59356.27 & +238.67 & $r$ & 20.62 & 0.27 & ZTF \\
59359.24 & +241.64 & $r$ & 20.20 & 0.23 & ZTF \\
59362.27 & +244.67 & $r$ & 20.68 & 0.27 & ZTF \\
59364.32 & +246.72 & $r$ & 20.46 & 0.17 & ZTF \\
59366.35 & +248.75 & $r$ & 20.19 & 0.12 & ZTF \\
59368.30 & +250.70 & $r$ & 20.31 & 0.11 & ZTF \\
59370.26 & +252.66 & $r$ & 20.74 & 0.18 & ZTF \\
59375.24 & +257.64 & $r$ & 20.38 & 0.14 & ZTF \\
59377.24 & +259.64 & $r$ & 20.45 & 0.13 & ZTF \\
59379.24 & +261.64 & $r$ & 20.53 & 0.17 & ZTF \\
59385.26 & +267.66 & $r$ & 20.74 & 0.29 & ZTF \\
59393.22 & +275.62 & $r$ & 20.82 & 0.18 & ZTF \\
59100.24 & -17.36 & $c$ & 19.89 & 0.20 & ATLAS \\
59108.22 & -9.38 & $c$ & 15.88 & 0.04 & ATLAS \\
59206.64 & +89.04 & $c$ & 16.60 & 0.01 & ATLAS \\
59228.65 & +111.05 & $c$ & 18.22 & 0.04 & ATLAS \\
59238.61 & +121.01 & $c$ & 20.75 & 0.31 & ATLAS \\
59256.56 & +138.96 & $c$ & 20.19 & 0.17 & ATLAS \\
59182.64 & +65.04 & $o$ & 15.79 & 0.01 & ATLAS \\
59184.65 & +67.05 & $o$ & 15.77 & 0.01 & ATLAS \\
59190.65 & +73.05 & $o$ & 15.89 & 0.01 & ATLAS \\
59214.60 & +97.00 & $o$ & 16.42 & 0.01 & ATLAS \\
59222.67 & +105.07 & $o$ & 16.68 & 0.03 & ATLAS \\
59242.63 & +125.03 & $o$ & 19.17 & 0.11 & ATLAS \\
59244.61 & +127.01 & $o$ & 19.28 & 0.20 & ATLAS \\
59246.61 & +129.01 & $o$ & 19.31 & 0.12 & ATLAS \\
59250.65 & +133.05 & $o$ & 19.36 & 0.09 & ATLAS \\
59252.65 & +135.05 & $o$ & 19.34 & 0.11 & ATLAS \\
\enddata
\tablenotetext{a}{Relative to $B$-band maximum (MJD 59117.6)}
\end{deluxetable}

\begin{deluxetable}{cccccc}[h!]
\tablecaption{Optical Photometry of SN~2020tlf \label{tbl:phot_table}}
\tablecolumns{6}
\tablewidth{0.45\textwidth}
\tablehead{
\colhead{MJD} &
\colhead{Phase\tablenotemark{a}} &
\colhead{Filter} & \colhead{Magnitude} & \colhead{Uncertainty} & \colhead{Instrument}
}
\startdata
59113.08 & -3.92 & $u$ & 14.46 & 0.02 & LCO \\
59208.49 & +91.49 & $u$ & 19.52 & 0.21 & LCO \\
59223.42 & +106.42 & $u$ & 20.19 & 0.13 & LCO \\
59227.53 & +110.53 & $u$ & 20.33 & 0.23 & LCO \\
59232.40 & +115.40 & $u$ & 20.48 & 0.24 & LCO \\
59251.48 & +134.48 & $u$ & 20.87 & 0.26 & LCO \\
59113.08 & -3.92 & $g$ & 14.76 & 0.01 & LCO \\
59191.52 & +74.52 & $g$ & 16.56 & 0.03 & LCO \\
59208.49 & +91.49 & $g$ & 16.95 & 0.02 & LCO \\
59217.47 & +100.47 & $g$ & 17.28 & 0.02 & LCO \\
59223.42 & +106.42 & $g$ & 17.45 & 0.03 & LCO \\
59227.54 & +110.54 & $g$ & 17.62 & 0.04 & LCO \\
59232.40 & +115.40 & $g$ & 17.90 & 0.11 & LCO \\
59251.49 & +134.49 & $g$ & 19.41 & 0.05 & LCO \\
59274.46 & +157.46 & $g$ & 20.60 & 0.20 & LCO \\
59276.28 & +159.28 & $g$ & 20.74 & 0.26 & LCO \\
59281.39 & +164.39 & $g$ & 20.89 & 0.24 & LCO \\
59287.33 & +170.33 & $g$ & 20.87 & 0.26 & LCO \\
59302.27 & +185.27 & $g$ & 20.82 & 0.30 & LCO \\
59131.46 & +14.46 & $r$ & 15.09 & 0.01 & LCO \\
59113.08 & -3.92 & $r$ & 15.20 & 0.02 & LCO \\
59191.53 & +74.53 & $r$ & 15.95 & 0.02 & LCO \\
59208.50 & +91.50 & $r$ & 16.30 & 0.02 & LCO \\
59217.48 & +100.48 & $r$ & 16.33 & 0.02 & LCO \\
59223.43 & +106.43 & $r$ & 16.65 & 0.02 & LCO \\
59227.54 & +110.54 & $r$ & 17.29 & 0.03 & LCO \\
59251.49 & +134.49 & $r$ & 19.05 & 0.05 & LCO \\
59274.46 & +157.46 & $r$ & 19.02 & 0.19 & LCO \\
59276.28 & +159.28 & $r$ & 19.33 & 0.15 & LCO \\
59281.39 & +164.39 & $r$ & 19.15 & 0.18 & LCO \\
59287.33 & +170.33 & $r$ & 19.68 & 0.18 & LCO \\
59302.27 & +185.27 & $r$ & 19.68 & 0.18 & LCO \\
59113.08 & -3.92 & $i$ & 15.43 & 0.02 & LCO \\
59191.53 & +74.53 & $i$ & 15.91 & 0.02 & LCO \\
59208.50 & +91.50 & $i$ & 16.34 & 0.03 & LCO \\
59217.48 & +100.48 & $i$ & 16.41 & 0.02 & LCO \\
59223.43 & +106.43 & $i$ & 16.63 & 0.03 & LCO \\
59227.54 & +110.54 & $i$ & 17.32 & 0.04 & LCO \\
59251.49 & +134.49 & $i$ & 19.14 & 0.14 & LCO \\
59276.28 & +159.28 & $i$ & 19.36 & 0.17 & LCO \\
59281.39 & +164.39 & $i$ & 19.62 & 0.19 & LCO \\
59287.33 & +170.33 & $i$ & 19.79 & 0.19 & LCO \\
59302.27 & +185.27 & $i$ & 19.89 & 0.26 & LCO \\
59131.45 & +14.45 & $B$ & 15.23 & 0.01 & Lulin \\
59131.45 & +14.45 & $V$ & 14.96 & 0.01 & Lulin \\
59131.45 & +14.45 & $g$ & 15.07 & 0.01 & Lulin \\
59131.46 & +14.46 & $r$ & 15.09 & 0.01 & Lulin \\
\enddata
\tablenotetext{a}{Relative to $B$-band maximum (MJD 59117.6)}
\end{deluxetable}

\begin{deluxetable}{cccccc}[h!]
\tablecaption{Optical Photometry of SN~2020tlf \label{tbl:phot_table}}
\tablecolumns{6}
\tablewidth{0.45\textwidth}
\tablehead{
\colhead{MJD} &
\colhead{Phase\tablenotemark{a}} &
\colhead{Filter} & \colhead{Magnitude} & \colhead{Uncertainty} & \colhead{Instrument}
}
\startdata
59109.76 & -7.24 & $v$ & 15.52 & 0.08 & \emph{Swift} \\
59111.48 & -5.52 & $v$ & 15.10 & 0.06 & \emph{Swift} \\
59113.41 & -3.59 & $v$ & 14.84 & 0.06 & \emph{Swift} \\
59115.74 & -1.26 & $v$ & 14.63 & 0.05 & \emph{Swift} \\
59117.60 & +0.60 & $v$ & 14.60 & 0.06 & \emph{Swift} \\
59119.20 & +2.20 & $v$ & 14.53 & 0.05 & \emph{Swift} \\
59121.18 & +4.18 & $v$ & 14.65 & 0.07 & \emph{Swift} \\
59129.08 & +12.08 & $v$ & 14.99 & 0.07 & \emph{Swift} \\
59133.60 & +16.60 & $v$ & 15.02 & 0.07 & \emph{Swift} \\
59135.33 & +18.33 & $v$ & 15.01 & 0.07 & \emph{Swift} \\
59138.81 & +21.81 & $v$ & 15.13 & 0.07 & \emph{Swift} \\
59141.24 & +24.24 & $v$ & 15.20 & 0.07 & \emph{Swift} \\
59146.07 & +29.07 & $v$ & 15.25 & 0.08 & \emph{Swift} \\
59153.72 & +36.72 & $v$ & 15.38 & 0.10 & \emph{Swift} \\
59156.24 & +39.24 & $v$ & 15.45 & 0.09 & \emph{Swift} \\
59159.95 & +42.95 & $v$ & 15.66 & 0.10 & \emph{Swift} \\
59162.34 & +45.34 & $v$ & 15.68 & 0.10 & \emph{Swift} \\
59165.19 & +48.19 & $v$ & 15.65 & 0.10 & \emph{Swift} \\
59172.04 & +55.04 & $v$ & 15.66 & 0.10 & \emph{Swift} \\
59176.08 & +59.08 & $v$ & 15.88 & 0.12 & \emph{Swift} \\
59180.13 & +63.13 & $v$ & 15.81 & 0.12 & \emph{Swift} \\
59184.12 & +67.12 & $v$ & 16.13 & 0.15 & \emph{Swift} \\
59188.70 & +71.70 & $v$ & 16.14 & 0.15 & \emph{Swift} \\
59192.76 & +75.76 & $v$ & 16.26 & 0.16 & \emph{Swift} \\
59196.53 & +79.53 & $v$ & 16.04 & 0.16 & \emph{Swift} \\
59198.79 & +81.79 & $v$ & 16.29 & 0.19 & \emph{Swift} \\
59208.81 & +91.81 & $v$ & 16.82 & 0.24 & \emph{Swift} \\
59216.11 & +99.11 & $v$ & 16.90 & 0.25 & \emph{Swift} \\
59220.56 & +103.56 & $v$ & 16.96 & 0.26 & \emph{Swift} \\
59224.74 & +107.74 & $v$ & >17.30 & -- & \emph{Swift} \\
59232.85 & +115.85 & $v$ & >17.31 & -- & \emph{Swift} \\
59240.95 & +123.95 & $v$ & >17.36 & -- & \emph{Swift} \\
59246.91 & +129.91 & $v$ & >17.36 & -- & \emph{Swift} \\
59247.79 & +130.79 & $v$ & >17.29 & -- & \emph{Swift} \\
59250.83 & +133.83 & $v$ & >17.40 & -- & \emph{Swift} \\
59263.97 & +146.97 & $v$ & >17.25 & -- & \emph{Swift} \\
59109.75 & -7.25 & $b$ & 15.34 & 0.06 & \emph{Swift} \\
59111.48 & -5.52 & $b$ & 14.89 & 0.05 & \emph{Swift} \\
59113.41 & -3.59 & $b$ & 14.57 & 0.04 & \emph{Swift} \\
59115.74 & -1.26 & $b$ & 14.47 & 0.04 & \emph{Swift} \\
59117.60 & +0.60 & $b$ & 14.47 & 0.04 & \emph{Swift} \\
59119.19 & +2.19 & $b$ & 14.50 & 0.04 & \emph{Swift} \\
59121.18 & +4.18 & $b$ & 14.60 & 0.05 & \emph{Swift} \\
59129.08 & +12.08 & $b$ & 14.95 & 0.05 & \emph{Swift} \\
59133.59 & +16.59 & $b$ & 15.09 & 0.06 & \emph{Swift} \\
59135.33 & +18.33 & $b$ & 15.18 & 0.06 & \emph{Swift} \\
59138.80 & +21.80 & $b$ & 15.36 & 0.06 & \emph{Swift} \\
59141.24 & +24.24 & $b$ & 15.36 & 0.06 & \emph{Swift} \\
59146.07 & +29.07 & $b$ & 15.62 & 0.07 & \emph{Swift} \\
59153.72 & +36.72 & $b$ & 15.88 & 0.09 & \emph{Swift} \\
\enddata
\tablenotetext{a}{Relative to $B$-band maximum (MJD 59117.6)}
\end{deluxetable}

\begin{deluxetable}{cccccc}[h!]
\tablecaption{Optical Photometry of SN~2020tlf \label{tbl:phot_table}}
\tablecolumns{6}
\tablewidth{0.45\textwidth}
\tablehead{
\colhead{MJD} &
\colhead{Phase\tablenotemark{a}} &
\colhead{Filter} & \colhead{Magnitude} & \colhead{Uncertainty} & \colhead{Instrument}
}
\startdata
59156.24 & +39.24 & $b$ & 15.81 & 0.07 & \emph{Swift} \\
59159.94 & +42.94 & $b$ & 16.02 & 0.08 & \emph{Swift} \\
59162.34 & +45.34 & $b$ & 15.99 & 0.08 & \emph{Swift} \\
59165.19 & +48.19 & $b$ & 16.25 & 0.10 & \emph{Swift} \\
59172.03 & +55.03 & $b$ & 16.34 & 0.10 & \emph{Swift} \\
59176.08 & +59.08 & $b$ & 16.45 & 0.12 & \emph{Swift} \\
59180.12 & +63.12 & $b$ & 16.53 & 0.12 & \emph{Swift} \\
59184.11 & +67.11 & $b$ & 16.93 & 0.17 & \emph{Swift} \\
59188.69 & +71.69 & $b$ & 17.04 & 0.19 & \emph{Swift} \\
59192.75 & +75.75 & $b$ & 17.18 & 0.19 & \emph{Swift} \\
59196.53 & +79.53 & $b$ & 17.22 & 0.24 & \emph{Swift} \\
59198.79 & +81.79 & $b$ & 17.71 & 0.35 & \emph{Swift} \\
59208.80 & +91.80 & $b$ & >17.94 & -- & \emph{Swift} \\
59216.11 & +99.11 & $b$ & >17.96 & -- & \emph{Swift} \\
59220.56 & +103.56 & $b$ & >17.99 & -- & \emph{Swift} \\
59224.73 & +107.73 & $b$ & >17.95 & -- & \emph{Swift} \\
59232.85 & +115.85 & $b$ & >17.97 & -- & \emph{Swift} \\
59240.95 & +123.95 & $b$ & >17.97 & -- & \emph{Swift} \\
59246.91 & +129.91 & $b$ & >17.99 & -- & \emph{Swift} \\
59247.79 & +130.79 & $b$ & >17.96 & -- & \emph{Swift} \\
59250.83 & +133.83 & $b$ & >18.03 & -- & \emph{Swift} \\
59253.81 & +136.81 & $b$ & >17.99 & -- & \emph{Swift} \\
59257.74 & +140.74 & $b$ & >18.02 & -- & \emph{Swift} \\
59263.97 & +146.97 & $b$ & >17.89 & -- & \emph{Swift} \\
59109.75 & -7.25 & $u$ & 15.04 & 0.04 & \emph{Swift} \\
59111.48 & -5.52 & $u$ & 14.61 & 0.04 & \emph{Swift} \\
59113.41 & -3.59 & $u$ & 14.39 & 0.04 & \emph{Swift} \\
59115.74 & -1.26 & $u$ & 14.31 & 0.04 & \emph{Swift} \\
59117.60 & +0.60 & $u$ & 14.30 & 0.04 & \emph{Swift} \\
59119.19 & +2.19 & $u$ & 14.37 & 0.04 & \emph{Swift} \\
59121.18 & +4.18 & $u$ & 14.53 & 0.04 & \emph{Swift} \\
59129.08 & +12.08 & $u$ & 15.04 & 0.04 & \emph{Swift} \\
59133.59 & +16.59 & $u$ & 15.35 & 0.05 & \emph{Swift} \\
59135.33 & +18.33 & $u$ & 15.45 & 0.05 & \emph{Swift} \\
59138.80 & +21.80 & $u$ & 15.74 & 0.06 & \emph{Swift} \\
59141.24 & +24.24 & $u$ & 15.89 & 0.06 & \emph{Swift} \\
59146.07 & +29.07 & $u$ & 16.28 & 0.07 & \emph{Swift} \\
59153.72 & +36.72 & $u$ & 16.78 & 0.11 & \emph{Swift} \\
59156.24 & +39.24 & $u$ & 16.97 & 0.11 & \emph{Swift} \\
59159.94 & +42.94 & $u$ & 17.06 & 0.11 & \emph{Swift} \\
59162.34 & +45.34 & $u$ & 17.42 & 0.14 & \emph{Swift} \\
59165.19 & +48.19 & $u$ & 17.65 & 0.18 & \emph{Swift} \\
59172.03 & +55.03 & $u$ & 18.01 & 0.23 & \emph{Swift} \\
59176.08 & +59.08 & $u$ & 17.99 & 0.23 & \emph{Swift} \\
59180.12 & +63.12 & $u$ & 18.12 & 0.26 & \emph{Swift} \\
59184.11 & +67.11 & $u$ & 18.42 & 0.35 & \emph{Swift} \\
59188.69 & +71.69 & $u$ & >18.49 & -- & \emph{Swift} \\
59192.75 & +75.75 & $u$ & >18.58 & -- & \emph{Swift} \\
59196.53 & +79.53 & $u$ & >18.38 & -- & \emph{Swift} \\
59198.79 & +81.79 & $u$ & >18.41 & -- & \emph{Swift} \\
\enddata
\tablenotetext{a}{Relative to $B$-band maximum (MJD 59117.6)}
\end{deluxetable}

\begin{deluxetable}{cccccc}[h!]
\tablecaption{Optical Photometry of SN~2020tlf \label{tbl:phot_table}}
\tablecolumns{6}
\tablewidth{0.45\textwidth}
\tablehead{
\colhead{MJD} &
\colhead{Phase\tablenotemark{a}} &
\colhead{Filter} & \colhead{Magnitude} & \colhead{Uncertainty} & \colhead{Instrument}
}
\startdata
59208.80 & +91.80 & $u$ & >18.56 & -- & \emph{Swift} \\
59216.11 & +99.11 & $u$ & >18.60 & -- & \emph{Swift} \\
59220.55 & +103.55 & $u$ & >18.62 & -- & \emph{Swift} \\
59224.73 & +107.73 & $u$ & >18.61 & -- & \emph{Swift} \\
59228.67 & +111.67 & $u$ & >18.04 & -- & \emph{Swift} \\
59232.85 & +115.85 & $u$ & >18.57 & -- & \emph{Swift} \\
59240.94 & +123.94 & $u$ & >18.63 & -- & \emph{Swift} \\
59246.91 & +129.91 & $u$ & >18.58 & -- & \emph{Swift} \\
59247.79 & +130.79 & $u$ & >18.54 & -- & \emph{Swift} \\
59250.83 & +133.83 & $u$ & >18.64 & -- & \emph{Swift} \\
59253.81 & +136.81 & $u$ & >18.59 & -- & \emph{Swift} \\
59257.73 & +140.73 & $u$ & >18.62 & -- & \emph{Swift} \\
59263.96 & +146.96 & $u$ & >18.49 & -- & \emph{Swift} \\
59109.75 & -7.25 & $w1$ & 15.02 & 0.04 & \emph{Swift} \\
59111.48 & -5.52 & $w1$ & 14.60 & 0.04 & \emph{Swift} \\
59113.41 & -3.59 & $w1$ & 14.38 & 0.04 & \emph{Swift} \\
59115.73 & -1.27 & $w1$ & 14.37 & 0.04 & \emph{Swift} \\
59117.60 & +0.60 & $w1$ & 14.48 & 0.04 & \emph{Swift} \\
59119.19 & +2.19 & $w1$ & 14.66 & 0.04 & \emph{Swift} \\
59121.18 & +4.18 & $w1$ & 14.89 & 0.04 & \emph{Swift} \\
59129.08 & +12.08 & $w1$ & 15.85 & 0.05 & \emph{Swift} \\
59133.59 & +16.59 & $w1$ & 16.41 & 0.06 & \emph{Swift} \\
59135.33 & +18.33 & $w1$ & 16.63 & 0.06 & \emph{Swift} \\
59138.80 & +21.80 & $w1$ & 17.20 & 0.08 & \emph{Swift} \\
59141.24 & +24.24 & $w1$ & 17.51 & 0.10 & \emph{Swift} \\
59146.07 & +29.07 & $w1$ & 17.89 & 0.13 & \emph{Swift} \\
59153.72 & +36.72 & $w1$ & 18.59 & 0.26 & \emph{Swift} \\
59156.24 & +39.24 & $w1$ & >19.20 & -- & \emph{Swift} \\
59159.94 & +42.94 & $w1$ & >19.24 & -- & \emph{Swift} \\
59162.34 & +45.34 & $w1$ & >19.23 & -- & \emph{Swift} \\
59165.18 & +48.18 & $w1$ & >19.22 & -- & \emph{Swift} \\
59172.03 & +55.03 & $w1$ & >19.27 & -- & \emph{Swift} \\
59176.08 & +59.08 & $w1$ & >19.26 & -- & \emph{Swift} \\
59180.12 & +63.12 & $w1$ & >19.21 & -- & \emph{Swift} \\
59184.11 & +67.11 & $w1$ & >19.24 & -- & \emph{Swift} \\
59188.69 & +71.69 & $w1$ & >19.24 & -- & \emph{Swift} \\
59192.75 & +75.75 & $w1$ & >19.31 & -- & \emph{Swift} \\
59196.53 & +79.53 & $w1$ & >19.11 & -- & \emph{Swift} \\
59198.79 & +81.79 & $w1$ & >19.14 & -- & \emph{Swift} \\
59204.96 & +87.96 & $w1$ & >18.93 & -- & \emph{Swift} \\
59208.80 & +91.80 & $w1$ & >19.25 & -- & \emph{Swift} \\
59216.11 & +99.11 & $w1$ & >19.25 & -- & \emph{Swift} \\
59220.55 & +103.55 & $w1$ & >19.27 & -- & \emph{Swift} \\
59224.73 & +107.73 & $w1$ & >19.23 & -- & \emph{Swift} \\
59228.67 & +111.67 & $w1$ & >19.20 & -- & \emph{Swift} \\
59232.85 & +115.85 & $w1$ & >19.27 & -- & \emph{Swift} \\
59240.94 & +123.94 & $w1$ & >19.27 & -- & \emph{Swift} \\
59246.91 & +129.91 & $w1$ & >19.24 & -- & \emph{Swift} \\
59247.79 & +130.79 & $w1$ & >19.21 & -- & \emph{Swift} \\
59250.83 & +133.83 & $w1$ & >19.25 & -- & \emph{Swift} \\
\enddata
\tablenotetext{a}{Relative to $B$-band maximum (MJD 59117.6)}
\end{deluxetable}

\begin{deluxetable}{cccccc}[h!]
\tablecaption{Optical Photometry of SN~2020tlf \label{tbl:phot_table}}
\tablecolumns{6}
\tablewidth{0.45\textwidth}
\tablehead{
\colhead{MJD} &
\colhead{Phase\tablenotemark{a}} &
\colhead{Filter} & \colhead{Magnitude} & \colhead{Uncertainty} & \colhead{Instrument}
}
\startdata
59253.81 & +136.81 & $w1$ & >19.27 & -- & \emph{Swift} \\
59257.73 & +140.73 & $w1$ & >19.29 & -- & \emph{Swift} \\
59263.96 & +146.96 & $w1$ & >19.11 & -- & \emph{Swift} \\
59109.76 & -7.24 & $m2$ & 15.00 & 0.04 & \emph{Swift} \\
59111.49 & -5.51 & $m2$ & 14.58 & 0.04 & \emph{Swift} \\
59113.42 & -3.58 & $m2$ & 14.38 & 0.04 & \emph{Swift} \\
59115.74 & -1.26 & $m2$ & 14.42 & 0.04 & \emph{Swift} \\
59117.60 & +0.60 & $m2$ & 14.60 & 0.04 & \emph{Swift} \\
59119.20 & +2.20 & $m2$ & 14.79 & 0.04 & \emph{Swift} \\
59121.18 & +4.18 & $m2$ & 15.08 & 0.04 & \emph{Swift} \\
59129.08 & +12.08 & $m2$ & 16.34 & 0.05 & \emph{Swift} \\
59133.60 & +16.60 & $m2$ & 17.07 & 0.06 & \emph{Swift} \\
59135.34 & +18.34 & $m2$ & 17.45 & 0.07 & \emph{Swift} \\
59138.81 & +21.81 & $m2$ & 18.05 & 0.11 & \emph{Swift} \\
59141.25 & +24.25 & $m2$ & 19.19 & 0.26 & \emph{Swift} \\
59146.08 & +29.08 & $m2$ & 19.42 & 0.32 & \emph{Swift} \\
59153.72 & +36.72 & $m2$ & >19.54 & -- & \emph{Swift} \\
59156.25 & +39.25 & $m2$ & >19.64 & -- & \emph{Swift} \\
59159.95 & +42.95 & $m2$ & >19.65 & -- & \emph{Swift} \\
59162.34 & +45.34 & $m2$ & >19.24 & -- & \emph{Swift} \\
59165.19 & +48.19 & $m2$ & >19.67 & -- & \emph{Swift} \\
59172.04 & +55.04 & $m2$ & >19.61 & -- & \emph{Swift} \\
59176.09 & +59.09 & $m2$ & >19.66 & -- & \emph{Swift} \\
59180.13 & +63.13 & $m2$ & >19.68 & -- & \emph{Swift} \\
59184.12 & +67.12 & $m2$ & >19.66 & -- & \emph{Swift} \\
59188.70 & +71.70 & $m2$ & >19.64 & -- & \emph{Swift} \\
59192.76 & +75.76 & $m2$ & >19.70 & -- & \emph{Swift} \\
59196.53 & +79.53 & $m2$ & >19.46 & -- & \emph{Swift} \\
59198.79 & +81.79 & $m2$ & >19.66 & -- & \emph{Swift} \\
59208.81 & +91.81 & $m2$ & >19.65 & -- & \emph{Swift} \\
59216.12 & +99.12 & $m2$ & >19.69 & -- & \emph{Swift} \\
59220.56 & +103.56 & $m2$ & >19.71 & -- & \emph{Swift} \\
59224.74 & +107.74 & $m2$ & >19.63 & -- & \emph{Swift} \\
59232.85 & +115.85 & $m2$ & >19.68 & -- & \emph{Swift} \\
59240.95 & +123.95 & $m2$ & >19.72 & -- & \emph{Swift} \\
59246.92 & +129.92 & $m2$ & >19.64 & -- & \emph{Swift} \\
59247.79 & +130.79 & $m2$ & >19.62 & -- & \emph{Swift} \\
59250.83 & +133.83 & $m2$ & >19.63 & -- & \emph{Swift} \\
59263.97 & +146.97 & $m2$ & >19.54 & -- & \emph{Swift} \\
59109.75 & -7.25 & $w2$ & 14.95 & 0.04 & \emph{Swift} \\
59111.48 & -5.52 & $w2$ & 14.55 & 0.04 & \emph{Swift} \\
59113.41 & -3.59 & $w2$ & 14.42 & 0.04 & \emph{Swift} \\
59115.74 & -1.26 & $w2$ & 14.53 & 0.04 & \emph{Swift} \\
59117.60 & +0.60 & $w2$ & 14.77 & 0.04 & \emph{Swift} \\
59119.19 & +2.19 & $w2$ & 15.06 & 0.04 & \emph{Swift} \\
59121.18 & +4.18 & $w2$ & 15.43 & 0.04 & \emph{Swift} \\
59129.08 & +12.08 & $w2$ & 16.77 & 0.06 & \emph{Swift} \\
59133.59 & +16.59 & $w2$ & 17.30 & 0.07 & \emph{Swift} \\
59135.33 & +18.33 & $w2$ & 17.62 & 0.08 & \emph{Swift} \\
59138.81 & +21.81 & $w2$ & 18.27 & 0.12 & \emph{Swift} \\
\enddata
\tablenotetext{a}{Relative to $B$-band maximum (MJD 59117.6)}
\end{deluxetable}

\begin{deluxetable}{cccccc}[h!]
\tablecaption{Optical Photometry of SN~2020tlf \label{tbl:phot_table_e}}
\tablecolumns{6}
\tablewidth{0.45\textwidth}
\tablehead{
\colhead{MJD} &
\colhead{Phase\tablenotemark{a}} &
\colhead{Filter} & \colhead{Magnitude} & \colhead{Uncertainty} & \colhead{Instrument}
}
\startdata
59141.24 & +24.24 & $w2$ & 19.23 & 0.25 & \emph{Swift} \\
59146.07 & +29.07 & $w2$ & 19.43 & 0.30 & \emph{Swift} \\
59153.72 & +36.72 & $w2$ & >19.55 & -- & \emph{Swift} \\
59156.24 & +39.24 & $w2$ & >19.71 & -- & \emph{Swift} \\
59159.94 & +42.94 & $w2$ & >19.74 & -- & \emph{Swift} \\
59162.34 & +45.34 & $w2$ & >19.73 & -- & \emph{Swift} \\
59165.19 & +48.19 & $w2$ & >19.73 & -- & \emph{Swift} \\
59172.03 & +55.03 & $w2$ & >19.75 & -- & \emph{Swift} \\
59176.08 & +59.08 & $w2$ & >19.74 & -- & \emph{Swift} \\
59180.13 & +63.13 & $w2$ & >19.73 & -- & \emph{Swift} \\
59184.12 & +67.12 & $w2$ & >19.71 & -- & \emph{Swift} \\
59188.69 & +71.69 & $w2$ & >19.70 & -- & \emph{Swift} \\
59192.76 & +75.76 & $w2$ & >19.78 & -- & \emph{Swift} \\
59196.53 & +79.53 & $w2$ & >19.52 & -- & \emph{Swift} \\
59198.79 & +81.79 & $w2$ & >19.68 & -- & \emph{Swift} \\
59208.80 & +91.80 & $w2$ & >19.73 & -- & \emph{Swift} \\
59216.11 & +99.11 & $w2$ & >19.73 & -- & \emph{Swift} \\
59220.56 & +103.56 & $w2$ & >19.78 & -- & \emph{Swift} \\
59224.73 & +107.73 & $w2$ & >19.69 & -- & \emph{Swift} \\
59232.85 & +115.85 & $w2$ & >19.77 & -- & \emph{Swift} \\
59240.95 & +123.95 & $w2$ & >19.75 & -- & \emph{Swift} \\
59246.91 & +129.91 & $w2$ & >19.70 & -- & \emph{Swift} \\
59247.79 & +130.79 & $w2$ & >19.69 & -- & \emph{Swift} \\
59250.83 & +133.83 & $w2$ & >19.72 & -- & \emph{Swift} \\
59253.81 & +136.81 & $w2$ & >19.71 & -- & \emph{Swift} \\
59257.74 & +140.74 & $w2$ & >19.31 & -- & \emph{Swift} \\
59263.97 & +146.97 & $w2$ & >19.59 & -- & \emph{Swift} \\
\enddata
\tablenotetext{a}{Relative to $B$-band maximum (MJD 59117.6)}
\end{deluxetable}

\end{document}